\documentclass[twocolumn,usegraphicx]{aastex63}

\usepackage{psfig}   
\usepackage{graphicx}
\usepackage{subfigure}

\shorttitle{Disc photoevaporation locales}
\shortauthors{R. J. Parker et al}

\begin{document}

\title{External photoevaporation of protoplanetary discs: does location matter?}

%\date{}
                             
%\pagerange{\pageref{firstpage}--\pageref{lastpage}} \pubyear{2019}

\correspondingauthor{Richard Parker}
\email{R.Parker@sheffield.ac.uk}

\author[0000-0002-1474-7848]{Richard J. Parker}
\altaffiliation{Royal Society Dorothy Hodgkin Fellow}
\affiliation{Department of Physics and Astronomy, The University of Sheffield, Hicks Building, Hounsfield Road, Sheffield, S3 7RH, UK}
%\nocollaboration{1}
\author{Hayley L. Alcock}
\affiliation{Department of Physics and Astronomy, The University of Sheffield, Hicks Building, Hounsfield Road, Sheffield, S3 7RH, UK}
%\nocollaboration{1}
\author{Rhana B. Nicholson}
\affiliation{Department of Physics and Astronomy, The University of Sheffield, Hicks Building, Hounsfield Road, Sheffield, S3 7RH, UK}
%\nocollaboration{1}
\author[0000-0002-6648-2968]{Olja Pani{\'c}}
\altaffiliation{Royal Society Dorothy Hodgkin Fellow}
\affiliation{School of Physics and Astronomy, E.C. Stoner Building, The University of Leeds, Leeds, LS2 9JT, UK}
%\nocollaboration{1}
\author{Simon P. Goodwin}
\affiliation{Department of Physics and Astronomy, The University of Sheffield, Hicks Building, Hounsfield Road, Sheffield, S3 7RH, UK}
%\nocollaboration{1}

\begin{abstract}
Many theoretical studies have shown that external photoevaporation from massive stars can severely truncate, or destroy altogether, the gaseous protoplanetary discs around young stars. In tandem, several observational studies report a correlation between the mass of a protoplanetary disc and its distance to massive ionising stars in star-forming regions, and cite external photoevaporation by the massive stars as the origin of this correlation. We present $N$-body simulations of the dynamical evolution of star-forming regions and determine the mass-loss in protoplanetary discs from external photoevaporation due to far ultraviolet (FUV) and extreme ultraviolet (EUV) radiation from massive stars. We find that projection effects can be significant, in that low-mass disc-hosting stars that appear close to the ionising sources may be fore- or background stars in the star-forming region. We find very little evidence in our simulations for a trend in increasing disc mass with increasing distance from the massive star(s), even when projection effects are ignored. Furthermore, the dynamical evolution of these young star-forming regions moves stars whose discs have been photoevaporated to far-flung locations, away from the ionising stars, and we suggest that  any correlation between disc mass and distance the ionising star is either coincidental, or due to some process other than external photoevaporation. 
\end{abstract}
%\vspace*{1cm}

%\begin{keywords}   
%methods: numerical -- protoplanetary discs -- photodissociation region (PDR) -- open clusters and associations: general 
%\end{keywords}

\section{Introduction}

Most stars are observed to form in groups where the stellar density significantly exceeds that in the local Solar neighborhood \citep{Lada03,Bressert10}. In these young star-forming regions, stars are frequently observed to have excess flux in the infrared portion of their SEDs \citep{Haisch01,Richert18}, which is interpreted as due to the presence of a protoplanetary disc. Furthermore, such protoplanetary discs have been directly imaged \citep{odell94,Brogan15} and can now have their masses (dust, and sometimes gas) measured \citep[e.g.][]{Mann14,Ansdell17,Eisner18}.

These protoplanetary discs rapidly evolve \citep{Haisch01,Richert18}, with the fraction of young stars in a given star-forming region hosting a disc dropping from around 80\,per cent at 1\,Myr to around 10--20\,per cent after 5\,Myr \citep[with the caveat that ages of young stars are notoriously difficult to determine;][]{Bell13,Soderblom14}. This either implies that planets are being formed rapidly, and/or  a combination of processes are destroying the discs.

Young star-forming regions often host massive stars ($>15$\,M$_\odot$), whose rapid internal evolution produces intense far ultra violet (FUV) and extreme ultra violet (EUV) radiation fields. Several authors {\citep{Storzer99,Hollenbach00,Scally01,Adams04,Fatuzzo08,Haworth18a,Nicholson19a,ConchaRamirez19,Winter19b,Parker21a} have shown that these radiation fields can destroy or reduce the masses of protoplanetary discs by photoevaporating the gas \citep[and to a much lesser extent, dust;][]{Haworth18a}. Massive stars are inherently rare, but for example in  a region containing 1000 low-mass stars one or two stars $>$20\,M$_\odot$ are expected from randomly sampling the stellar Initial Mass Function (IMF). Several authors \citep{Parker07,Maschberger08} have argued that the only limit on the mass of a star that can form is the molecular cloud mass, so occasionally a low-mass star-forming region containing only tens to hundreds of stars could  produce high-mass stars (and observational examples of this appear to exist).

When one or two massive stars are present, they are expected to significantly alter the properties of protoplanetary discs around stars due to photoevaporation within a sphere of influence that usually extends out to $\sim$0.5\,pc \citep[e.g.][]{Scally01}. Many observational studies have measured the masses of protoplanetary discs as a function of projected distance from the photoionising massive star(s). Many of these studies report a positive correlation between the mass of a protoplanetary disc and the projected distance from the massive star, with the interpretation being that the closer discs are more susceptible to destruction from the radiation fields \citep{Guarcello07,Fang12,Guarcello14,Mann14,Mann15,Ansdell17,Eisner18,vanTerwisga19}.

However, observations are two dimensional projections of the inherent three dimensional distribution, and so disc-hosting stars that are in the fore- and background of the star-forming region may appear to be close to the massive star(s) in two dimensions. Furthermore, dense, young star-forming regions are often dynamically old. Both theory \citep[e.g.][]{McMillan07,Girichidis11,Bate12,Kuznetsova15} and observations \citep[e.g][]{Gomez93,Larson95,Cartwright04,Gutermuth09,Buckner19} indicate that star-formation results in substructured stellar distributions in star-forming regions. In these regions, the important dynamical timescale is not related to the size scale of the entire region, but rather that of the clumps of (proto)stars in the filaments and substructure \citep{Allison10}, which are typically of order 0.1\,pc \citep{Andre14} and so can be many (local) crossing times old.  

In star-forming regions with these initial conditions, the influence of external processes \citep[both from direct interactions, e.g.][]{Parker12a,Zheng15,Vincke16} and photoevaporation \citep[e.g.][]{Scally01,Adams04,Parker21a} have a heightened effect on the disruption of planetary systems. A combination of violent and two-body relaxation constantly processes the population of stars in any fixed field of view, so stars that have their discs destroyed or altered may not stay close to the ionising sources, and conversely, disc-bearing stars may move from the outskirts to the centre.

In this paper, we focus on both of these issues. We first discuss the strength of the correlations reported in the observational data (Section~\ref{obs}) before discussing the effects of two dimensional projection on a synthetic star cluster with a population of stars that have been affected by photoevaporation (Section~\ref{projection}). We then describe $N$-body simulations and a post-processing disc photoevaporation analysis (Section~\ref{simulation}), before examining the combined effects of projection issues and stellar dynamics in Section~\ref{results}. We provide a discussion in Section~\ref{discuss} and we conclude in Section~\ref{conclude}.

\section{Observational data}
\label{obs}

We will compare the protoplanetary disc properties of stars in simulated star-forming regions with several observed samples of discs in the Orion Nebula Cluster (\citet{Mann14} and \citet{Eisner18}), the Orion Molecular cloud \citep{vanTerwisga19}, NGC\,2024 \citep{Mann15,vanTerwisga20} and the $\sigma$~Orionis star-forming region \citep{Ansdell17}.

A correlation between the mass of the protoplanetary disc and the projected distance to the most luminous ionising source was reported in the ONC by \citet{Mann14}, and to a lesser extent by \citet{Eisner18}. A similar correlation was reported for $\sigma$~Orionis by \citet{Ansdell17} and \citet{vanTerwisga19} found that the discs closer to the centre of the Trapezium cluster were less massive than discs further from the centre of the Trapezium by a factor of five. Finally, \citet{Mann15} report that there is no correlation of disc mass with distance from the ionising star in NGC2024, which they attribute being due to the weaker radiation field in NGC\,2024 compared to that in the ONC. In these observations, the projected distances range from hundredths of a pc to several pc. Estimates for the ages of these regions vary \citep[e.g.][]{Bell13}, but generally the ONC is thought to be between 1--2\,Myr \citep{DaRio10,Jeffries11,Reggiani11b}, NGC\,2024 is slightly younger \citep[0.5--1\,Myr,][]{Meyer96,Levine06}, and $\sigma$~Orionis is thought to be slightly older \citep[3--5\,Myr,][]{Oliveira04}.

In Fig.~\ref{obs_data} we show the the disc masses as a function of projected distance from the brightest ionising star in each star-forming region. In the datasets of the ONC \citep{Mann14,Eisner18} and the Orion Molecular Cloud \citep{vanTerwisga19}, this corresponds to the distance from the position of $\theta^1$~Ori~C; in the  $\sigma$~Orionis  star-forming region \citep{Ansdell17} this corresponds to the distance from the position of the star $\sigma$~Ori, and in NGC\,2024 \citep{Mann15,vanTerwisga20} this corresponds to the distance from the position of IRS~2b. The disc masses are calculated by taking the reported dust masses from sub-mm continuum observations and assuming a gas-to-dust ratio of 100.

The data from \citet{Mann14} for the ONC are shown by the orange circles in Fig.~\ref{obs_data}, and the data for the ONC from \citet{Eisner18} are shown by the green stars. Data for the Orion Molecular cloud \citep{vanTerwisga19} are shown by the grey diamonds. Data for NGC\,2024 from \citet{Mann15} are shown by the raspberry squares, and data for NGC\,2024 from \citet{vanTerwisga20} are shown by the yellow pentagons. Finally, the data for $\sigma$~Orionis \citep{Ansdell17} are shown by the blue triangles, with the darker points indicating an unambiguous detection of gas in the disc. 

\begin{figure}
\begin{center}
\rotatebox{270}{\includegraphics[scale=0.4]{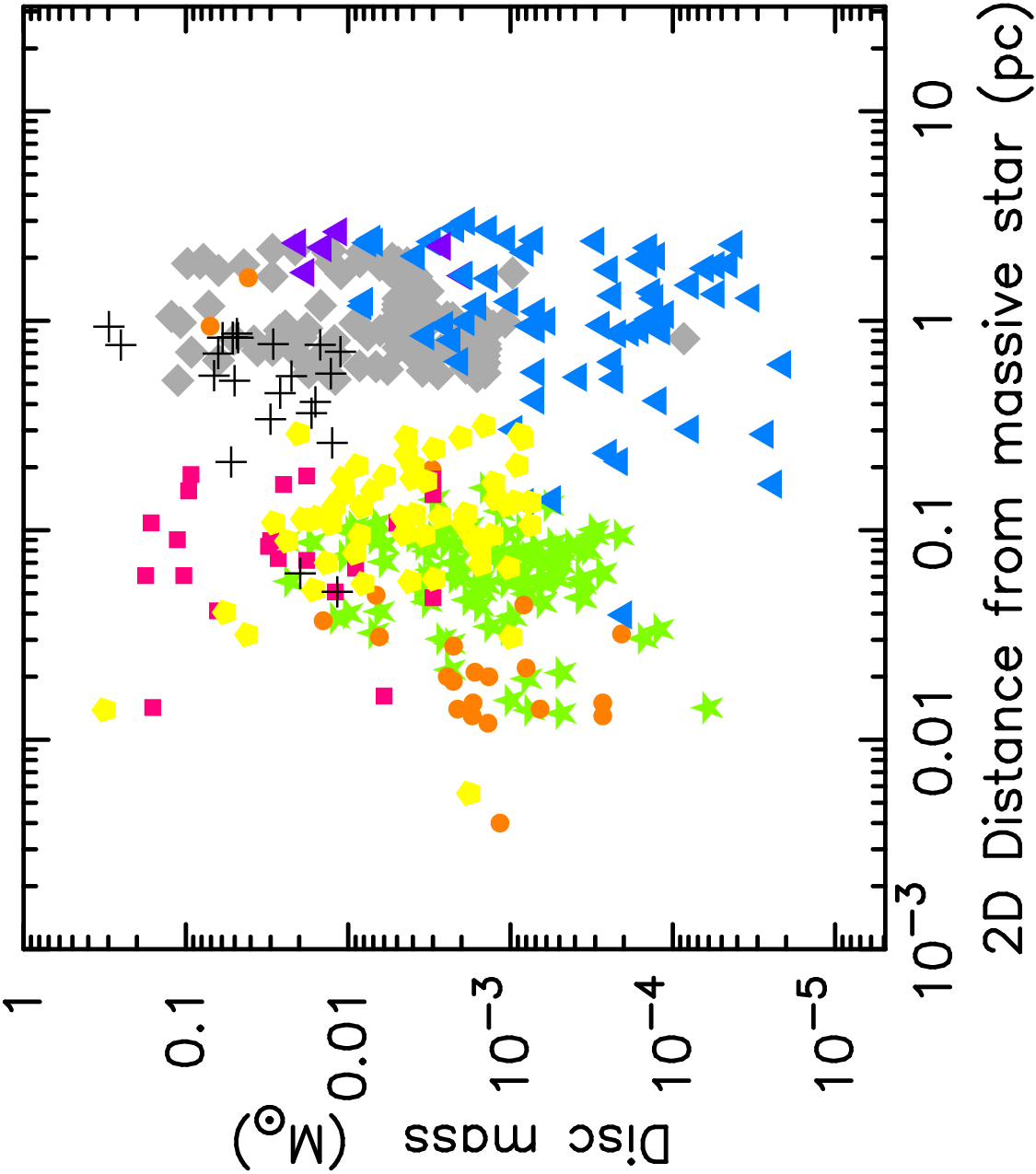}}
\end{center}
\caption{Disc masses as a function of distance from the brightest ionising star in the 1 -- 2\,Myr old ONC (\citealp[][the orange circles]{Mann14}; \citealp[][the green stars]{Eisner18}; and \citealp[][the grey diamonds]{vanTerwisga19}), the brightest ionising source in the 0.5 -- 1\,Myr old NGC\,2024 (\citealp[][the raspberry squares]{Mann15}; and \citealp[][the yellow pentagons]{vanTerwisga20}) and the brightest ionising star in the 3 -- 5\,Myr old $\sigma$~Orionis region \citep[][ the blue triangles, where the darker points indicate unambiguous detection of gas within the discs]{Ansdell17}. The black plus symbols are synthetic data, where the disc masses are 10\,per cent of the host star mass, and the distance to the massive star is randomly drawn from a uniform distribution between 0 and 1\,pc.} 
\label{obs_data}
\end{figure}

\begin{table*}
  \caption{Spearman rank $\rho$ tests for correlations between the disc mass and the distance to ionising stars of star-forming regions. We indicate age of each region, the Spearman rank coefficient $\rho$, as well as the $p$-value for rejecting the null hypothesis that there is no correlation. The final column in the table indicates the symbol shape and colour in Figs.~1, 3, 5--7, 9 and 11--12.}
  \begin{center}
    \begin{tabular}{cccccc}
      \hline
Region & Ref. & Age & Spearman $\rho$ & $p$-value & Symbol in Figs.~1, 3, 5--7, 9, 11--12\\
\hline
ONC & \citet{Mann14} & 1 -- 2\,Myr & 0.55 & $8\times10^{-3}$ & Orange circle\\
ONC & \citet{Eisner18} & 1 -- 2\,Myr & 0.16 & 0.13 & Green star\\
\hline
Orion MC & \citet{vanTerwisga19} & 1 -- 2\,Myr & 0.25 & $4\times10^{-3}$ & Grey diamond\\
\hline 
 ONC \& Orion MC &   \emph{As rows 1--3 above} & 1 -- 2\,Myr &  0.54 & $7\times10^{-20}$ & \emph{As rows 1--3 above} \\
%combined & & & & \\
\hline
NGC\,2024 & \citet{Mann15} & 0.5 -- 1\,Myr & -0.12 & 0.58 & Raspberry square\\
NGC\,2024 & \citet{vanTerwisga20} & 0.5 -- 1\,Myr & -0.20 & 0.12 & Yellow pentagon\\
\hline 
$\sigma$~Ori & \citet{Ansdell17} & 3 -- 5\,Myr & 0.34 & $4\times10^{-3}$ & Blue/purple triangle \\
      \hline
    \end{tabular}
  \end{center}
  \label{observations}
\end{table*}

In order to determine the strength of any correlation in the datasets, we use the Spearman rank coefficient test,  which determines whether a correlation exists between two variables. The Spearman test returns a value $\rho$ between $-1$ and $+1$ and an  associated $p$-value to assess the significance of the correlation (i.e.\,\,the probability of obtaining a correlation with the rank coefficient $\rho$ from random chance, under the null hypothesis that there is no correlation). Positive values for $\rho$ towards unity represent a strong positive correlation, negative values of $\rho$ represent a negative correlation, whereas $\rho$ values around zero indicate no correlation.

We summarise the Spearman coefficient, $\rho$, for all of the datasets in Fig.~\ref{obs_data} in Table~\ref{observations}. Significant ($p$-values less than 0.01) positive correlations are present in the \citet{Mann14} ONC sample and the \citet{vanTerwisga20} data for the extended Orion Molecular Cloud, but not in the \citet{Eisner18} data for the ONC. The three datasets from Orion probe different physical distance ranges from the ionising stars at the centre of the ONC, and for completeness we calculate the Spearman coefficient for all three datasets. If we combine both the ONC and the OMC data (under the assumption that the photoionising star(s) at the centre of the ONC are responsible for the disc evolution across these different distance scales), the Spearman $\rho$ coefficient is 0.54, with an associated $p$-value of $7 \times 10^{-20}$. In NGC\,2024 the Spearman rank coefficients are negative, although not significantly so. In $\sigma$~Ori there appears to be a significant positive correlation.

% ALL ONC  0.538354516       6.90285582E-20

%Only the disc sample in the ONC from \citet{Mann14} displays a significant correlation between disc mass and distance to the ionising star according to the Pearson $r$-coefficient ($r = 0.83$). In contrast, $r = 0.09$ for the disc sample in \citet[][which are also discs in the ONC]{Eisner18} and $r = 0.10$ for the more distant discs in the \citet{vanTerwisga19} data. The $r$-coefficient for all three datasets combined is $r = 0.31$. For NGC\,2024 \citep{Mann15}, $r = -0.12$. The discs in $\sigma$~Ori have $r = 0.37$, which is only a marginal correlation at best. 

% Spearman rank Eisner ONC  0.159551606      0.126601696    
 %Pearson r statistic   9.93220434E-02  0.343530715       9.96506140E-02
 
 %Spearman rank van Terwisga Orion  0.248196378       4.25809622E-03
 %Pearson r statistic  0.101231962      0.249938905      0.101579897    
 
 %Spearman rank Mann ONC  0.551880300       7.74831418E-03
 %Pearson r statistic  0.834236622       1.39920598E-06   1.20191121    
 
 %Spearman rank sigma Ori  0.335069895       4.01384244E-03
 %Pearson r statistic  0.365781277       1.57928921E-03  0.383544028    
 
 %Spearman rank Mann NGC2024 -0.123268455      0.584706545    
 %Pearson r statistic -0.116443560      0.605820179     -0.116974168    
 
% Spearman rank van Terwisga NGC2024 -0.208336040      0.119902909    
 %Pearson r statistic -0.269406348       4.27086875E-02 -0.276223570    

Interestingly, the shapes of the positively correlated distributions are consistent with the idea that the disc masses are proportional to the host star masses. For example, if we draw stellar masses from a standard \citet{Maschberger13} IMF of the form
\begin{equation}
p(m) \propto \left(\frac{m}{\mu}\right)^{-\alpha}\left(1 + \left(\frac{m}{\mu}\right)^{1 - \alpha}\right)^{-\beta},
\label{maschberger_imf}
\end{equation}
and set the disc masses to be 10\,per cent of the host star mass, then we produce distributions such as the one shown by the black plus symbols in Fig.~\ref{obs_data}. (Note that in Equation~\ref{maschberger_imf},  $\mu = 0.2$\,M$_\odot$ is the average stellar mass, $\alpha = 2.3$ is the \citet{Salpeter55} power-law exponent for higher mass stars, and $\beta = 1.4$ describes the slope of the IMF for low-mass objects, and we sample masses between 0.1 -- 50\,M$_\odot$.)
We draw the same number of stars as in the \citet{Mann14} sample, i.e.\,\,22. If all distances from the ionising star are equally probable, the distribution of black points in Fig.~\ref{obs_data} is simply the stellar mass function placed on its side and reduced by a factor of ten. In this particular realisation the Spearman rank coefficient is $\rho = 0.46$, with a $p$-value of $3.2\times10^{-2}$, which although larger than the $p$-values calculated for the observed samples, could imply a modest correlation. In 100 repeats of this experiment, we find positive correlations 7\,per cent of the time, and negative correlations 5\,per cent of the time. We therefore have demonstrated that before considering any photoevaporation effects, any correlation (if it exists -- it is only present in three samples out of seven observed disc populations) could simply be due to low-mass stars (and hence lower-mass discs) being more common and hence more likely to be observed at closer distances to the ionising star(s) than more massive stars (with more massive discs). However, we note that increasing the number of star-disc systems leads to fewer significant correlations (simply because increasing the number of objects causes a reduction in the number of random correlations caused by low-$N$ sampling), suggesting that the reported trends in the observational data may also be hampered by small number statistics.

\section{Projection Effects}
\label{projection}

Observations of protoplanetary disc locations in star-forming regions are usually only available in two dimensions, and so the position of a host star and its protoplanetary disc is projected from three dimensions into two on the sky. This is an important consideration when discussing the occurrence rate of protoplanetary discs and their individual masses as a function of distance from photoionising massive stars.

In this section we determine how important projection effects may be on a synthetic star cluster where we have not modelled any previous dynamical evolution and we have assumed that the cluster is dense enough so that significant photoevaporation has already occurred. We have assumed a simplified geometry, where the cluster has a uniform number density profile from the centre to the outskirts (Fig.~\ref{projection_effects-a}). In this panel, stars that have had their discs affected by photoevaporation are shown by the red crosses. It is immediately obvious from this plot that the two-dimensional projection of fore- and background stars that are not within the photoevaporative zone ($<0.3$\,pc from the massive star(s)) means that some of these objects would mistakenly be classed as being close to the photoionising massive star(s). 

\begin{figure*}
  \begin{center}
\hspace*{-1.0cm}\subfigure[]{\label{projection_effects-a}\rotatebox{270}{\includegraphics[scale=0.25]{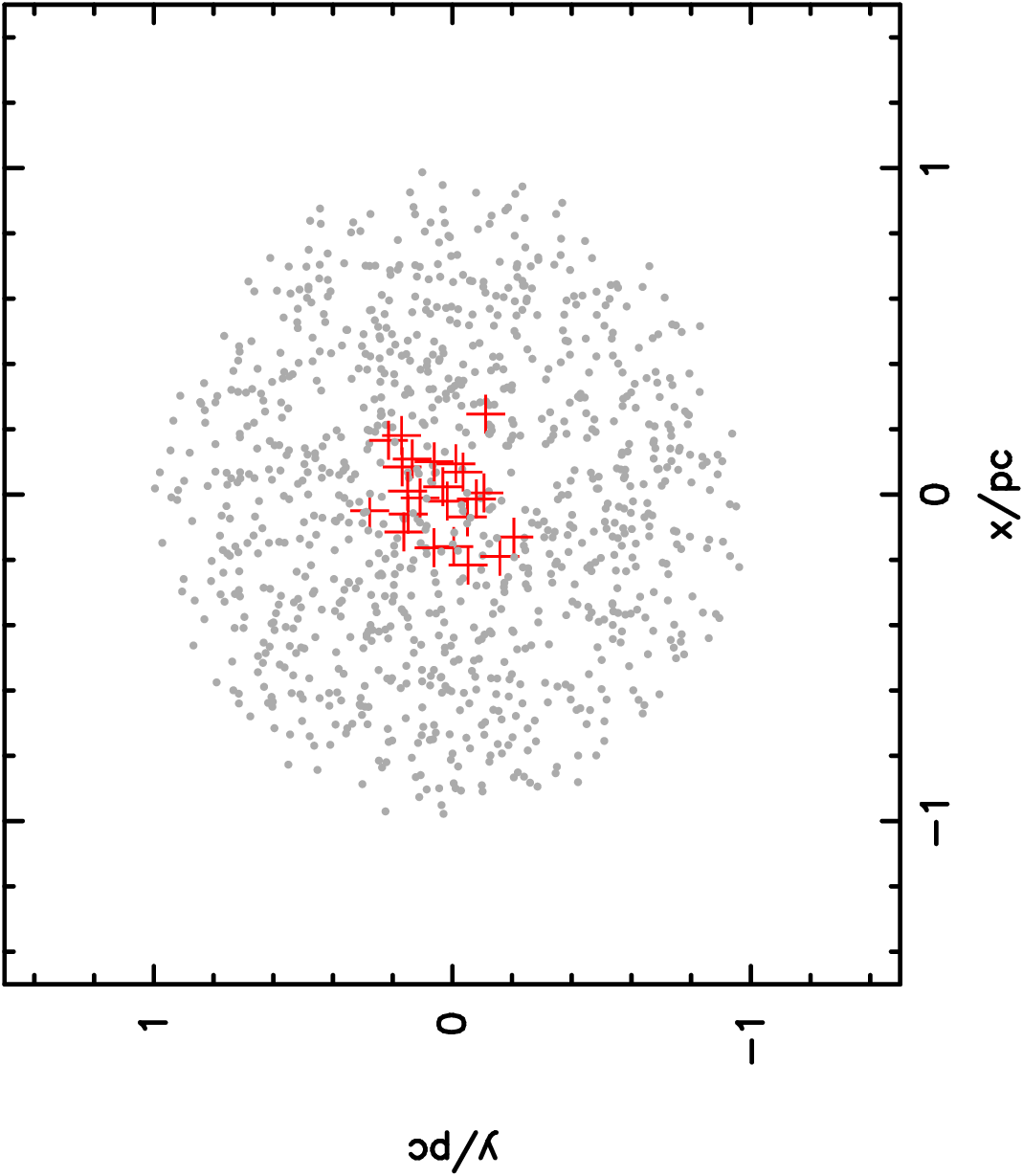}}}
\hspace*{0.3cm}
\subfigure[]{\label{projection_effects-b}\rotatebox{270}{\includegraphics[scale=0.27]{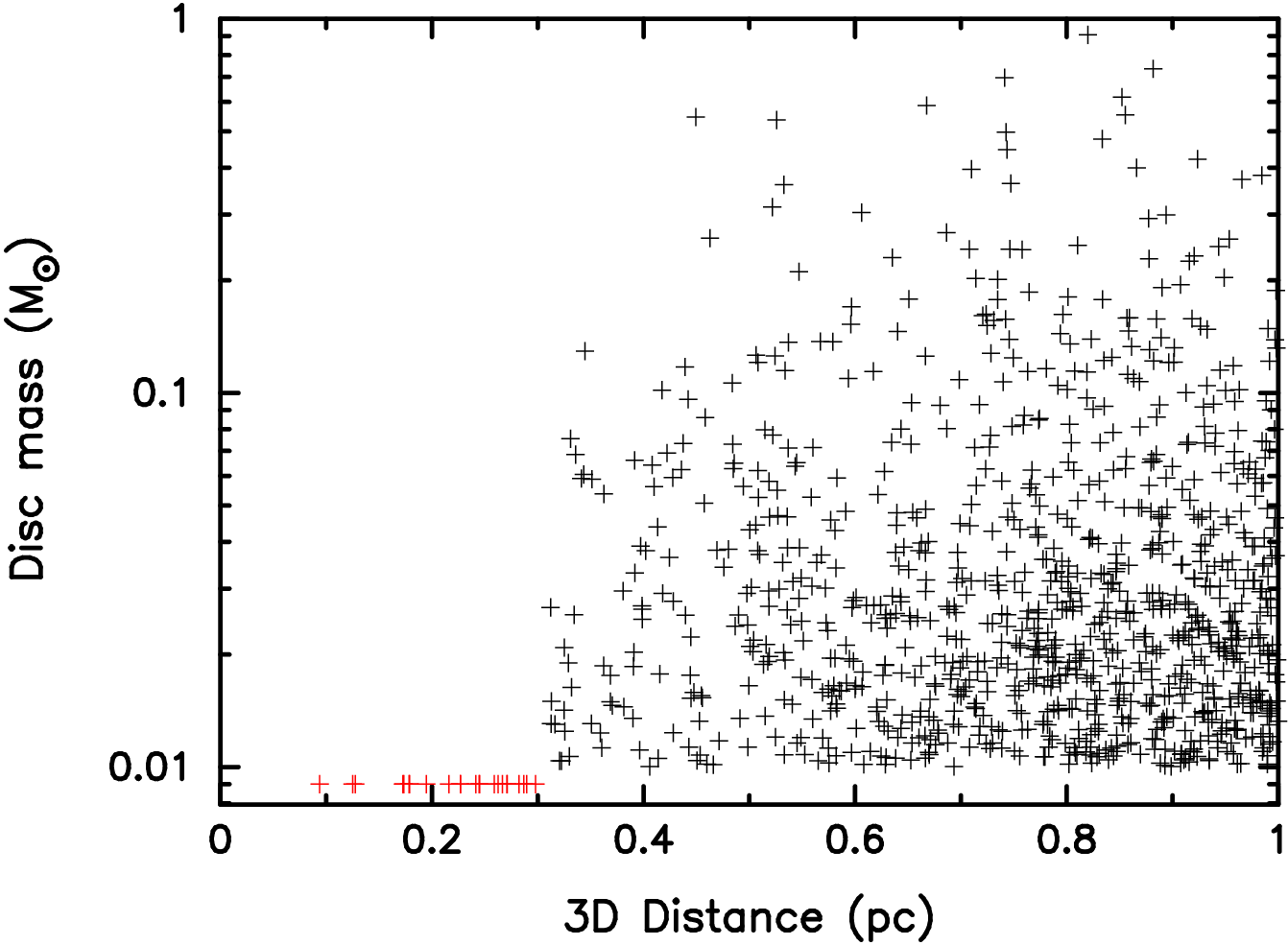}}}
\hspace*{0.3cm}\subfigure[]{\label{projection_effects-c}\rotatebox{270}{\includegraphics[scale=0.27]{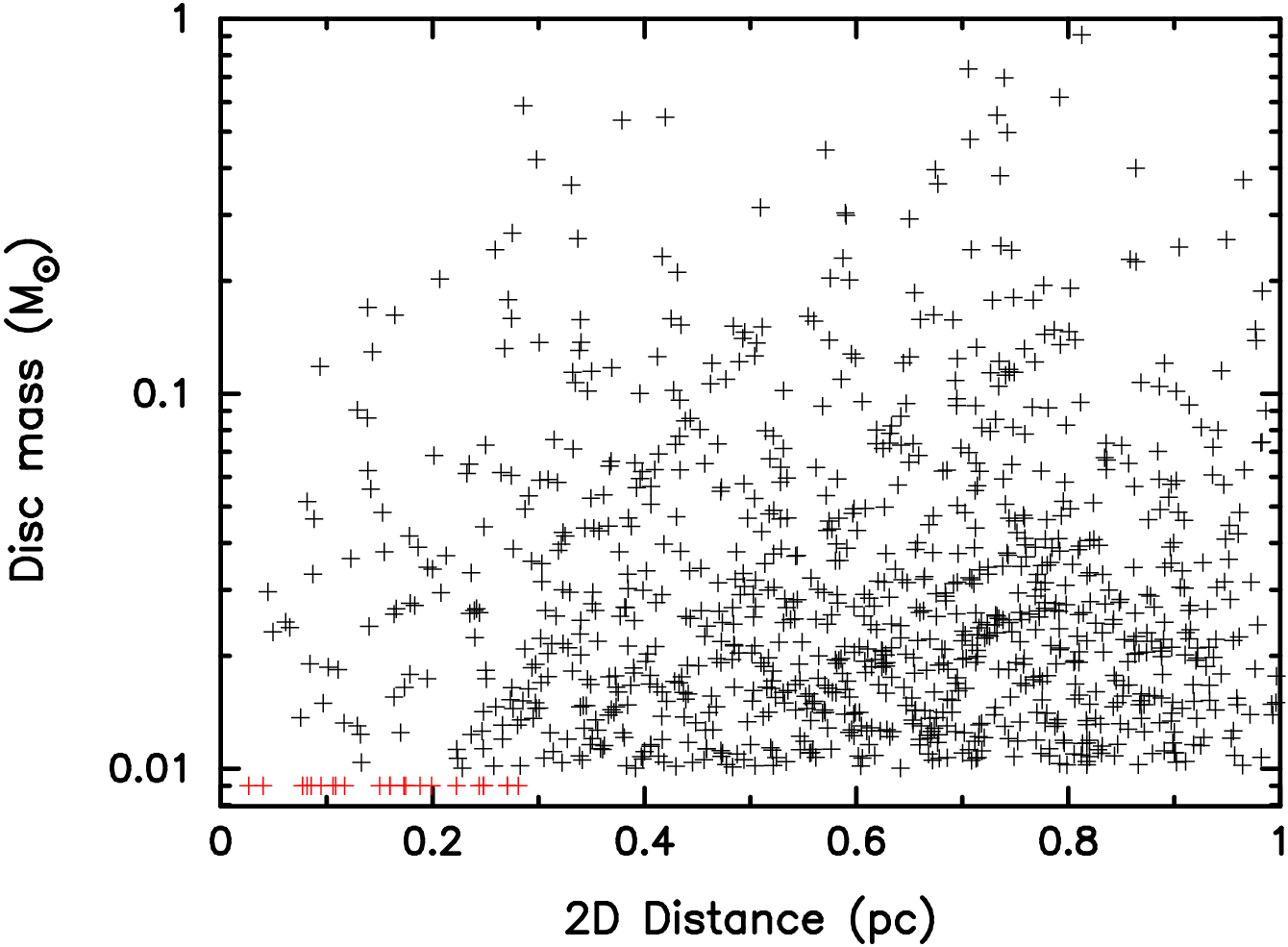}}}  
\hspace*{0.3cm}
\subfigure[]{\label{projection_effects-d}\rotatebox{270}{\includegraphics[scale=0.27]{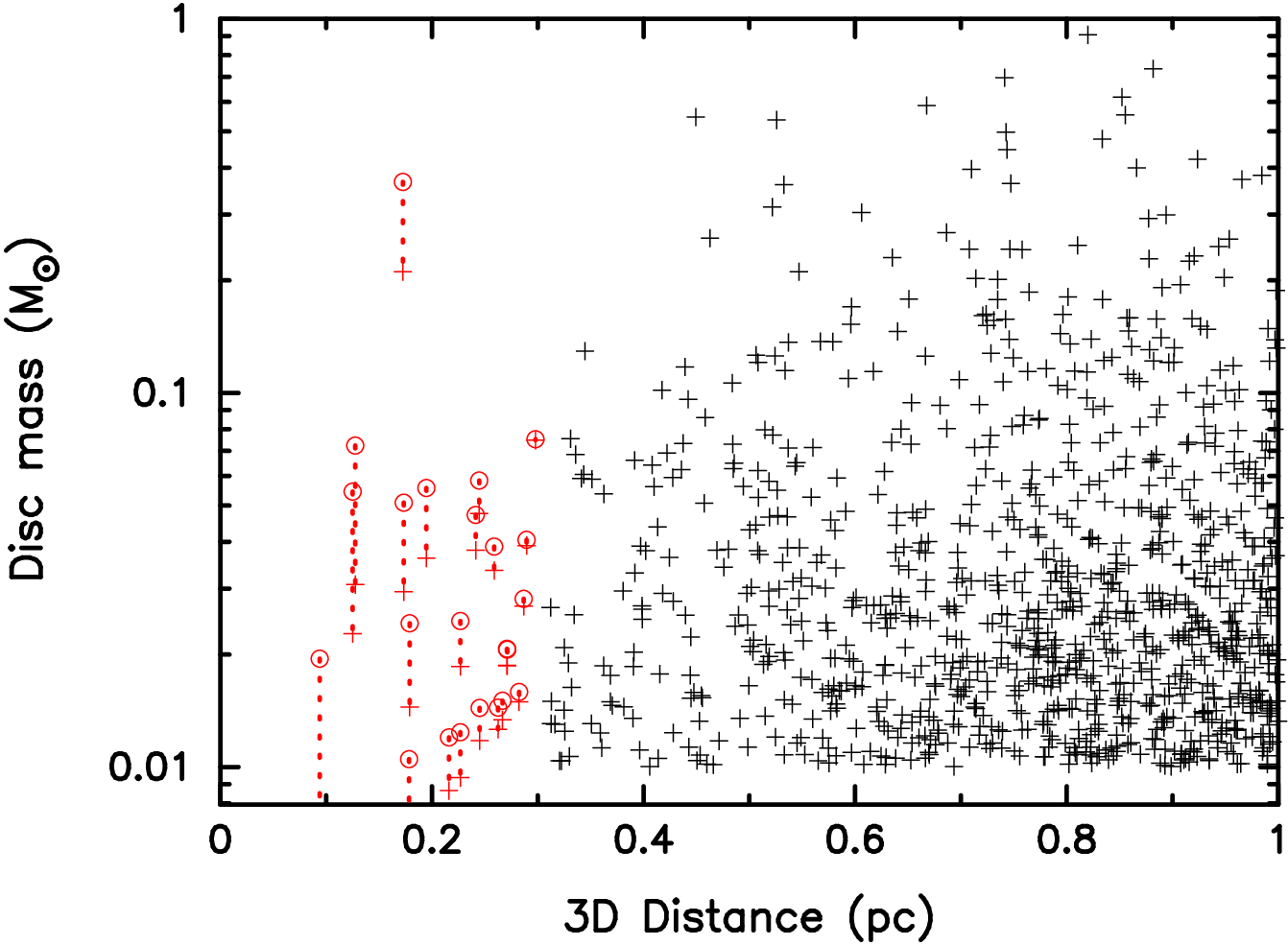}}}
\hspace*{0.3cm}\subfigure[]{\label{projection_effects-e}\rotatebox{270}{\includegraphics[scale=0.27]{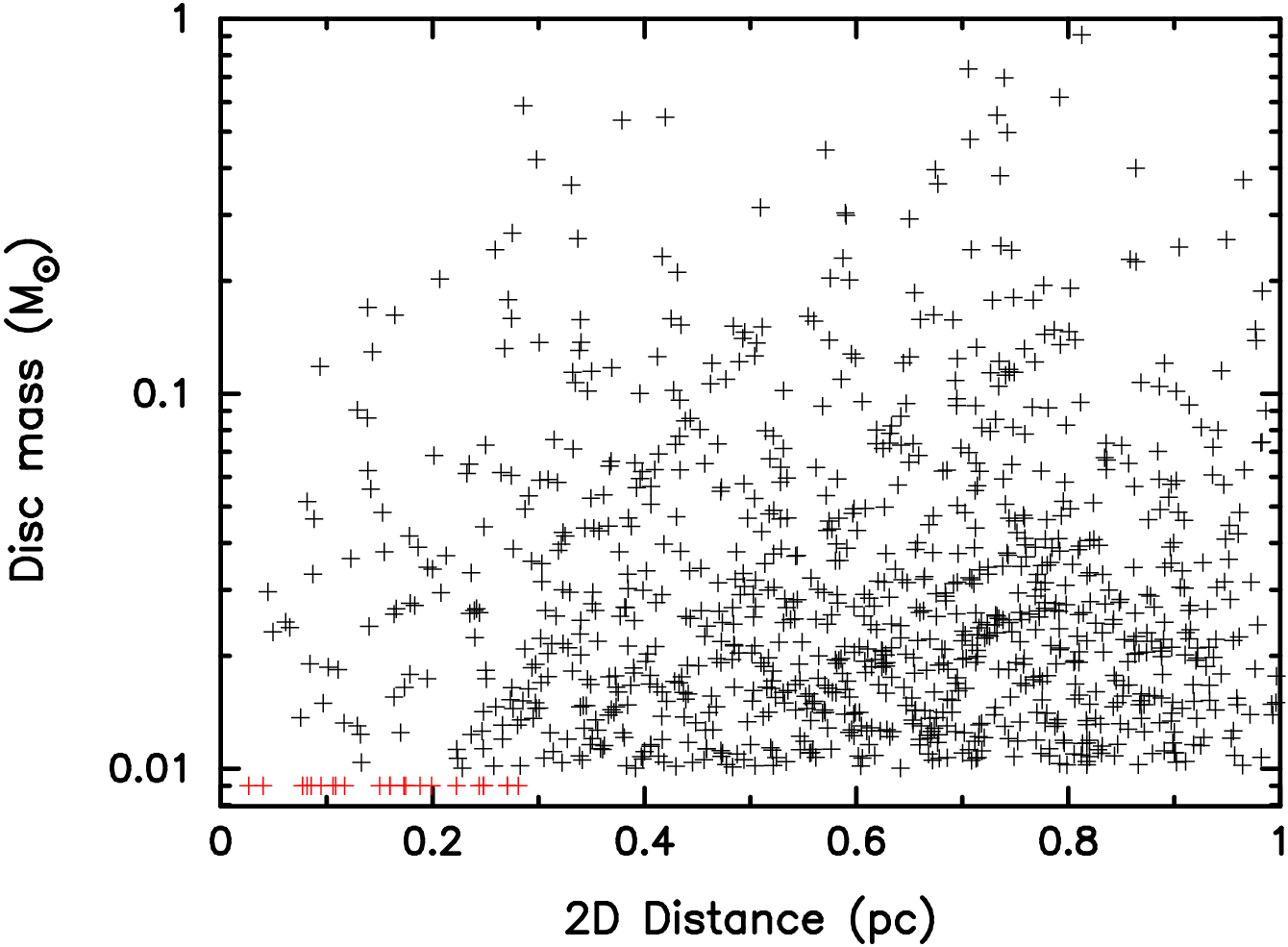}}}  

\caption{Illustration of potential projection effects in analysing photoevaporation and destruction of protoplanetary discs. In panel (a) we show the two dimensional positions of stars placed randomly in a uniform density sphere. The `photoevaporative zone' extends from the centre of the cluster out to a radius of 0.3\,pc. In all panels the red plus signs indicate the positions of stars whose discs are affected by photoevaporation. In panels (b) and (c) we show the effects of projection where we assume the discs are completely destroyed if they reside within 0.3\,pc of the massive star. In panel (b) we show the disc masses against the three dimensional distance from the the massive star, and in panel (c) we show the projected two dimensional distance. In panels (d) and (e) we show the effects of projection where we assume the discs lose mass by a factor which is inversely proportional to their distance from the most massive star. In panel (d) we show the disc masses against the three dimensional distance from  the massive star, and in panel (e) we show the projected two dimensional distance.  }
\label{projection_effects}
  \end{center}
\end{figure*}

We present two scenarios for the severity of the photoevaporation.  In both cases, we draw stellar masses randomly from the \citet{Maschberger13} IMF (Equation~\ref{maschberger_imf}) and we sample this distribution in the mass range 0.1 -- 50\,M$_\odot$ and randomly place these stars within a uniform density sphere. We assume that the photoionising massive star is at the centre of the cluster. This so-called `mass segregation' is observed in many star-forming regions that host massive stars, and can be either primordial \citep{Bonnell98} or occur from subsequent dynamical evolution \citep{Allison10}. Disc masses are set to be 10\,per cent of the stellar mass. 

In the first scenario, photoevaporation has destroyed all of the discs within a 0.3\,pc radius of the most massive star. (Note that we assume dynamics are unimportant for the moment.) We show the disc mass as a function of three dimensional distance from the photoionising massive star in Fig.~\ref{projection_effects-b}, where the stars with discs that have been destroyed by photoevaporation are shown by the red points. The two dimensional projection places fore- and background stars close to the massive star(s), so in this projection we would see a mixture of stars with and without discs in the photoevaporative zone (Fig.~\ref{projection_effects-c}).

In the second scenario, we change the original disc masses by a factor that is linearly proportional to the distance from the ionising star. Stars beyond 0.3\,pc retain their original disc masses. In Fig.~\ref{projection_effects-d}, which is a three dimensional projection, we show the original disc masses within 0.3\,pc by the open circles, and draw a line to the new disc masses which are shown by the red plus symbols. The two dimensional projection is shown in Fig.~\ref{projection_effects-e}, and it is clearly impossible to distinguish discs that have been affected by photoevaporation from those that have not in this plot.

We have repeated this thought experiment for star-forming regions with different geometries (including very substructured fractal distributions, and much more centrally concentrated clusters than the one we present here) and unsurprisingly find the same result. We therefore expect projection effects to confuse our analysis of photoevaporation, both  in observed star-forming regions and in simulated star-forming regions where we allow the regions to evolve dynamically but are projecting the positions of the stars into a two-dimensional field of view. %, and the protoplanetary discs to gradually lose mass depending on their distance to the photoionising massive star(s). 

\section{$N$-body evolution of star-forming regions}
\label{simulation}

In this section we describe the set-up and implementation of our $N$-body simulations of the dynamical evolution of star-forming regions, as well as the post-processing disc mass-loss due to photoevaporation. Our simulations are very similar to those presented in \citet*{Parker21a} and for full details we refer the interested reader to that paper, as well as \citet{Parker14b}, which describes the initial conditions for the positions and velocities of the stars in detail.

\subsection{Simulation set-up}

We set up star-forming regions with $N = 1500$ stars, with masses drawn from the probability distribution (Equation~\ref{maschberger_imf}) in \citet{Maschberger13}. We sample this IMF in the mass range 0.1 -- 50\,M$_\odot$.  

Observations of the early stages of star-formation indicate that stars form in filaments, and where these filaments intersect, hubs, or subclusters, of stars form \citep{Myers11,Andre10,Kuhn14}. This results in an initially substructured spatial distribution \citep{Cartwright04,Kuhn14}. Furthermore, the velocities of young stars (and pre-stellar cores) are small on local scales, but increase with increasing distance \citep{Larson81}. 

In order to mimic these spatially and kinematically substructured distributions, we initialise  the stellar positions and velocities in our simulations using the fractal generator described in \citet{Goodwin04a}. The full details of this method are given in \citet{Goodwin04a,Allison10,Parker14b} and we refer the interested reader to those papers. In brief, the amount of spatial substructure is set by the fractal dimension, $D$. In this work we adopt $D = 2.0$, which results in a moderate amount of substructure and is consistent with dynamical models of the early evolution of the ONC \citep{Allison11}. A lower fractal dimension (corresponding to more substructure) for a constant radius would increase the local density and lead to more extreme photoevaporation \citep{Nicholson19a}. However, because we will compare our results to observations of proplyds in the ONC, we choose a moderate amount of substructure from constraints on the amount of dynamical evolution that can have occurred in the ONC \citep{Allison10,Allison12,Parker14b}.

The amount of kinematic structure is also set by the fractal dimension. Lower fractal dimensions produce more kinematic substructure, whereas higher fractal dimensions produce less. Our choice of $D = 2.0$ means the regions have a moderate amount of kinematic substructure. The velocities are then scaled to a global virial ratio, $\alpha_{\rm vir}$, and in this work we adopt $\alpha_{\rm vir} = 0.3$. This means the star-forming regions are initially subvirial and will collapse into their gravitational potential. This subvirial collapse causes a single, roughly spherical star cluster to rapidly form, and then expand due to two-body relaxation. An imprint of the collapse may be seen in the velocity dispersion, which would appear supervirial (warm, or hot), despite the cluster being in virial equilibrium \citep{Parker16b}. This pseudo-supervirial expansion may be observed in real star-forming regions and clusters \citep{Bravi18,Kounkel18,Kuhn19}, although if the velocity dispersion is instead a signature of actual supervirial expansion this suggests that gas removal has significantly altered the gravitational potential of the region in question \citep{Tutukov78,Goodwin97,Goodwin06,Baumgardt07,Shukirgaliyev18}.

We place the stars randomly at the positions and velocities in the fractal distribution, and produce 20 realisations of the same initial conditions, where we vary the random number seed used to select stellar masses, positions and velocities. The star-forming regions have radii of either 1\,pc or 5\,pc, which is simply the radii of the fractal distributions. For simplicity, we do not include primordial binary or multiple systems in the simulations, and due to the short timeframe on which discs evolve \citep{Haisch01,Richert18} and photoevaporation affects those discs (both timescales are less than 5\,Myr), we do not include stellar evolution.

The total mass of each region varies due to the stochastic sampling of the IMF, but is in the range 440 -- 630\,M$_\odot$ for twenty realisations of the same initial conditions. The fractal dimension is $D = 2.0$ in each simulation and with this value the simulations with radii $r = 1$\,pc have initial median stellar densities of $10^4$\,M$_\odot$\,pc$^{-3}$. Our simulations with radii  $r = 5$\,pc have initial median stellar densities of $10^2$\,M$_\odot$\,pc$^{-3}$.

Note that these median densities are higher than the volume-averaged stellar density due to the substructure inherent in the fractal distributions. For star-forming regions of comparable radii and masses, the median stellar density is higher for regions with substructure. In \citet{Parker21a} we showed the evolution of the median density in simulations with similar initial conditions to those we present here. Typically, in such dense regions the median density drops by a factor of $\sim$100 in the first few Myr, due to the violent relaxation these regions undergo. 

Interestingly, when the initial amount of substructure is different, but the median density is comparable, the less substructured regions cause more disc destruction via photoevaporation \citep[][ their fig.~3]{Parker21a}. The reason for this is that to achieve comparable median stellar density, the radii of non-substructured regions have to be much smaller than for substructured regions. This means that the massive stars in the smoother regions are initially closer on average to disc-hosting stars, and the star-forming regions can collapse into a deeper potential well (thus facilitating higher long-term densities and consequently higher levels of disc destruction).

We follow the dynamical evolution in the simulations using the \texttt{kira} integrator within the \texttt{Starlab} package \citep{Zwart99,Zwart01}. This implements a 4$^{\rm th}$-order Hermite scheme with block timestepping to evolve the regions, and we run the simulations until they reach an age of 10\,Myr. We do not include a background gas potential in the simulations. Whilst the abrupt removal of gas has been shown to alter the dynamical evolution of star-forming regions \citep[e.g.][]{Tutukov78,Lada84,Goodwin97,Baumgardt07}, recent state-of-the-art hybrid simulations suggest that the effects of removing gas left over from star formation are much more subtle (\citealp{Sills18}{, cf.\,\, \citealp{Smith11}). 

\subsection{Disc properties}

The discs are not directly modelled in our simulations. For each star with mass $m < 3.0$\,M$_\odot$, we assign a disc with mass that is 10\,per cent of the host star's mass and in a different set of simulations we assign a disc that is 1\,per cent of the host star mass. This spans the range of disc masses expected from different magnetic fields and angular momentum distributions in molecular clouds. However, in our models these `discs' are an abstract construct; we do not include the extra mass in the $N$-body integration and the disc does not interact with the host star. Furthermore, when mass is removed from each disc due to photoevaporation (see below), we do not model the disc's internal response (change in surface density, radius, etc.).

We set the initial disc radii to be 100\,au in all simulations, and the radius of an individual disc is allowed to decrease if the disc loses mass due to external photoevaporation (see below), but does not change due to any internal disc physics (e.g.\,\,viscous evolution or internal photoevaporation from the host star).

\subsection{Photoevaporation}

Massive stars emit significant power in the form of far ultraviolet (FUV) luminosity, where the individual photon energies are $h\nu < 13.6\,{\rm eV}$ and extreme ultraviolet (EUV) luminosity, where the individual photon energies are $h\nu > 13.6\,{\rm eV}$. Several authors \citep[e.g.][]{Storzer99,Hollenbach00,Haworth18b} have shown that FUV radiation is more destructive to protoplanetary discs than EUV radiation.

  To implement disc mass loss from photoevaporation due to FUV radiation, we use the \texttt{FRIED} grid from \citet{Haworth18b}, which takes the disc radius, disc mass and FUV luminosity flux received by the star
  \begin{equation}
    F_{\rm FUV} = \frac{L_{\rm FUV}}{4\pi d^2},
    \label{fuv_flux}
  \end{equation}
  expressed in terms of the background FUV flux in the interstellar medium, $G_0 = 1.6 \times 10^{-3}$ erg\,cm$^{-2}$\,s$^{-1}$ \citep{Habing68}, and computes the mass-loss from the disc due to the FUV radiation. In Equation~\ref{fuv_flux}, $d$ is the distance from the FUV emitting star, and $L_{\rm FUV}$ is the FUV luminosity of that star. We take $L_{\rm FUV}$ as a function of stellar mass from fig.~1 in \citet{Armitage00}, who used the stellar atmosphere models of \citet{Buser92}.

To a lesser extent, the disc also loses mass due to EUV radiation, and we subtract mass from the disc, $\dot{M}_{\rm EUV}$, according to the prescription in \citet{Johnstone98}: %We follow the photoevaporation of discs using the prescription in \citet{Scally01}, which implements the effects of both FUV and EUV radiation described in \citet{Storzer99} and \citet{Hollenbach00}. The FUV field around a massive star is effective out to a radius of around 0.3\,pc, beyond which it does not contribute to the mass loss due to photoevaporation. Within this 0.3\,pc radius, the mass loss due to FUV radiation (photon energies $h\nu < 13.6\,{\rm eV}$) is dependent only on the radius of the disc $r_d$ and is given by
%\begin{equation}
%\dot{M}_{\rm FUV} \simeq 2 \times 10^{-9} r_d \,\,{\rm M_\odot \,yr}^{-1}.
%\label{fuv_equation}
%\end{equation}
%The disc also loses mass due to EUV radiation (photon energies $h\nu < 13.6\,{\rm eV}$), and this does depend on the distance from the massive star(s) $d$ as well as the radius: 
\begin{equation}
\dot{M}_{\rm EUV} \simeq 8 \times 10^{-12} r^{3/2}_d\sqrt{\frac{\Phi_i}{d^2}}\,\,{\rm M_\odot \,yr}^{-1}.
\label{euv_equation}
\end{equation}
Here, $\Phi_i$ is the  ionizing EUV photon luminosity from each massive star in units of $10^{49}$\,s$^{-1}$ and is dependent on the stellar mass according to the observations of \citet{Vacca96} and \citet{Sternberg03}, the disc radius $r_d$ is expressed in units of au and the distance to the massive star $d$ is in pc.

  We subtract mass from the discs according to the FUV-induced mass-loss rate in the \texttt{FRIED} grid and the EUV-induced mass-loss rate from Equation~\ref{euv_equation}. Models of mass loss in discs usually assume the mass is removed from the edge of the disc (where the surface density is lowest) and we would expect the radius of the disc to decrease in this scenario. We employ a very simple way of reducing the radius by assuming the surface density of the disc at 1\,au, $\Sigma_{\rm 1\,au}$, from the host star remains constant during mass-loss \citep[see also][]{Haworth18b}. If
  \begin{equation}
\Sigma_{\rm 1\,au} = \frac{M_d}{2\pi r_d {\rm 1\,au}},
  \end{equation}
  where $M_d$ is the disc mass, and $r_d$ is the radius of the disc, then if the surface density at 1\,au remains constant, a reduction in mass due to photoevaporation will result in the disc radius decreasing by a factor equal to the disc mass decrease, so that
  \begin{equation}
r_d(t_k) = \frac{M_d(t_k)}{M_d(t_{k-1})}r_d(t_{k-1}),
\label{rescale_disc}
\end{equation} 
where  $r_d(t_k)$ and $M_d(t_k)$ are the disc radius and disc mass after photoevaporation in a given time interval, and $r_d(t_{k-1})$ and $M_d(t_{k-1})$ are  disc radius and disc mass before photoevaporation. In the case where the disc radius decreases due to mass loss, the smaller radius acts to slow down the rate of photoevaporation. However, if the $G_0$ field is high enough, a disc can still be completely destroyed (i.e.\,\,lose all of its mass).  

We do not include any other internal evolution of the discs, e.g.\,\,viscous evolution or inner truncation from internal photoevaporation from the host star, nor do we allow mass-loss due to accretion from the disc onto the host star. In practice, both viscous evolution and accretion onto the star will act to deplete the disc (viscous spreading increases the disc radius and decreases the disc surface density, making it more susceptible to external photoevaporation). In \citet{Parker21a} we showed that viscous spreading and accretion onto the star both contribute to many more discs being destroyed, and on much more rapid timescales (within 1\,Myr when both accretion and viscous spreading are enabled, compared to $\sim$5\,Myr without). 

For further details of the evolution of discs in these $N$-body simulations, we refer the interested reader to \citet{Parker21a}, which also includes a discussion of the adopted timestep ($10^{-3}$Myr) in the disc evolution calculations.

%We follow the mass loss in the discs due to photoevaporation by subtracting mass from our discs according to equations~\ref{fuv_equation}~and~\ref{euv_equation}. The mass loss due to EUV radiation depends on the distance to the massive star(s), $d$, whereas the mass loss due to FUV radiation is only invoked if the disc hosting stars are closer than 0.3\,pc from the massive star(s). The only effect we investigate is the total mass-loss from the disc and the discs do not respond to this photoevaporation through any kind of internal evolution. Once the mass-loss is equal to or more than the original disc mass at 0\,Myr, we class the stellar host as a `disc-less star' \citep[note that our discs can have masses below the ALMA sensitivity limit of $\sim 5 \times 10^{-6}$M$_\odot$;][]{Ansdell17}.

%This simple photoevaporation prescription produces similar results to more detailed models described in \citet{Facchini16,Haworth18b,Winter18b,Winter19}, and for the purposes of this paper the subtle differences between the methods are not relevant.

\section{Results}
\label{results}

In this section we first focus on simulations with high initial stellar densities (such as those suggested for the Orion Nebula Cluster). We then look at lower density simulations (commensurate with the observed present-day density in the $\sigma$~Orionis cluster), before comparing our results with distributions of disc-bearing and disc-less stars in the Orion star-forming region.

\subsection{Photoevaporation in high-density simulations}

\begin{figure*}
  \begin{center}
\setlength{\subfigcapskip}{10pt}
%\hspace*{-1.0cm}\subfigure[0\,Myr]{\label{sim_2_2D_disc_10percent-a}\rotatebox{270}{\includegraphics[scale=0.2]{disc_mass_ion_dist_Or_C0p3F2p01pSmFS10_08_Fe_fuv+euv_0Myr.ps}}}
%\hspace*{0.3cm}
%\subfigure[0.5\,Myr]{\label{sim_2_2D_disc_10percent-b}\rotatebox{270}{\includegraphics[scale=0.2]{disc_mass_ion_dist_Or_C0p3F2p01pSmFS10_08_Fe_fuv+euv_0p5Myr.ps}}}
%\hspace*{0.3cm}\subfigure[1\,Myr]{\label{sim_2_2D_disc_10percent-c}\rotatebox{270}{\includegraphics[scale=0.2]{disc_mass_ion_dist_Or_C0p3F2p01pSmFS10_08_Fe_fuv+euv_1Myr.ps}}}
%\hspace*{0.3cm}\subfigure[4\,Myr]{\label{sim_2_2D_disc_10percent-d}\rotatebox{270}{\includegraphics[scale=0.2]{disc_mass_ion_dist_Or_C0p3F2p01pSmFS10_08_Fe_fuv+euv_4Myr.ps}}}
%\hspace*{-1.0cm}\subfigure[0\,Myr]{\label{sim_2_2D_disc_10percent-e}\rotatebox{270}{\includegraphics[scale=0.2]{disc_hist_ion_dist_Or_C0p3F2p01pSmFS10_08_Fe_fuv+euv_0Myr.ps}}}
%\hspace*{0.3cm}
%\subfigure[0.5\,Myr]{\label{sim_2_2D_disc_10percent-f}\rotatebox{270}{\includegraphics[scale=0.2]{disc_hist_ion_dist_Or_C0p3F2p01pSmFS10_08_Fe_fuv+euv_0p5Myr.ps}}}
%\hspace*{0.3cm}\subfigure[1\,Myr]{\label{sim_2_2D_disc_10percent-g}\rotatebox{270}{\includegraphics[scale=0.2]{disc_hist_ion_dist_Or_C0p3F2p01pSmFS10_08_Fe_fuv+euv_1Myr.ps}}}
%\hspace*{0.3cm}\subfigure[4\,Myr]{\label{sim_2_2D_disc_10percent-h}\rotatebox{270}{\includegraphics[scale=0.2]{disc_hist_ion_dist_Or_C0p3F2p01pSmFS10_08_Fe_fuv+euv_4Myr.ps}}}  

\hspace*{-1.0cm}\subfigure[0\,Myr]{\label{sim_2_2D_disc_10percent-a}\rotatebox{270}{\includegraphics[scale=0.2]{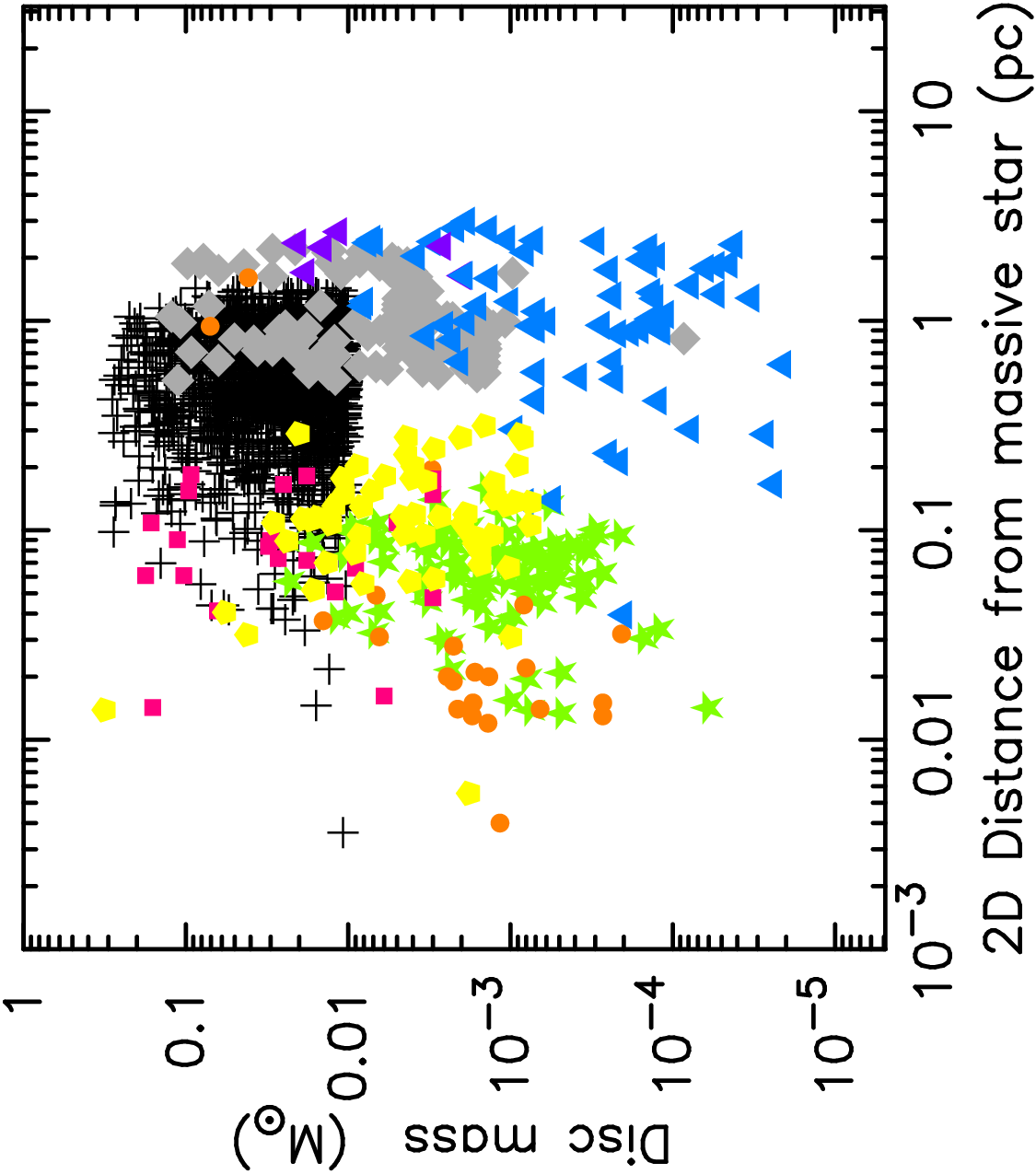}}}
\hspace*{0.3cm}
\subfigure[0.5\,Myr]{\label{sim_2_2D_disc_10percent-b}\rotatebox{270}{\includegraphics[scale=0.2]{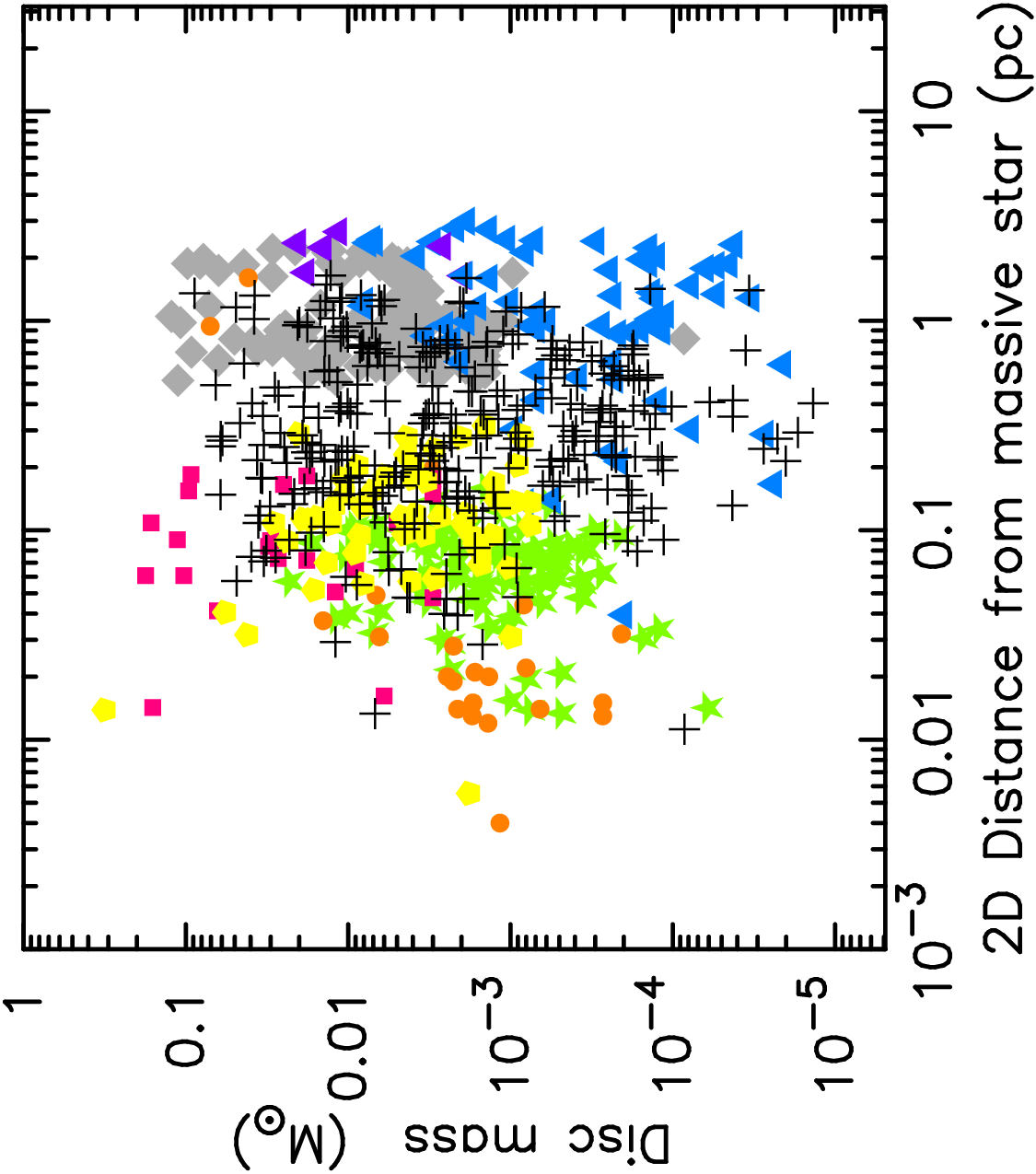}}}
\hspace*{0.3cm}\subfigure[1\,Myr]{\label{sim_2_2D_disc_10percent-c}\rotatebox{270}{\includegraphics[scale=0.2]{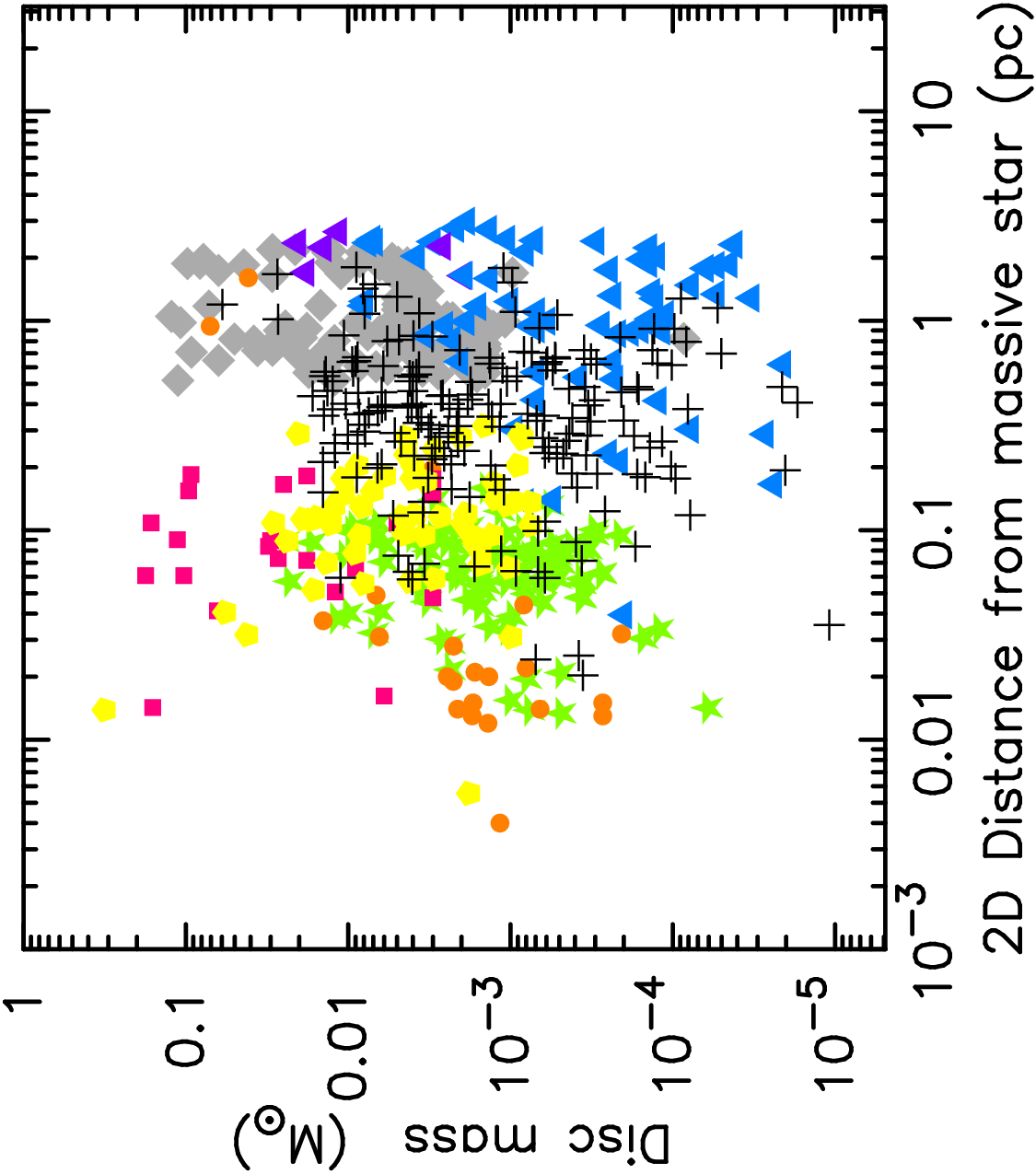}}}
\hspace*{0.3cm}\subfigure[4\,Myr]{\label{sim_2_2D_disc_10percent-d}\rotatebox{270}{\includegraphics[scale=0.2]{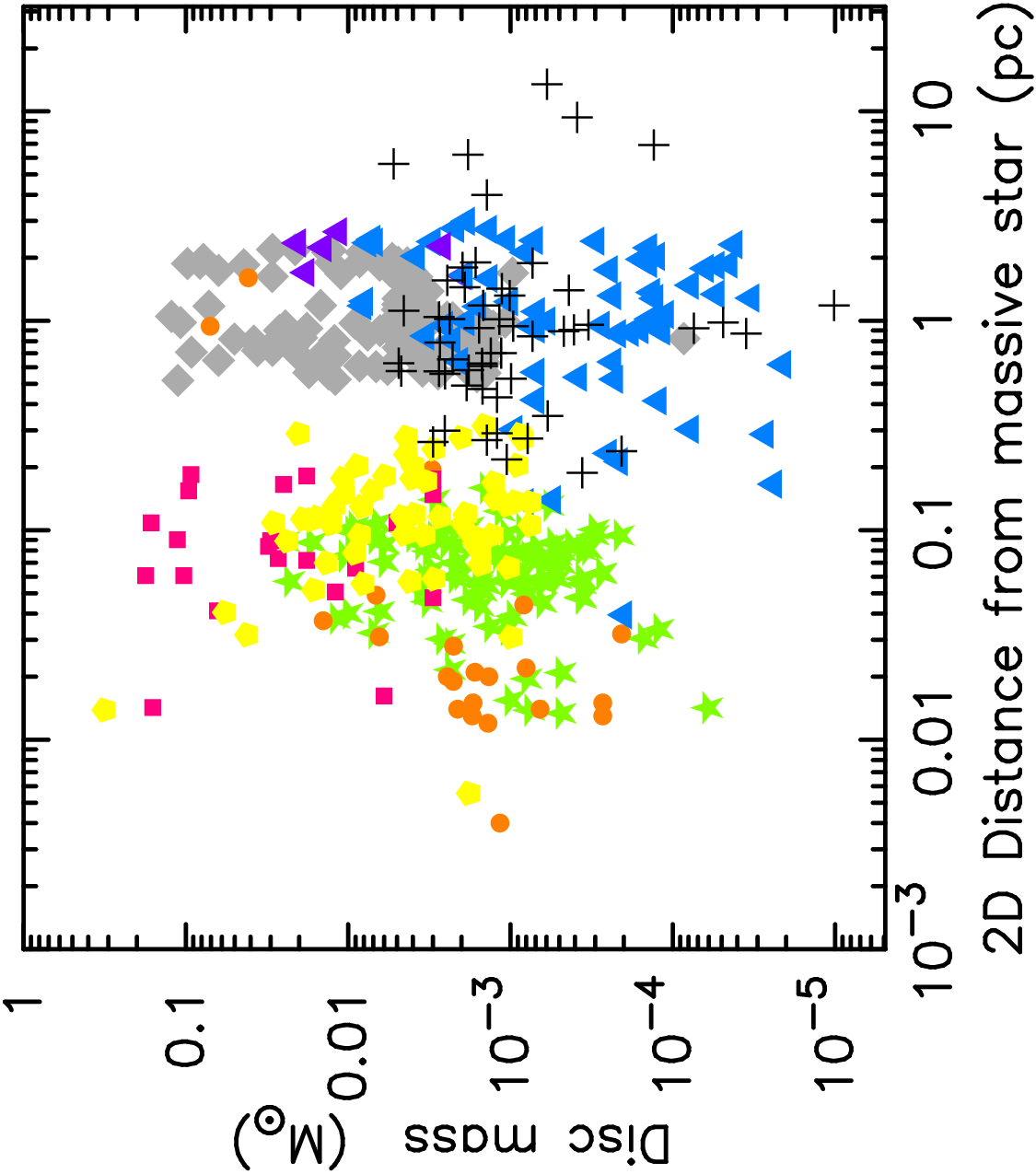}}}
\hspace*{-1.0cm}\subfigure[0\,Myr]{\label{sim_2_2D_disc_10percent-e}\rotatebox{270}{\includegraphics[scale=0.2]{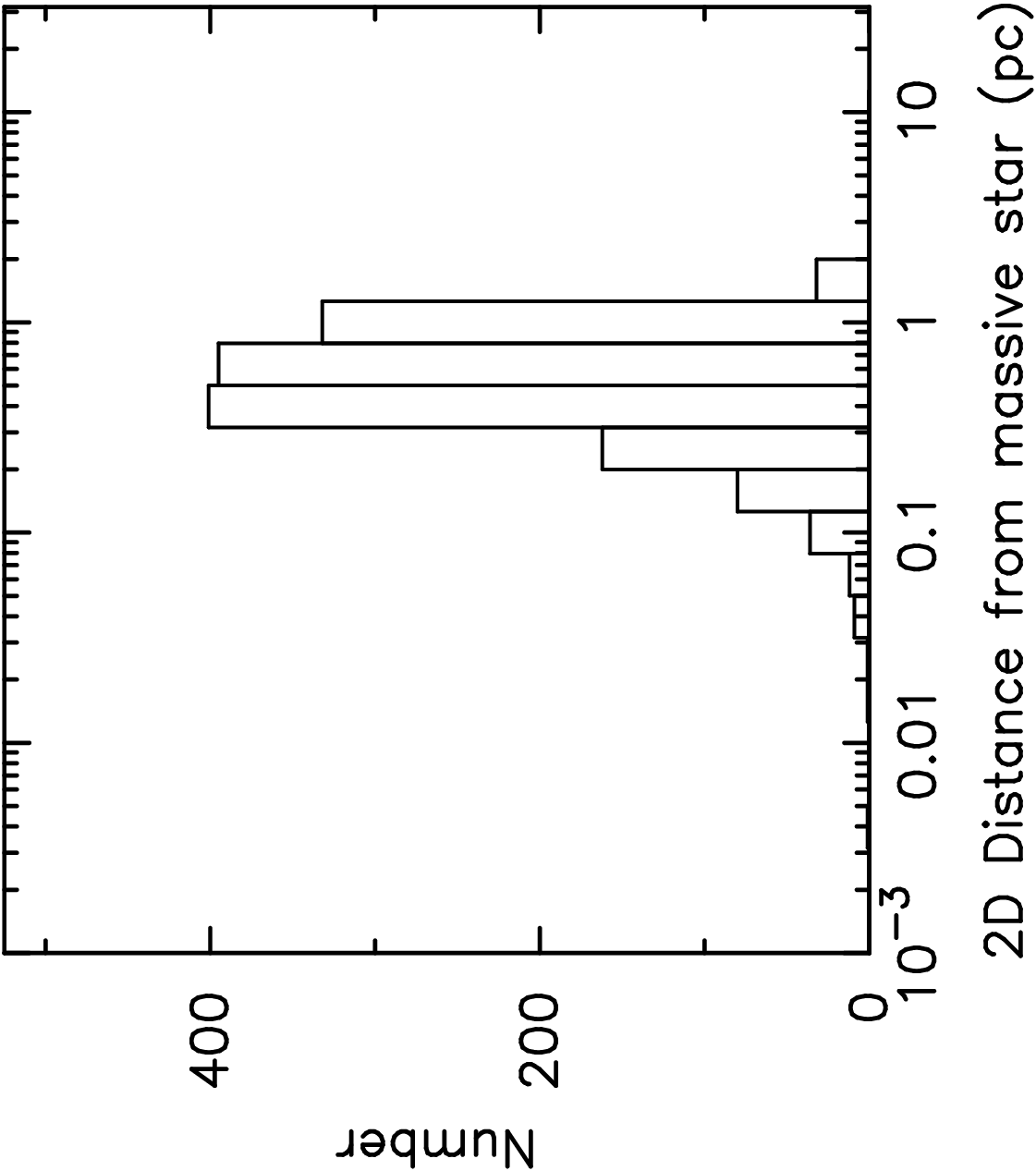}}}
\hspace*{0.3cm}
\subfigure[0.5\,Myr]{\label{sim_2_2D_disc_10percent-f}\rotatebox{270}{\includegraphics[scale=0.2]{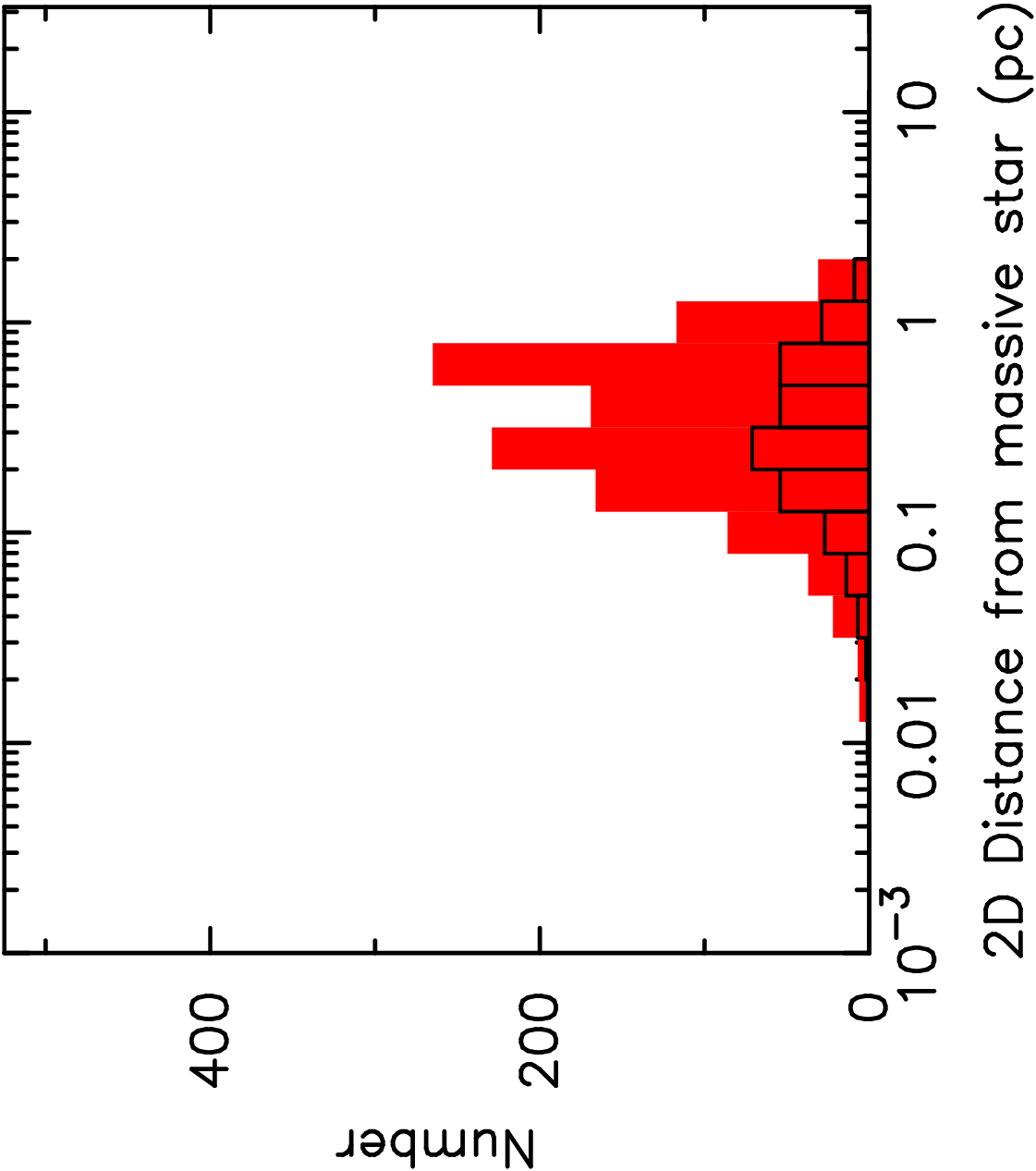}}}
\hspace*{0.3cm}\subfigure[1\,Myr]{\label{sim_2_2D_disc_10percent-g}\rotatebox{270}{\includegraphics[scale=0.2]{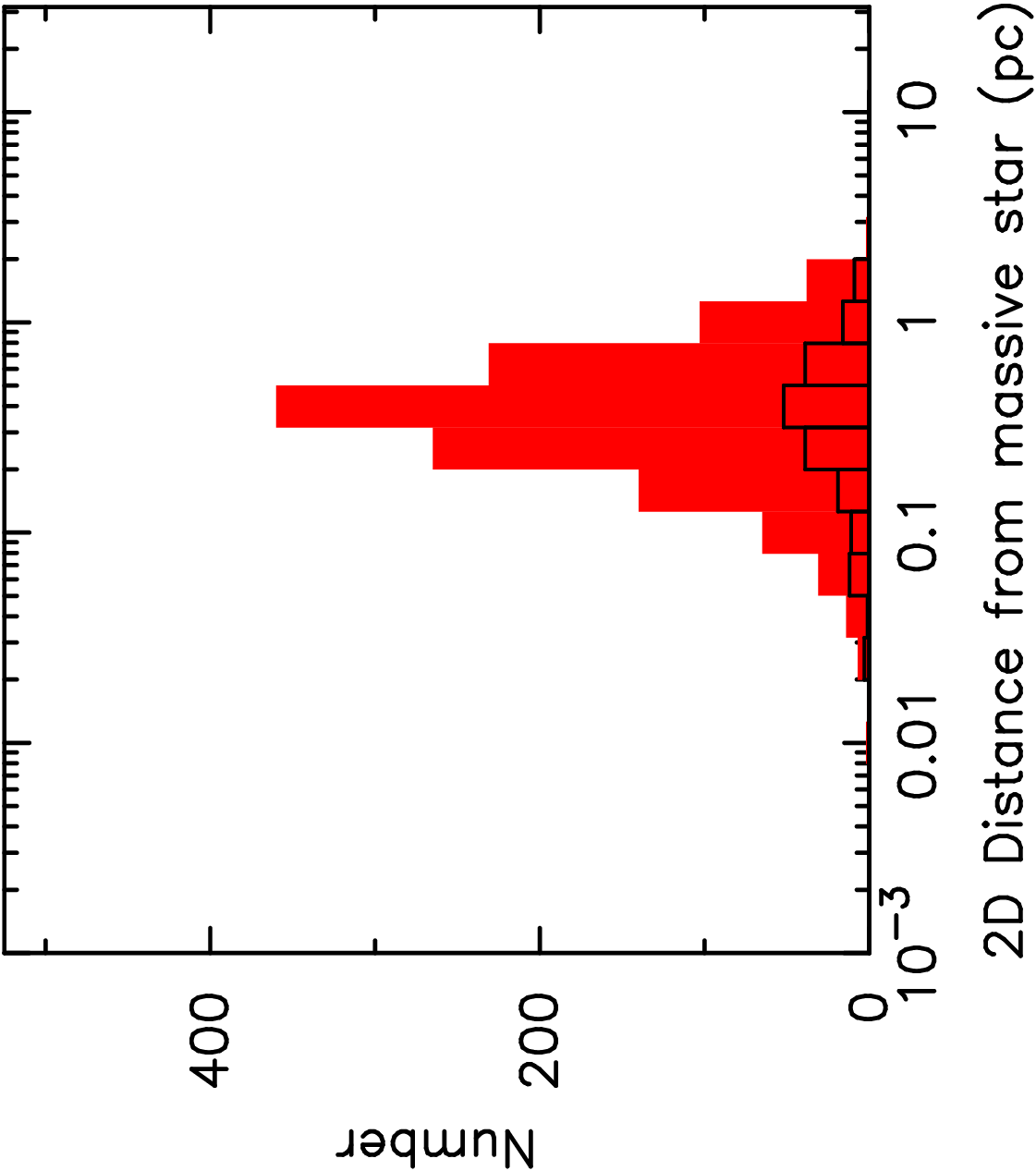}}}
\hspace*{0.3cm}\subfigure[4\,Myr]{\label{sim_2_2D_disc_10percent-h}\rotatebox{270}{\includegraphics[scale=0.2]{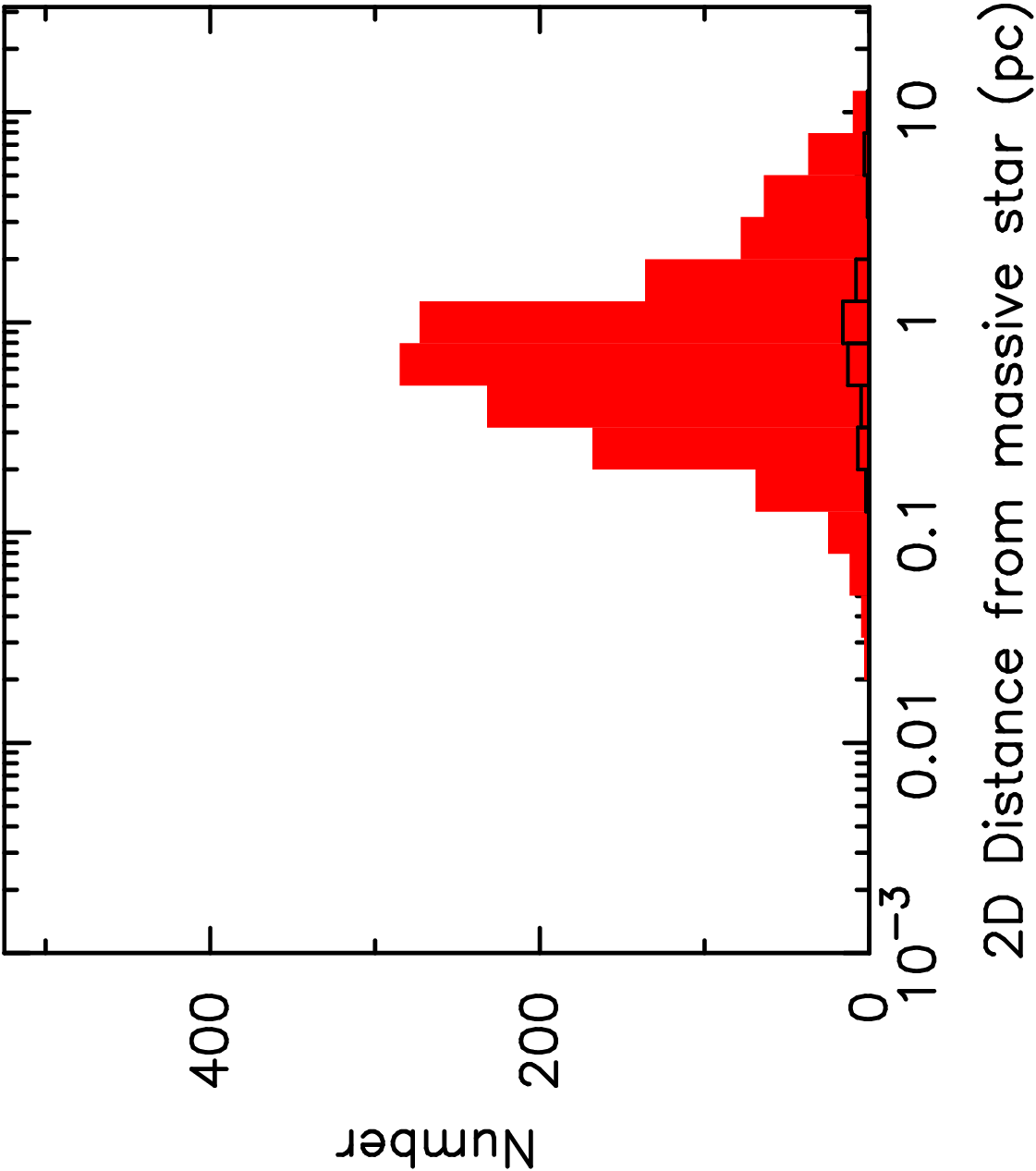}}}  
\caption{Photoevaporation of protoplanetary discs in high density star-forming regions ($10^4$\,stars\,pc$^{-3}$) where the initial disc mass is 10\,per cent of the host star's mass, and the distance to the ionising star is calculated in two dimensions. The radius of the star-forming region is 1\,pc. {\it Top row:} The plus symbols show the disc mass plotted against the distance to the massive ionising star for each low-mass star with a disc, using only two dimensions to calculate the projected distance. The masses and the respective distances from $\theta^1$~Ori~C of observed protoplanetary discs in the ONC by \citet[][orange circles]{Mann14}, and \citet[][green stars]{Eisner18}; and the Orion Molecular Cloud \citep[][the grey diamonds]{vanTerwisga19} are shown. We also show the masses and projected distances from IRS~2b of discs in NGC\,2024 by \citet[][the raspberry squares]{Mann15}  and \citet[][the yellow pentagons]{vanTerwisga20}.  Finally, we show the masses and projected distances from the ionising star $\sigma$~Ori in this eponymous star-forming region \citep{Ansdell17} by the blue triangles. Protoplanetary discs in  $\sigma$~Ori that have CO detections (i.e.\,\,gas content) are shown by the purple triangles. {\it Bottom row:} The number of remaining discs (open histogram) and the number of stars whose discs have been destroyed by photoevaporation (red histogram) as a function of the two dimensional projected distance from the ionising star. }
%Simulation 2, 10\% disc mass, 2D, distance to most massive star. 
\label{sim_2_2D_disc_10percent}
  \end{center}
\end{figure*}

\subsubsection{Dynamical evolution of the star-forming regions}

The dynamical evolution of the star-forming regions is very similar to that described in \citet{Parker14b}, \citet{Parker21a} and \citet{Nicholson19a}. The most massive stars are initially placed randomly in the substructure. Due to the high degree of spatial and kinematic substructure, as well as a subvirial energy ratio, the regions violently relax to form smooth, centrally concentrated star clusters after $\sim$1\,Myr. During this process, the massive stars migrate to the cluster centre via dynamical mass segregation \citep{Allison10,Parker14b}. In the following analysis we calculate the distances from the most massive star, but this is also usually close to the location of the cluster centre (defined as the mean position of all stars).

We will first describe the results for one realisation of the initial conditions, before discussing different realisations of the same simulations which, whilst formally statistically identical, contain different numbers of massive stars (with slightly different masses due to randomly sampling the stellar IMF), as well as (random) differences in the initial positions and velocities of the stars. Slight differences in the numbers of massive stars will lead to slightly different FUV fields, which in turn can translate into more, or fewer, discs being destroyed \citep[see][]{Parker21a}.

\begin{figure*}
  \begin{center}
\setlength{\subfigcapskip}{10pt}
\hspace*{-1.0cm}\subfigure[0\,Myr]{\label{2D_disc_10percent_av-a}\rotatebox{270}{\includegraphics[scale=0.2]{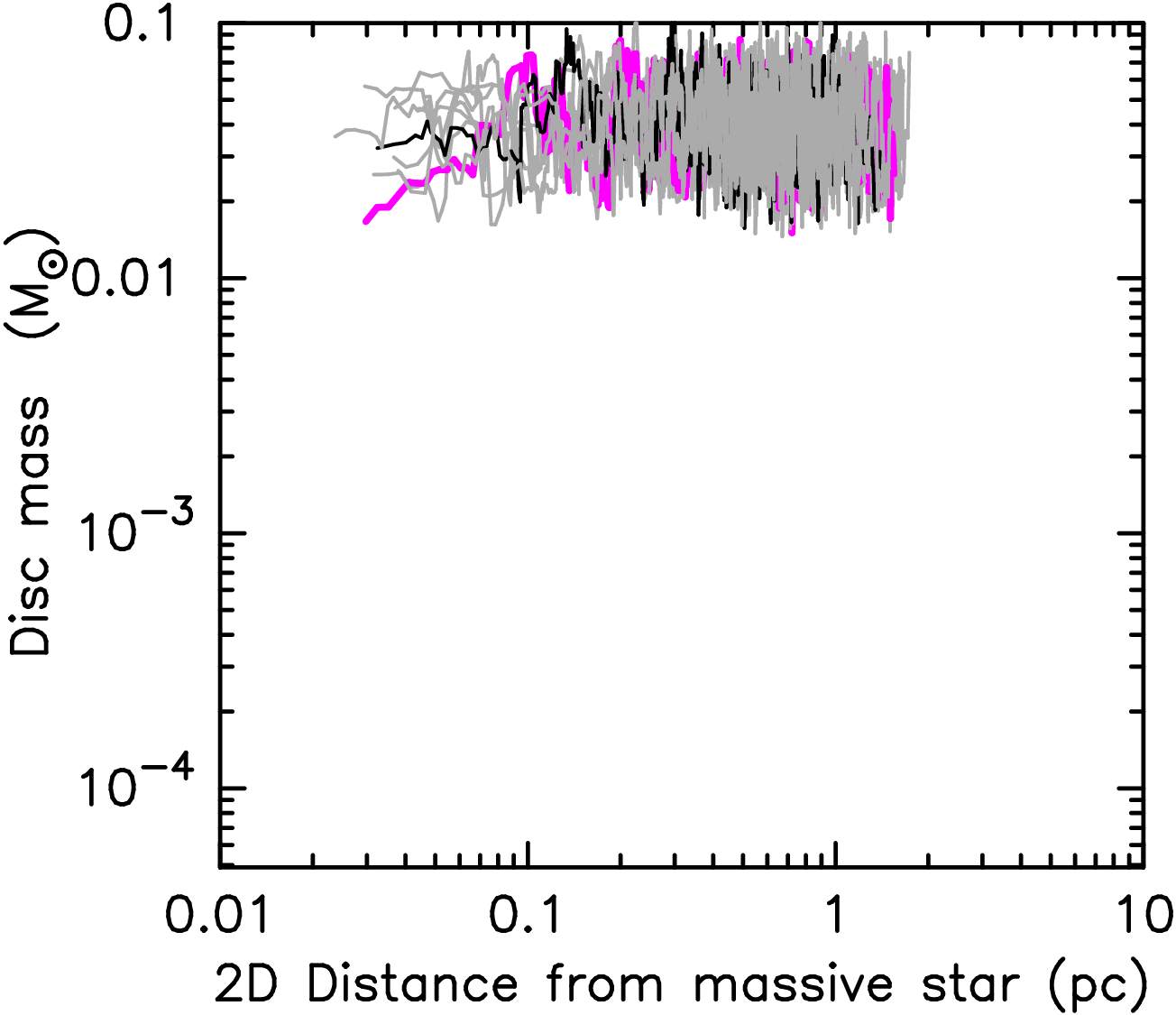}}}
\hspace*{0.1cm}
\subfigure[0.5\,Myr]{\label{2D_disc_10percent_av-b}\rotatebox{270}{\includegraphics[scale=0.2]{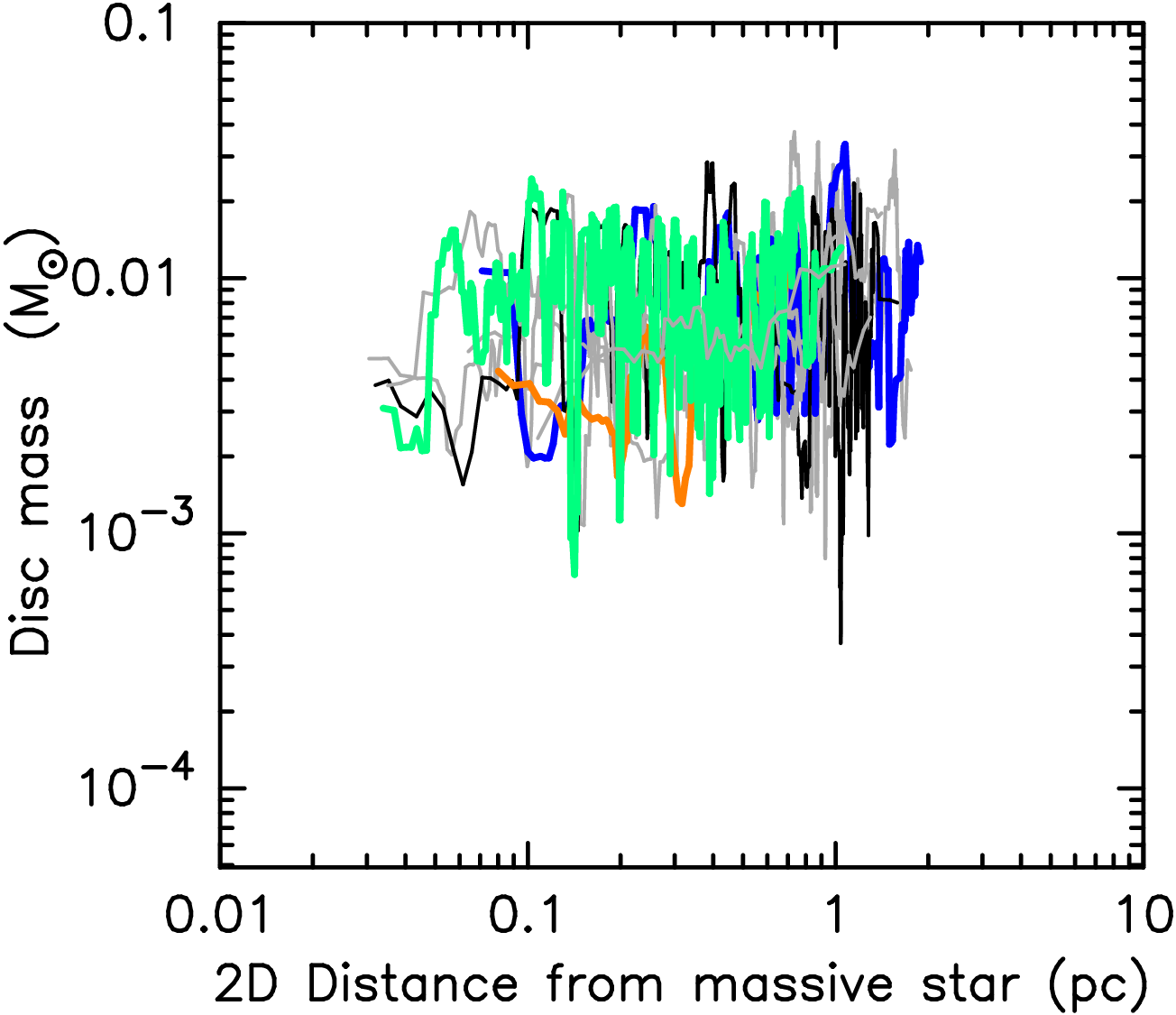}}}
\hspace*{0.1cm}\subfigure[1\,Myr]{\label{2D_disc_10percent_av-c}\rotatebox{270}{\includegraphics[scale=0.2]{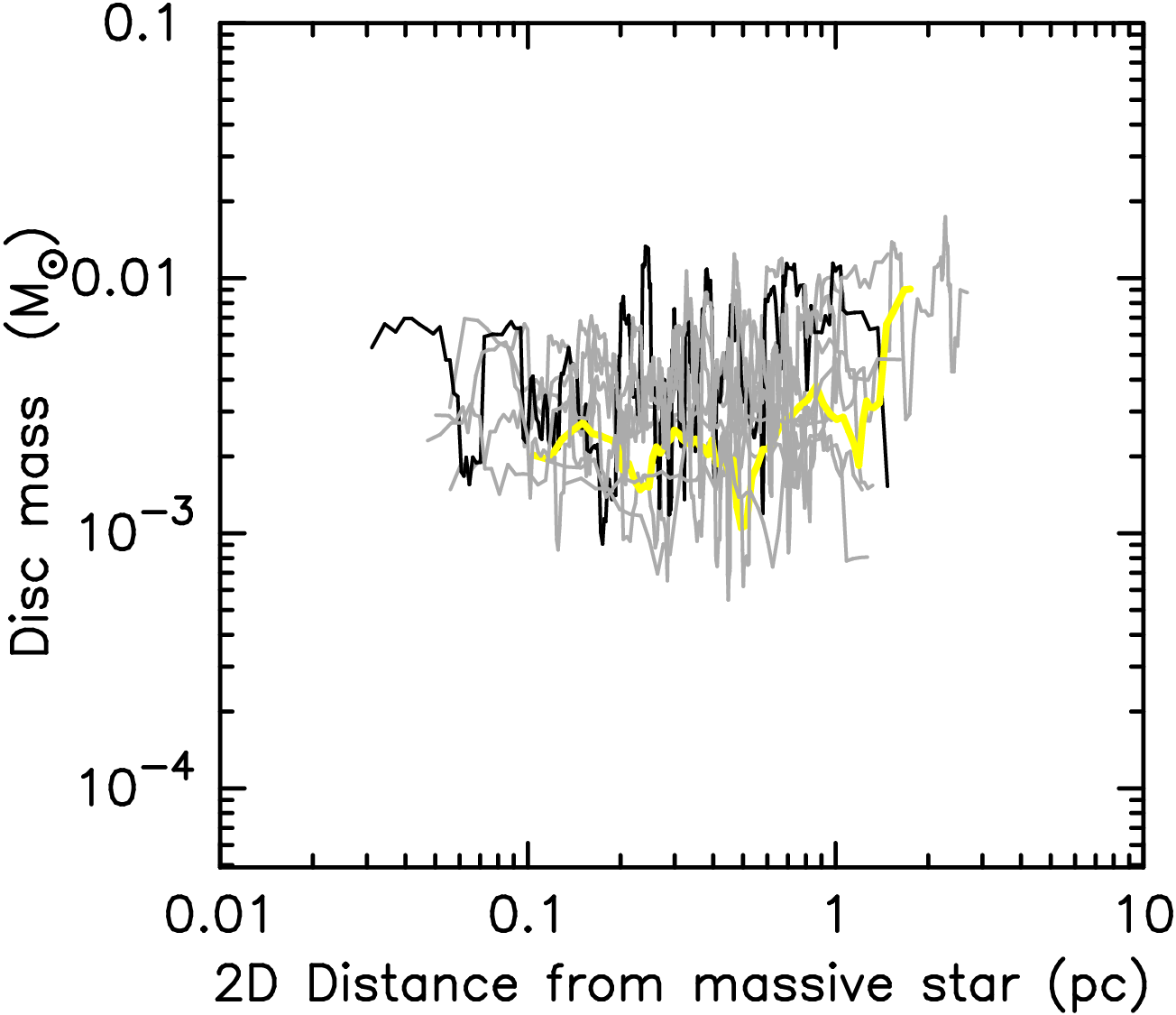}}}
\hspace*{0.1cm}\subfigure[4\,Myr]{\label{2D_disc_10percent_av-d}\rotatebox{270}{\includegraphics[scale=0.2]{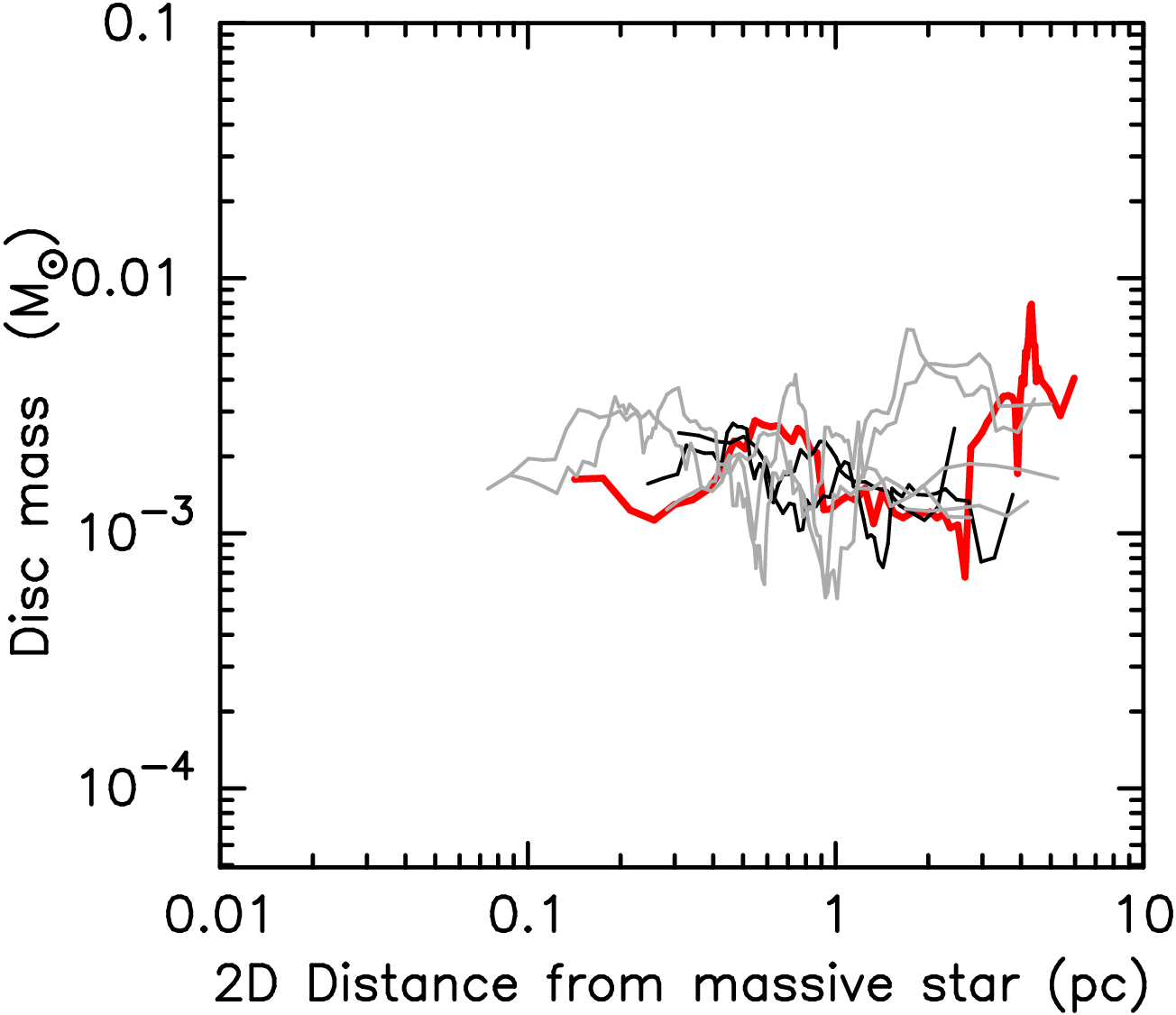}}}
\caption{Average disc mass as a function of distance from the most massive ionising star in the high-density simulations. Each line shows a rolling average of the disc masses of bins of ten stars within the ordered list of distances to the ionising star. For clarity, we only show the first ten simulations (not all twenty) but we report on the statistics for all twenty in the text. Thin grey lines indicate simulations where there is no correlation according to the Spearman rank test. The thick coloured lines indicate a positive correlation according to the Spearman test (with a $p$-value $<$0.1), i.e.\,\,the disc masses increase with increasing distance from the ionising star. The thin black lines indicate a negative correlation according to the Spearman test (with a $p$-value $<$0.1), i.e.\,\,the disc masses decrease with increasing distance from the ionising star. We show snapshots at 0, 0.5, 1 and 4\,Myr, and colours are the same at different snapshots, i.e.\,\,any long-term correlation would be indicated by similarly coloured lines.}
%Simulation 2, 10\% disc mass, 2D, distance to most massive star. 
\label{2D_disc_10percent_averages}
  \end{center}
\end{figure*}

\subsubsection{10\,per cent disc masses, 2D projection}

We start by looking at the effects of photoevaporation where the initial disc mass is set to be 10\,per cent of the host star's mass. The initial distribution (at $t = 0$\,Myr) is shown in Fig.~\ref{sim_2_2D_disc_10percent-a}, where we plot the disc mass as a function of the two-dimensional distance of the host star from the most massive star. (This distance is calculated using the $x$ and $y$ coordinates only.)

As the region evolves, the discs experience mass loss due to photoevaporation, so that after 4\,Myr of evolution the majority of the discs are destroyed (leading to fewer black datapoints from the simulations as time increases). From an initial population of 1462 discs in this simulation, after 0.5\,Myr there are only  301 discs, after 1\,Myr there are 181 discs and after 4\,Myr there are 53 discs. Whilst this mass loss and consequent destruction of discs may seem drastic, it is typical of many of the \texttt{FRIED} models presented by \citet{Haworth18b}. As an example,  radiation fields of $10^4G_0$ are typical in our very dense regions, and in the \texttt{FRIED} models such high $G_0$ values garner mass-loss rates of $\sim 10^{-6}$M$_\odot$\,yr$^{-1}$, meaning that in 0.1\,Myr a disc could lose 0.1\,M$_\odot$ of material. Whilst not all mass-loss rates are so drastic, this serves to illustrate why the disc masses are so strongly depleted in dense star-forming regions \citep[see also][]{Parker21a}.% We show an example of a disc evolving in different $G_0$ fields in {\bf  Appendix~\ref{appendix}}.

We see very little correlation between the mass of the discs and the distance of the host star from the ionising source \emph{at any snapshot in time}. %We find typical values for the Pearson $r$-coefficient of $r <0.2$, and only very occasionally find $r>0.5$. 
Interestingly, and as detailed in Section~\ref{obs}, observations suggest that, in some star-forming regions, there is a correlation between disc mass and projected distance from the massive stars. The orange circles are data from the \citet{Mann14} ALMA study of discs in the ONC, and these authors claim a strong correlation with disc masses increasing the further they are from the massive stars (where the disc masses are inferred from the dust content of the discs using sub-mm continuum observations, assuming a gas-to-dust ratio of 100). \citet{Ansdell17} report similar results from the $\sigma$~Orionis star-forming region (shown by the blue triangles in Fig.~\ref{sim_2_2D_disc_10percent-a}--\ref{sim_2_2D_disc_10percent-d}. Other authors \citep{Guarcello07,Guarcello10} report similar results in more distant star-forming regions, but \citet{Mann15} find no correlation between  mass of the discs and the distance of the host star from the ionising source in NGC\,2024 (raspberry squares).

We also plot histograms of the distance (again in 2D only) from the massive star of low-mass stars that host a disc (open histogram) and those that have had their discs destroyed (red histogram) in Figs.~\ref{sim_2_2D_disc_10percent-e}--\ref{sim_2_2D_disc_10percent-h}. These panels indicate that on average, a star closer to the ionising massive star is less likely to host a disc, but because stars move in and out of the cluster centre, disc-less stars can also be located at large distances from the ionising star(s).

In this particular simulation, the five most massive stars are 23, 18, 17, 15 \& 15\,M$_\odot$. The number of massive stars, and their exact masses, will result in a specific radiation field for a given stellar density, and this field may increase or decrease in a different realisation of the simulation  if the mass function is stochastically sampled. Additionally, the stars have different initial positions and velocities as set by a random number generator. The combination of these factors could mean that some simulations might display a correlation between the disc masses and distance to the main ionising sources, whereas others would not. 
 
 In Fig.~\ref{2D_disc_10percent_averages} we show the rolling average of disc mass as a function of distance from the most massive stars for each simulation (for clarity, we only show the first ten realisations of the twenty simulations, but we will report on all the simulations in the following text), in incremental bins containing ten stars. Where the Spearman rank test reports a significant ($p$-value $<0.1$) positive correlation\footnote{This $p$-value threshold is somewhat arbitrarily chosen, but we note that in two of the simulations that display a significant correlation, their $p$-values are less than $10^{-2}$ and in one case less than $10^{-4}$. These values are similar to those found in the observational data (see Section~\ref{obs}), but we cannot find any significant differences in either the mass functions, overall dynamical evolution or the (related) photoevaporation history between these simulations and others in our set of twenty.}, we show the rolling average by a coloured line. Where there is a significant negative correlation ($p$-value $<0.1$), we show the rolling average by a black line. Where there is no significant correlation, we show the rolling average by a light grey line.  
 
 First, we note that the random sampling of the IMF results in a correlation of disc mass with distance to the ionising star in one simulation (the magenta line in Fig.~\ref{2D_disc_10percent_av-a}) out of twenty (and also one simulation displays a negative correlation) \emph{before} any photoevaporation has occurred. Then, at 0.5\,Myr, 5/20 simulations display a positive correlation (and one is negative), at 1\,Myr 2/20 display a positive correlation and 2/20 display a negative correlation. Finally, at 4\,Myr 1/20 shows a positive correlation and 3/20 display a negative correlation. 
 
 We make five further points related to Fig.~\ref{2D_disc_10percent_averages}. First, the majority of simulations (always more than 70\,per cent) do not show a correlation. Secondly, those that do show a correlation tend to be transient -- the same simulation displays no correlation, or a negative correlation at earlier or later times. Third,  if we can obtain a significant negative correlation of decreasing disc mass with distance from the ionising source -- which cannot be due to anything other than stochastic dynamics -- it is not clear why one would attach any physical meaning to a significant positive correlation.  Fourth, we note that the results do not depend on the exact numbers of massive stars, or their individual masses. For example in Fig.~\ref{2D_disc_10percent_averages}, the simulation that displays a significant positive correlation (the red line) contains stars with masses 19, 19, 19, 12 and 11\,M$_\odot$. One of the simulations that displays a negative correlation contains stars with masses 22, 21, 15, 12 and 11\,M$_\odot$ (i.e.\,\,very similar), whereas another simulation that displays a negative correlation has stars with masses 34, 18, 11, 8 and 7\,M$_\odot$ (which would likely produce more EUV radiation, but less FUV radiation). Finally, we note that even though some simulations display a positive correlation, the rolling average fluctuates so much it is difficult to discern a strong correlation compared to other simulations where the Spearman test does not return a significant correlation.    
 
 In Appendix~\ref{appendix:mass-function} we show further scatter plots of the disc mass and distance to the most massive star, and indicate the masses of the five most massive stars in each. We conclude that the results do not depend on the exact numbers of massive stars drawn randomly from the IMF.

\subsubsection{10\,per cent disc masses, 3D projection}

\begin{figure*}
  \begin{center}
    \setlength{\subfigcapskip}{10pt}
   \hspace*{-1.0cm}\subfigure[0\,Myr]{\label{sim_2_3D_disc_10percent-a}\rotatebox{270}{\includegraphics[scale=0.2]{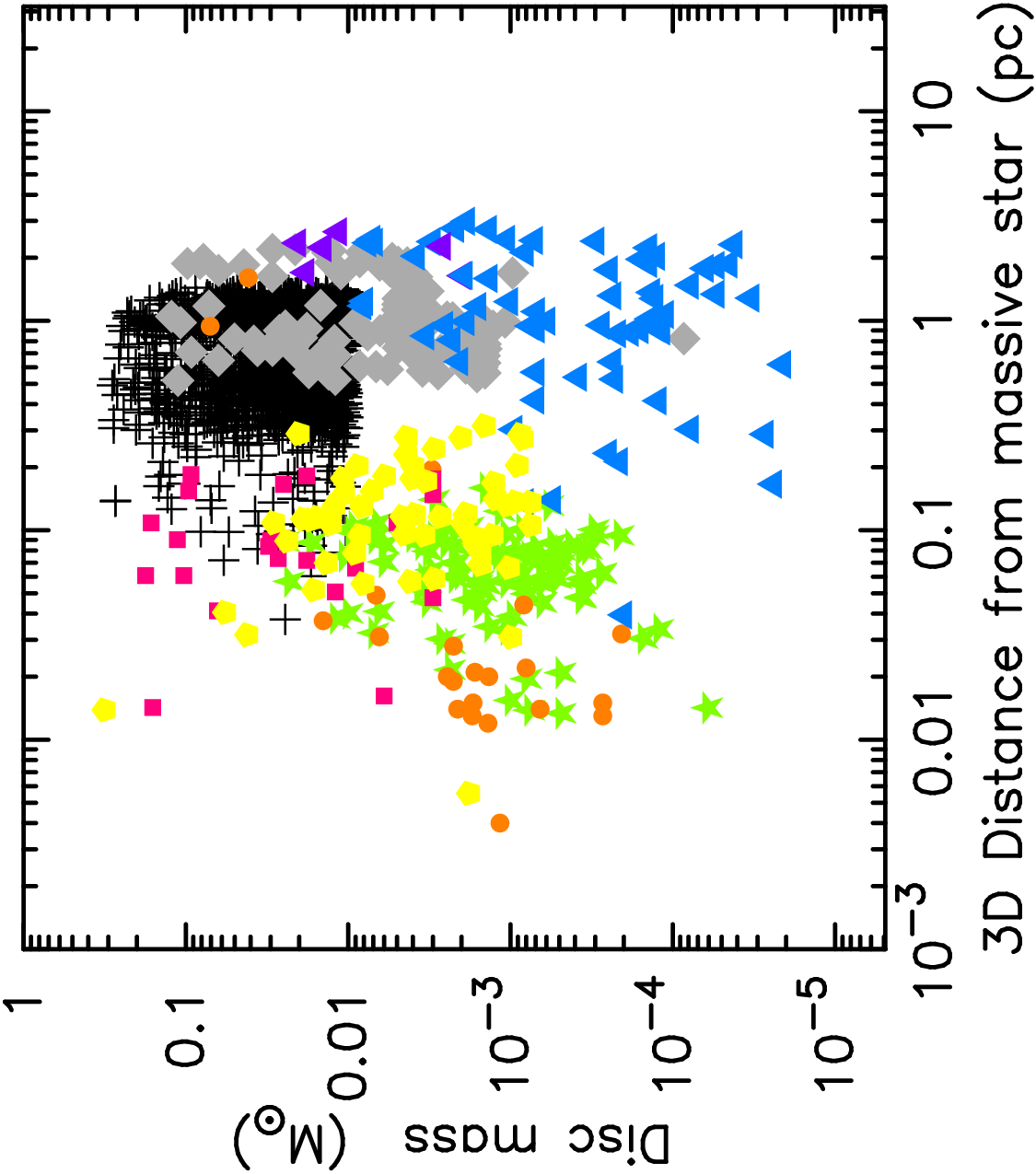}}}
\hspace*{0.3cm}
\subfigure[0.5\,Myr]{\label{sim_2_3D_disc_10percent-b}\rotatebox{270}{\includegraphics[scale=0.2]{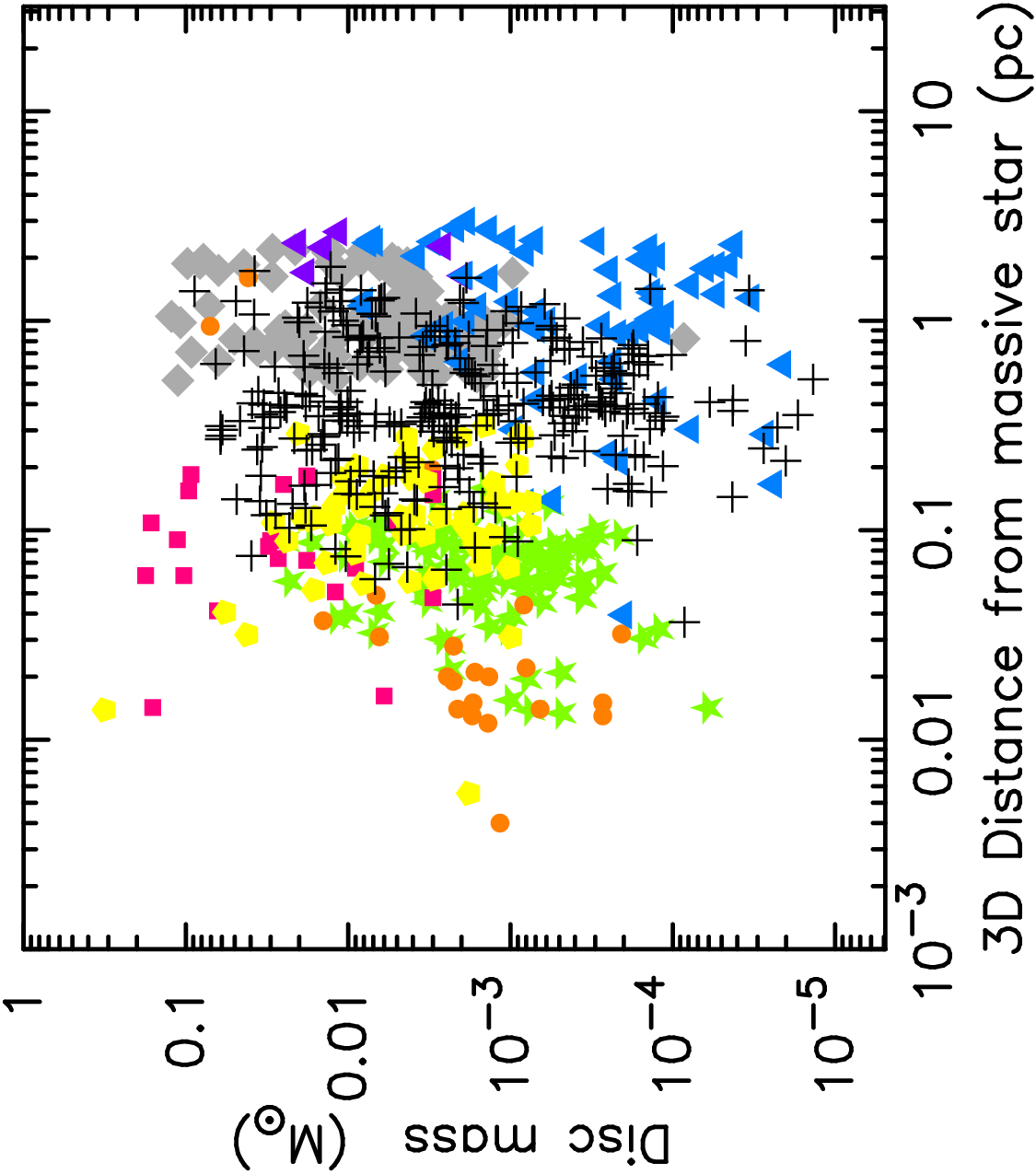}}}
\hspace*{0.3cm}\subfigure[1\,Myr]{\label{sim_2_3D_disc_10percent-c}\rotatebox{270}{\includegraphics[scale=0.2]{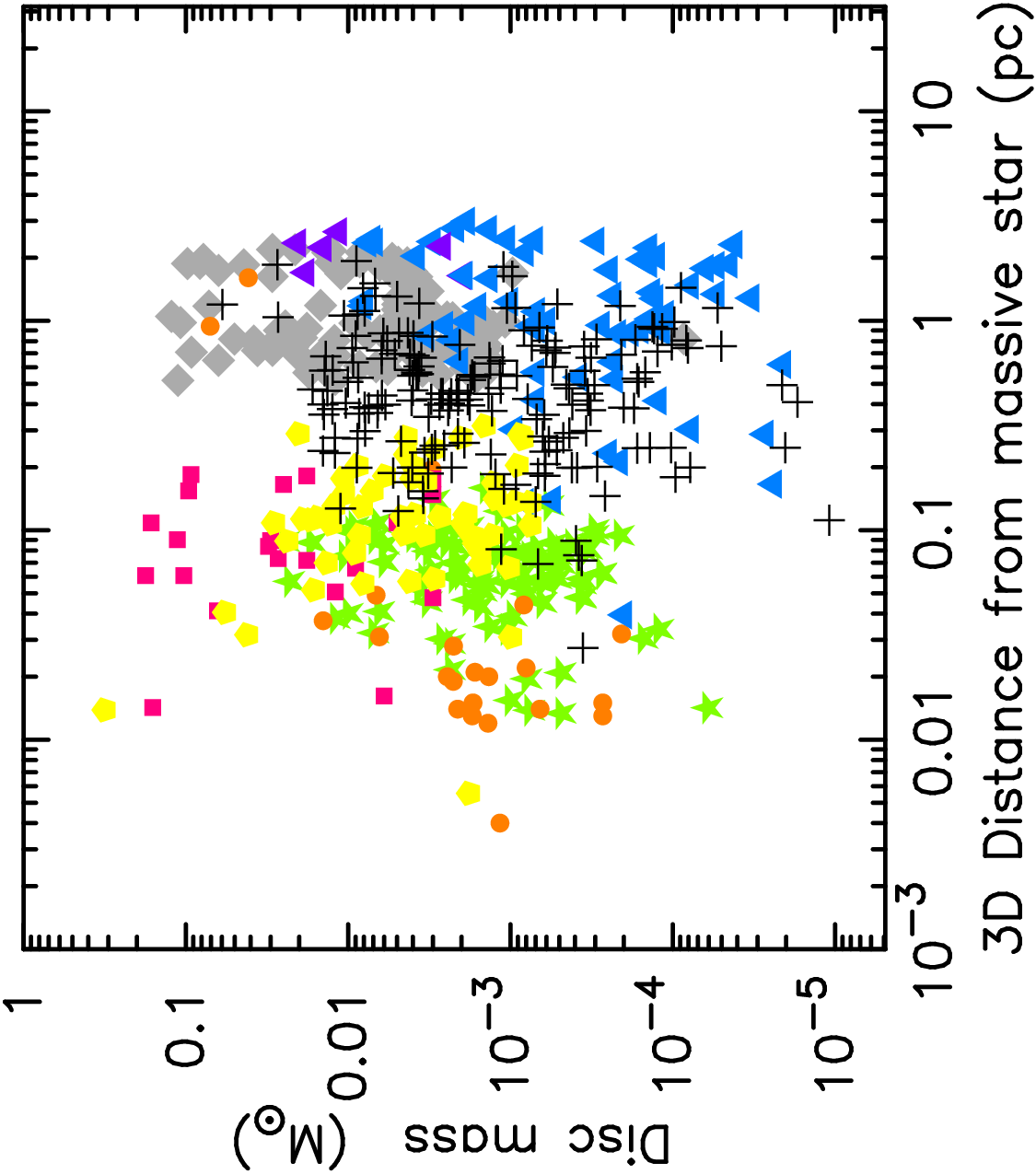}}}
\hspace*{0.3cm}\subfigure[4\,Myr]{\label{sim_2_3D_disc_10percent-d}\rotatebox{270}{\includegraphics[scale=0.2]{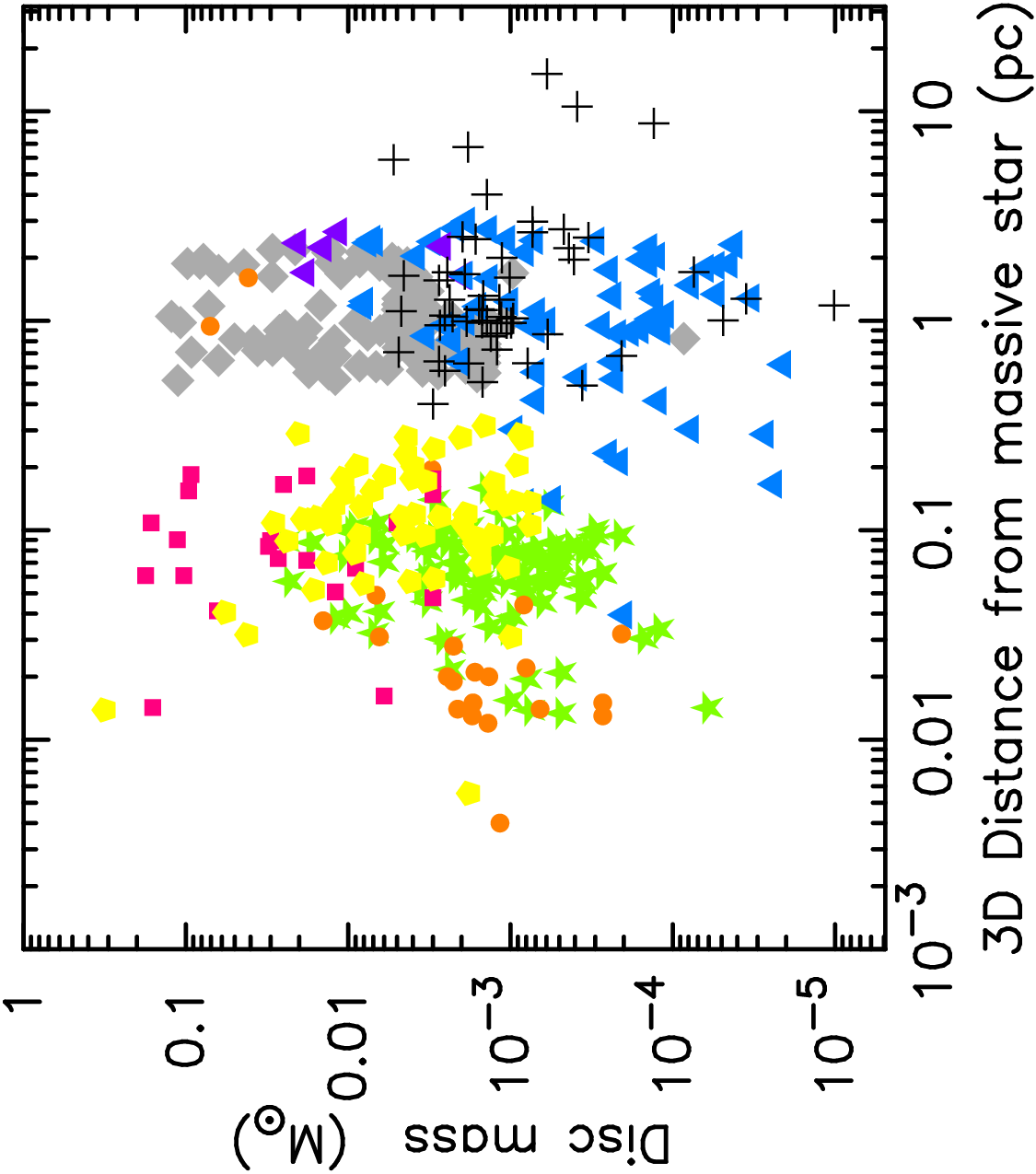}}}
\hspace*{-1.0cm}\subfigure[0\,Myr]{\label{sim_2_3D_disc_10percent-e}\rotatebox{270}{\includegraphics[scale=0.2]{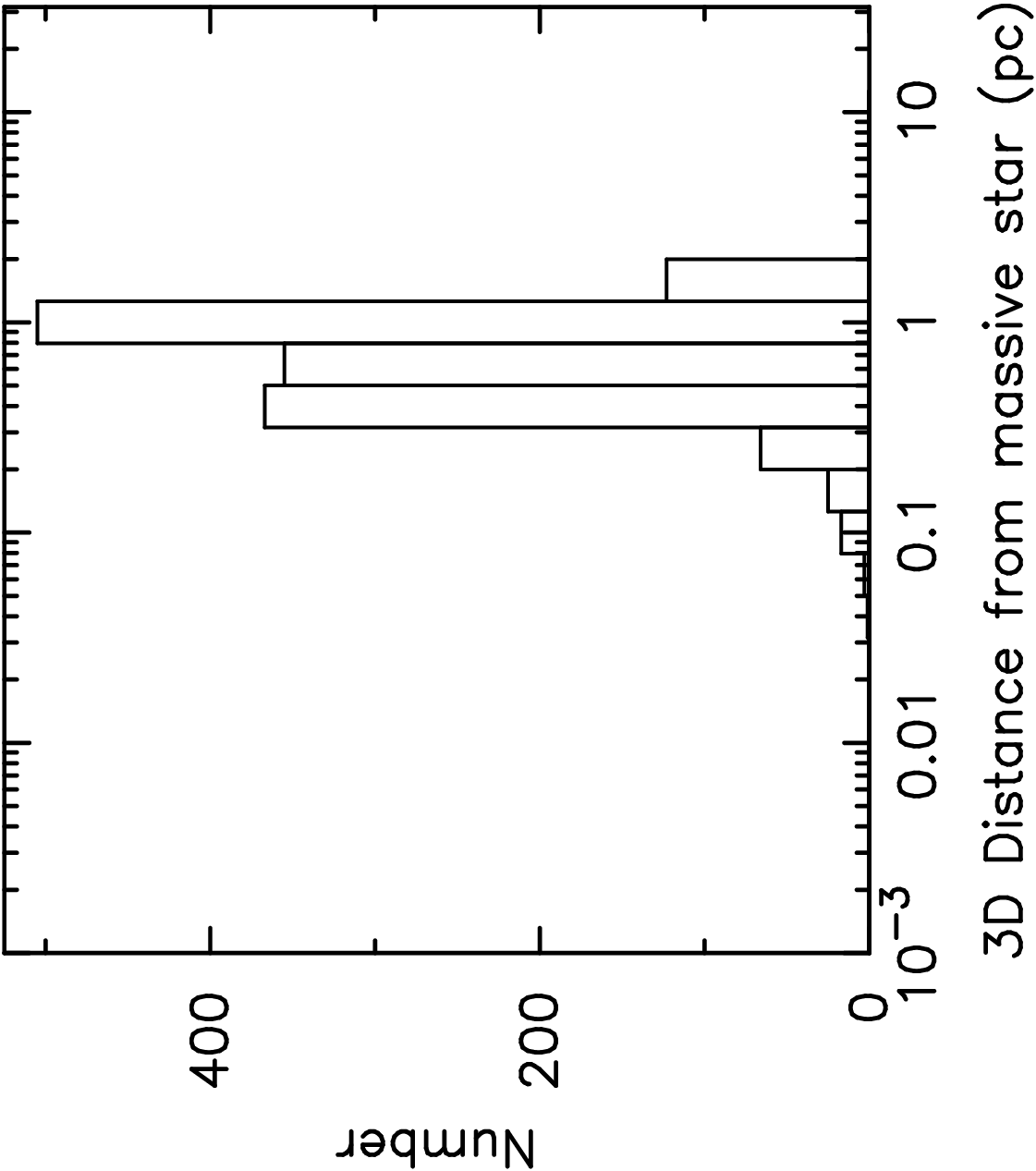}}}
\hspace*{0.3cm}
\subfigure[0.5\,Myr]{\label{sim_2_3D_disc_10percent-f}\rotatebox{270}{\includegraphics[scale=0.2]{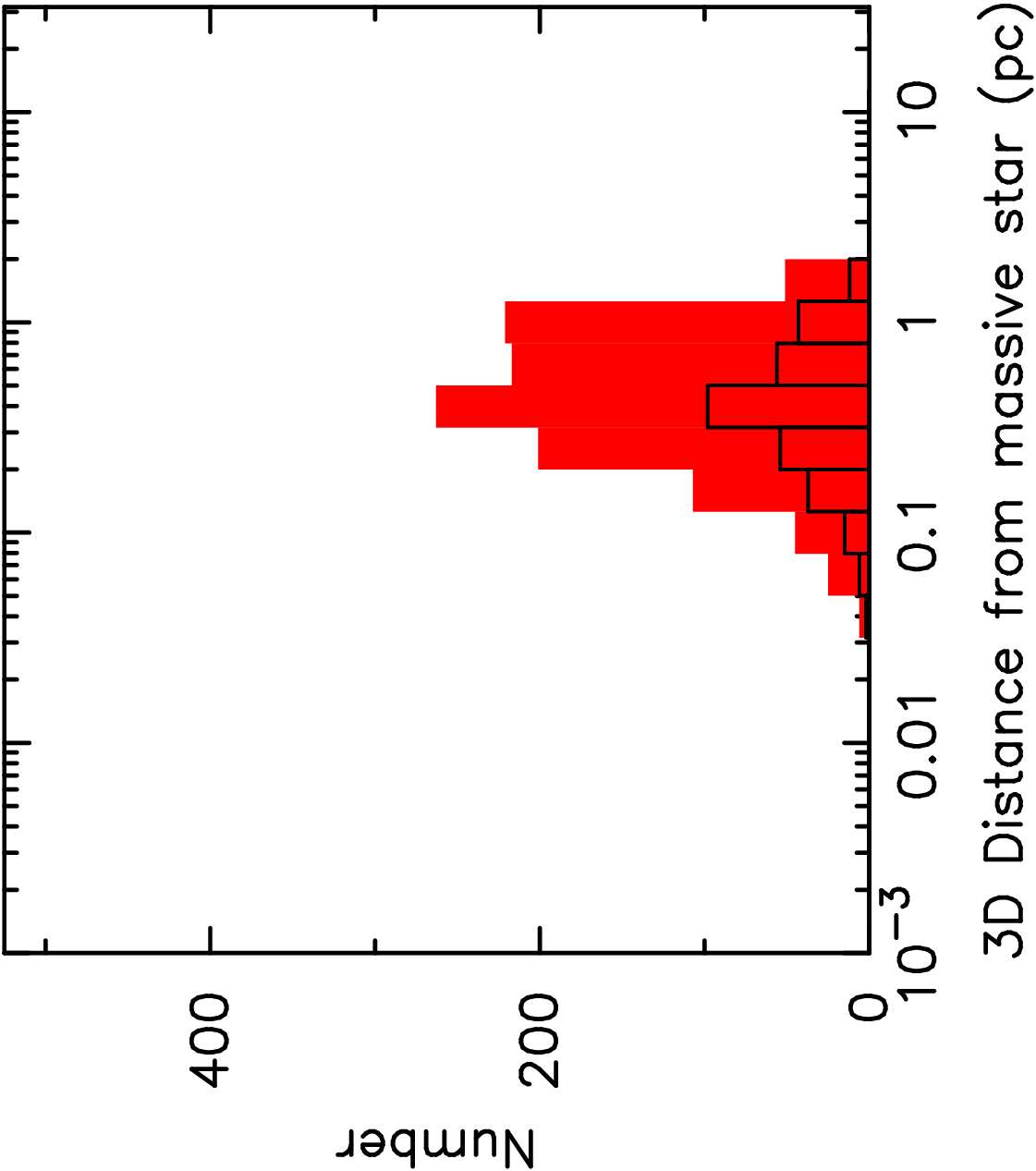}}}
\hspace*{0.3cm}\subfigure[1\,Myr]{\label{sim_2_3D_disc_10percent-g}\rotatebox{270}{\includegraphics[scale=0.2]{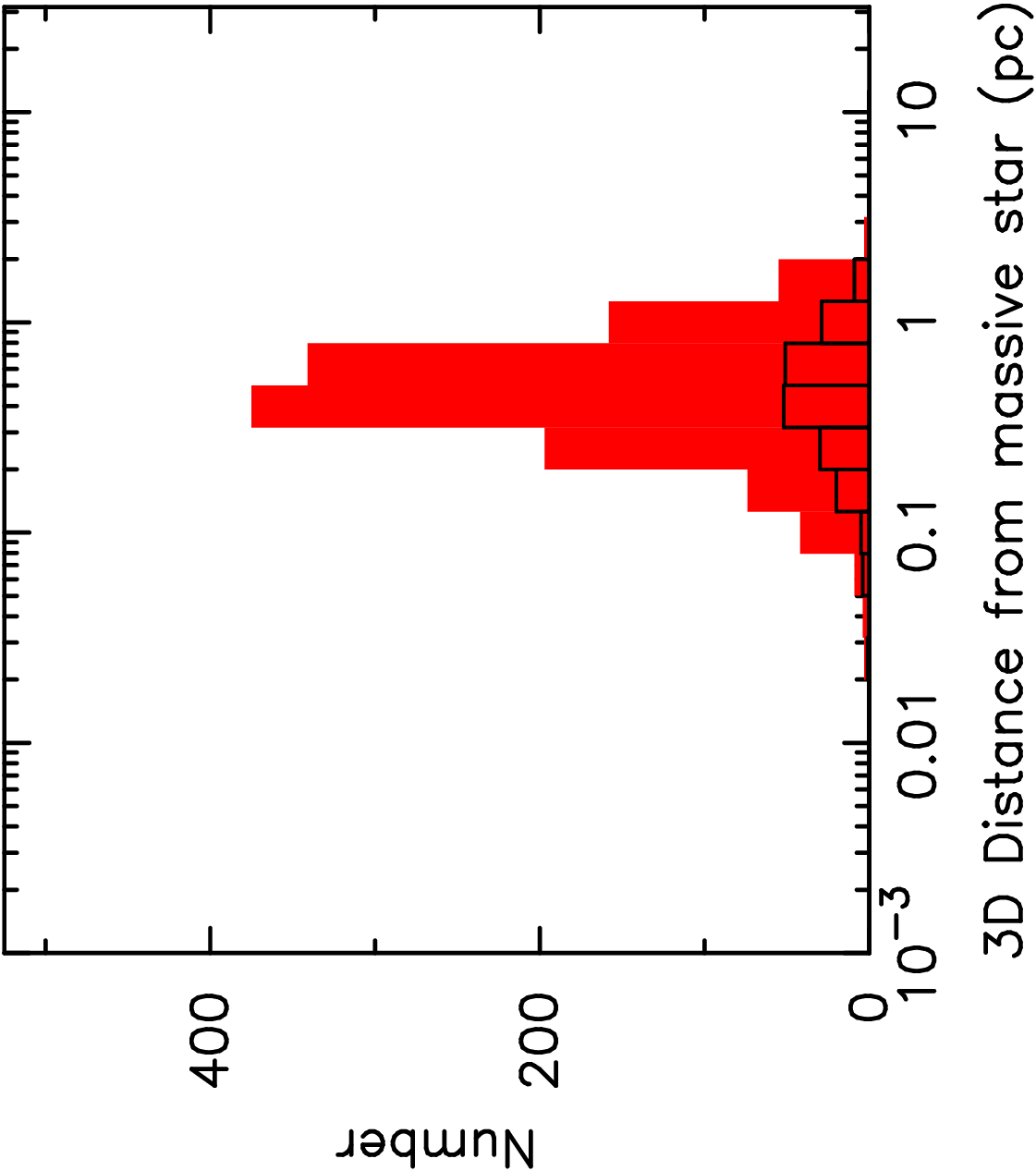}}}
\hspace*{0.3cm}\subfigure[4\,Myr]{\label{sim_2_3D_disc_10percent-h}\rotatebox{270}{\includegraphics[scale=0.2]{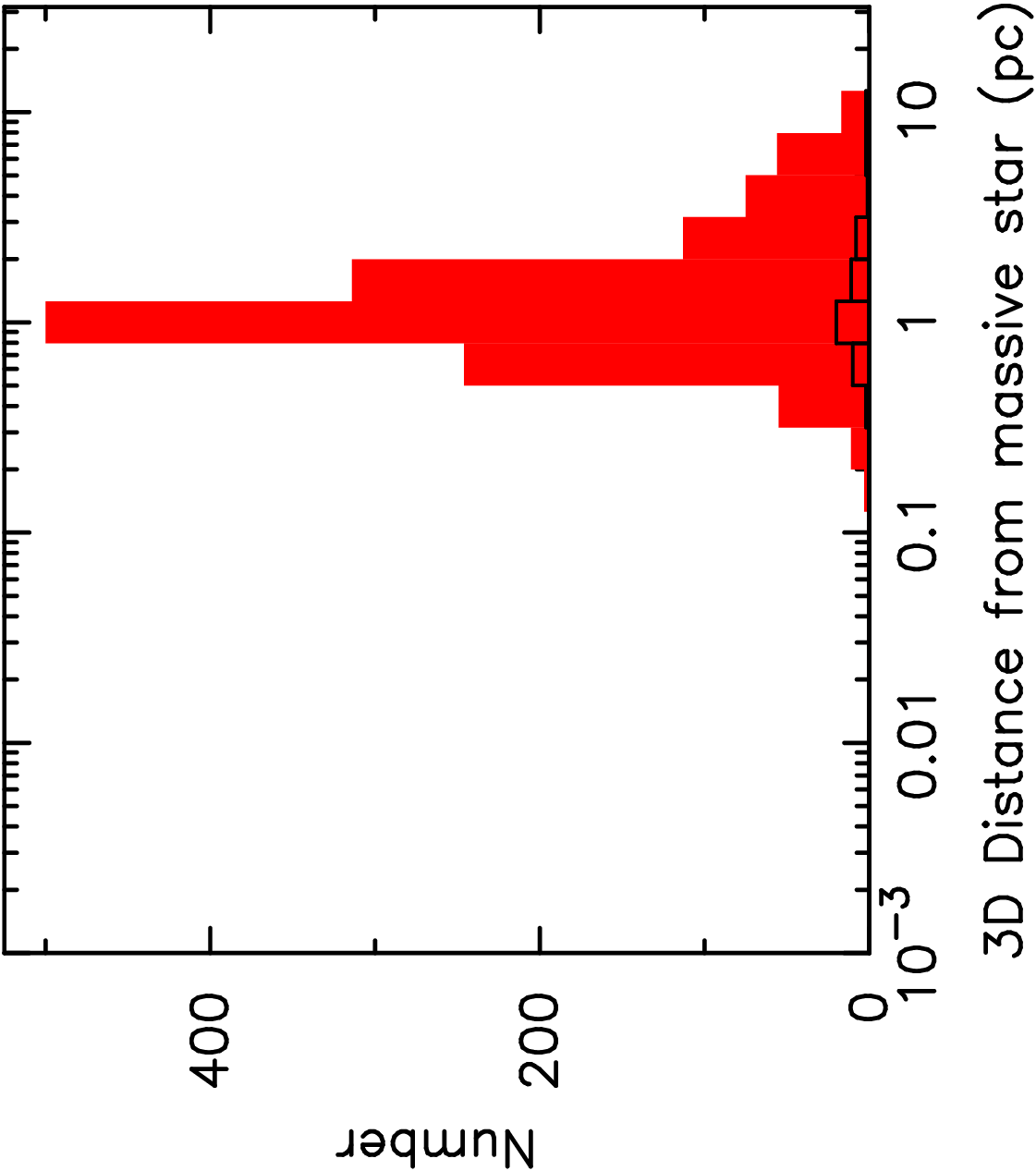}}} 
\caption{Photoevaporation of protoplanetary discs in high density star-forming regions ($10^4$\,stars\,pc$^{-3}$) where the initial disc mass is 10\,per cent of the host star's mass, and the distance to the ionising star is calculated in three dimensions. The radius of the star-forming region is 1\,pc. {\it Top row:} The plus symbols show the disc mass plotted against the distance to the massive ionising star for each low-mass star with a disc, using all three dimensions to calculate the distance. Stars whose discs have been completely evaporated are removed from the plot (hence the decreasing population of black points with increasing age). The masses (inferred from the dust content) and the respective distances from $\theta^1$~Ori~C of observed protoplanetary discs in the ONC  by \citet[][orange circles]{Mann14}, and \citet[][green stars]{Eisner18}; and the Orion Molecular Cloud \citep[][the grey diamonds]{vanTerwisga19} are shown. We also show the masses and projected distances from IRS~2b of discs in NGC\,2024 by \citet[][the raspberry squares]{Mann15} and \citet[][the yellow pentagons]{vanTerwisga20}.  Finally, we show the masses and projected distances from the ionising star $\sigma$~Ori in this eponymous star-forming region \citep{Ansdell17} by the blue triangles. Protoplanetary discs in  $\sigma$~Ori that have CO detections (i.e.\,\,gas content) are shown by the purple triangles. {\it Bottom row:} The number of remaining discs (open histogram) and the number of stars whose discs have been destroyed by photoevaporation (red histogram) as a function of the three dimensional projected distance from the ionising star. }
%Simulation 2, 10\% disc mass, 3D, distance to most massive star.
\label{sim_2_3D_disc_10percent}
  \end{center}
\end{figure*}

In Fig.~\ref{sim_2_3D_disc_10percent} we show the same simulation as in Fig.~\ref{sim_2_2D_disc_10percent}, but this time we use the full three dimensional information to calculate the positions of low-mass stars (including those with discs) with respect to the ionising massive star. Blinking between these two figures highlights the problem with projecting distances in two dimensions. %As shown in Fig.~\ref{projection_effects-c}, if the effects of photoevaporation abruptly end at some distance from the massive star(s), then this feature is washed out in a two-dimensional projection.

After 4\,Myr of dynamical evolution and mass-loss from discs due to photoevaporation, all of the discs within 0.5\,pc of the cluster centre have lost the gas content of their discs (shown by the absence of black points near the ionising star in Fig.~\ref{sim_2_3D_disc_10percent-d}). In the two-dimensional projection, this boundary is blurred (Fig.~\ref{sim_2_2D_disc_10percent}), with discs appearing to be much closer to the ionising stars in 2D projection (though none are as close to the ionising star as some of the observed discs in the ONC).}% in In our simulations, FUV radiation acts within 0.3\,pc of the most massive star but is ineffective beyond this distance. This can be seen in Figs.~\ref{sim_2_3D_disc_10percent-b}--\ref{sim_2_3D_disc_10percent-d}, where there is a sharp cut-off in stars hosting discs within 0.3\,pc of the most massive star. In the two-dimensional projection, this boundary is blurred (Fig.~\ref{sim_2_2D_disc_10percent}),  with the hints of a trend between disc mass and distance from most massive star in Figs.~\ref{sim_2_2D_disc_10percent-b}~and~\ref{sim_2_2D_disc_10percent-c} not present in the full three-dimensional projection (Figs.~\ref{sim_2_3D_disc_10percent-b}~and~\ref{sim_2_3D_disc_10percent-c}).

The histograms of destroyed discs (red) and surviving discs (black, open) show a similar trend to the two dimensional data (Figs.~\ref{sim_2_3D_disc_10percent-e}--\ref{sim_2_3D_disc_10percent-h}), where stars closest to the ionising sources are less likely to have discs, but disc-less stars can also be found at large distances from the massive stars.\\

For the remainder of our analysis, we will only consider two dimensional projections of the data, as this is the information observational studies are currently party to. However, we must bear in mind the caveat that some fore- and background stars will be present in our projections.

\subsubsection{1\,per cent disc masses, 2D projection}

When we follow the evolution of protoplanetary discs with initial masses of only 1\,per cent of that of the host star, we see that discs are rapidly destroyed in our dense star-forming regions \citep[cf.][]{Nicholson19a}. In Fig.~\ref{sim_2_2D_disc_1percent} we show the positions of disc-hosting stars with respect to the ionising massive stars, and very few discs survive beyond ages of 0.5  - 1\,Myr  (fewer than ten from an initial population of $\sim1460$ discs, depending on the initial conditions). This suggests that if protoplanetary discs are to form gas giant planets in young star-forming regions where photoevaporation is active, they require much higher initial masses (of order 10\,per cent of the host star's mass). %1461 initially

\begin{figure*}
  \begin{center}
\setlength{\subfigcapskip}{10pt}
%\hspace*{-1.0cm}\subfigure[0\,Myr]{\label{sim_2_2D_disc_1percent-a}\rotatebox{270}{\includegraphics[scale=0.2]{disc_mass_ion_dist_Or_C0p3F2p01pSmFS10_14_Fe_fuv+euv_0Myr_1percent.ps}}}
%\hspace*{0.3cm}
%\subfigure[0.5\,Myr]{\label{sim_2_2D_disc_1percent-b}\rotatebox{270}{\includegraphics[scale=0.2]{disc_mass_ion_dist_Or_C0p3F2p01pSmFS10_14_Fe_fuv+euv_0p5Myr_1percent.ps}}}
%\hspace*{0.3cm}\subfigure[1\,Myr]{\label{sim_2_2D_disc_1percent-c}\rotatebox{270}{\includegraphics[scale=0.2]{disc_mass_ion_dist_Or_C0p3F2p01pSmFS10_14_Fe_fuv+euv_1Myr_1percent.ps}}}
%\hspace*{0.3cm}\subfigure[4\,Myr]{\label{sim_2_2D_disc_1percent-d}\rotatebox{270}{\includegraphics[scale=0.2]{disc_mass_ion_dist_Or_C0p3F2p01pSmFS10_14_Fe_fuv+euv_4Myr_1percent.ps}}}  

\hspace*{-1.0cm}\subfigure[0\,Myr]{\label{sim_2_2D_disc_1percent-a}\rotatebox{270}{\includegraphics[scale=0.2]{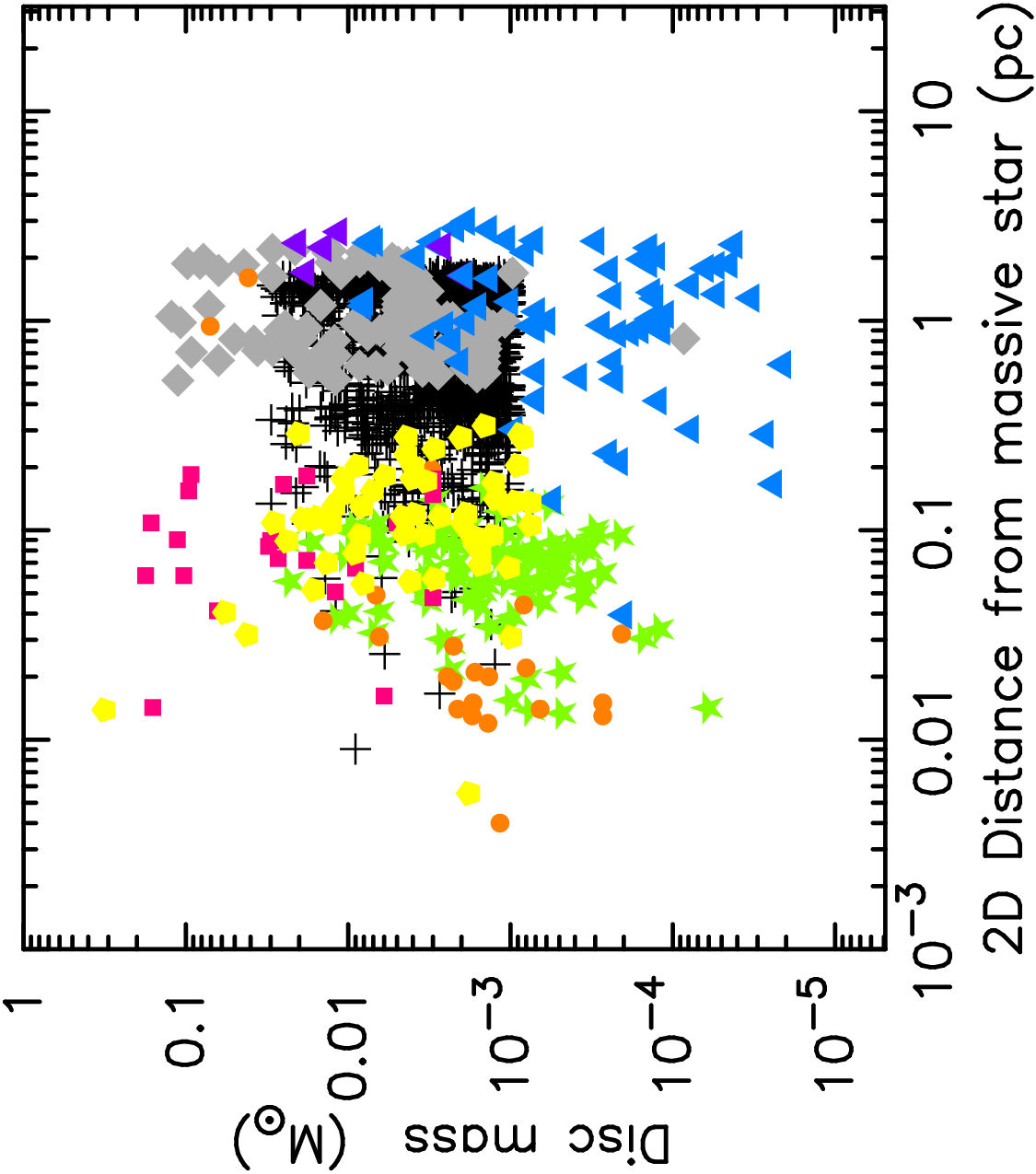}}}
\hspace*{0.3cm}
\subfigure[0.5\,Myr]{\label{sim_2_2D_disc_1percent-b}\rotatebox{270}{\includegraphics[scale=0.2]{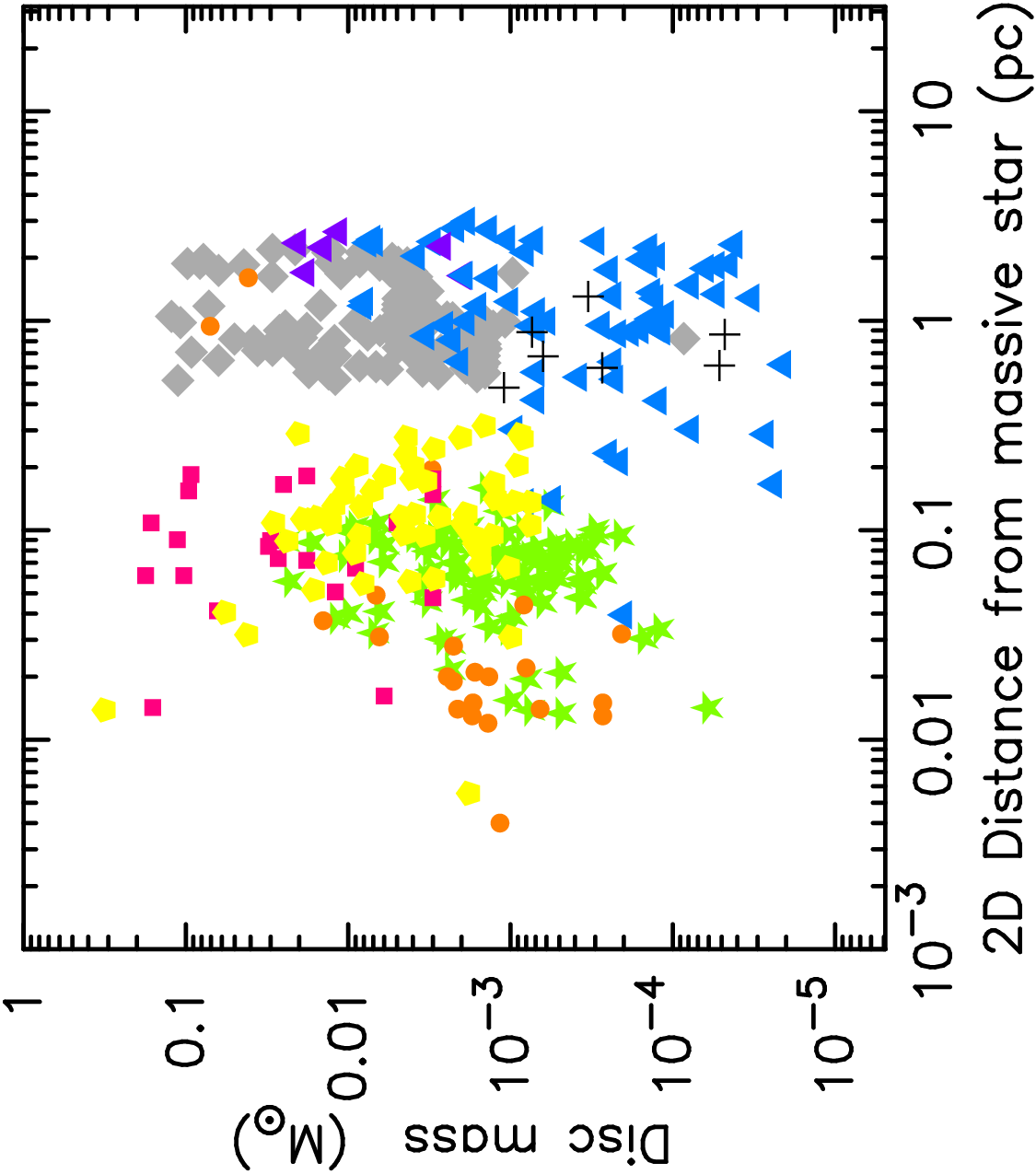}}}
\hspace*{0.3cm}\subfigure[1\,Myr]{\label{sim_2_2D_disc_1percent-c}\rotatebox{270}{\includegraphics[scale=0.2]{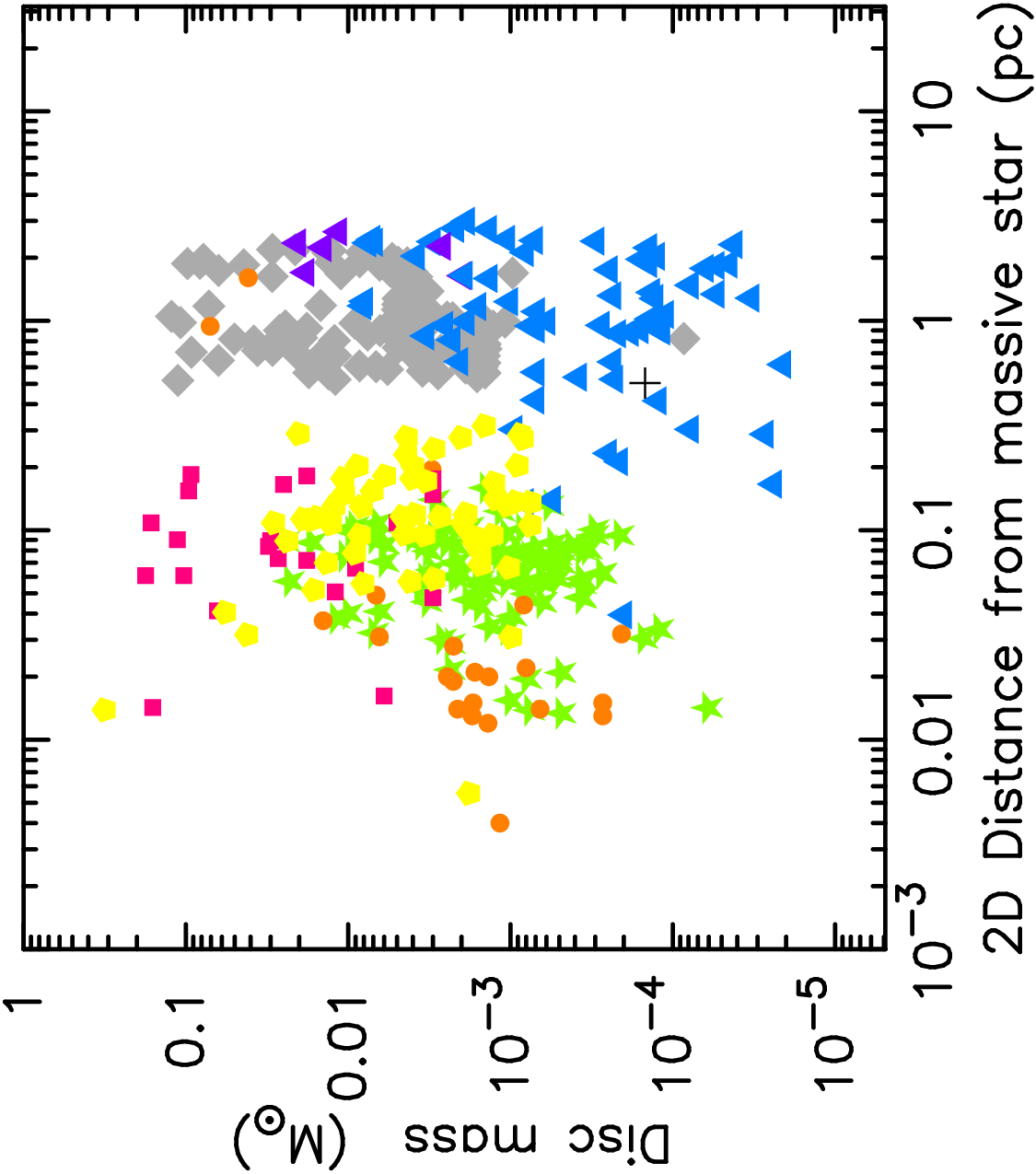}}}
\hspace*{0.3cm}\subfigure[4\,Myr]{\label{sim_2_2D_disc_1percent-d}\rotatebox{270}{\includegraphics[scale=0.2]{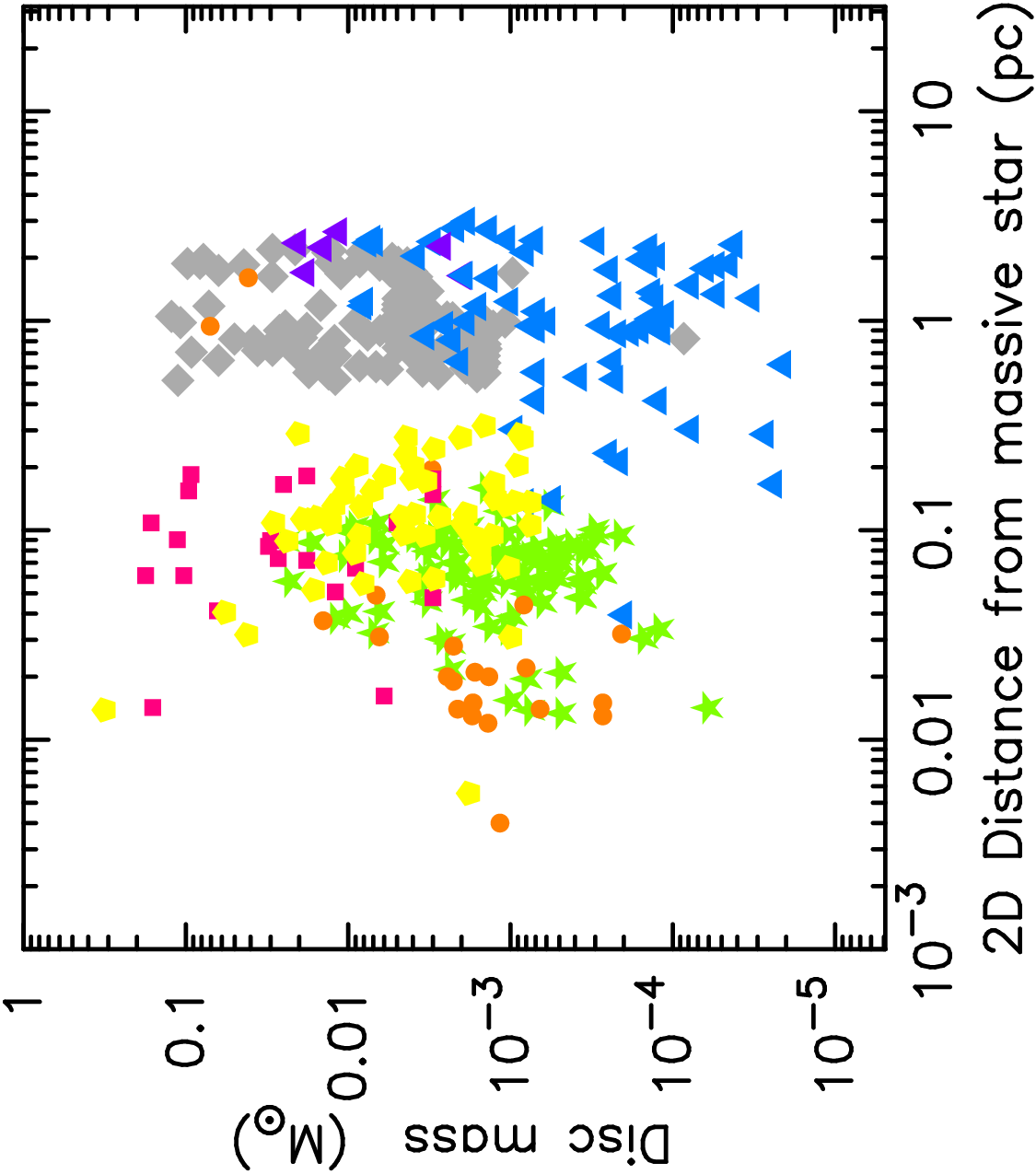}}}  
\caption{Photoevaporation of protoplanetary discs in high density star-forming regions ($10^4$\,stars\,pc$^{-3}$)  where the initial disc mass is 1\,per cent of the host star's mass. The radius of the star-forming region is 1\,pc. The plus symbols show the disc mass plotted against the distance to the massive ionising star for each low-mass star with a disc, using only two dimensions to calculate the projected distance. Stars whose discs have been completely evaporated are removed from the plot (hence the decreasing population of black points with increasing age).  The masses (inferred from the dust content) and the respective distances from $\theta^1$~Ori~C of observed protoplanetary discs in the ONC by \citet[][orange circles]{Mann14}, and \citet[][green stars]{Eisner18}; and the Orion Molecular Cloud \citep[][the grey diamonds]{vanTerwisga19} are shown. We also show the masses and projected distances from IRS~2b of discs in NGC\,2024 by \citet[][the raspberry squares]{Mann15} and \citet[][the yellow pentagons]{vanTerwisga20}.  Finally, we show the masses and projected distances from the ionising star $\sigma$~Ori in this eponymous star-forming region \citep{Ansdell17} by the blue triangles. Protoplanetary discs in  $\sigma$~Ori that have CO detections (i.e.\,\,gas content) are shown by the purple triangles.}
%Simulation 2, 1\% disc mass, 2D, distance to most massive star. 
\label{sim_2_2D_disc_1percent}
  \end{center}
\end{figure*}

\subsection{Photoevaporation in low-density simulations} 

\subsubsection{Dynamical evolution of the star-forming regions}

Our low-density simulations take longer to relax and form a star cluster, and this process has not reached completion after the first 4\,Myr of dynamical evolution. As a result the most massive star is not necessarily in the centre of the star-forming region  when we determine the distance of each low-mass disc-bearing star to the ionising source. These lower density regions still contain several (typically 1 -- 5) O-type stars, as well as up to $\sim$10 B-type stars, but have a much lower stellar density ($\sim$100\,stars\,pc$^{-3}$). This results in a background radiation field of between 100 -- 1000\,$G_0$, which is a factor of at least ten lower than in the high density simulations. The masses of the five most massive stars in the simulation we use to illustrate the destruction of discs in low-density environments are 19, 19, 19, 12 and 11\,M$_\odot$. In Appendix~\ref{appendix:mass-function} we show that the results -- as for the high density regions -- are not dependent on the exact realisation of the Initial Mass Function.

\subsubsection{10\,per cent disc masses, 2D projection}

When the discs are set to be 10\,per cent of the host star's mass  in our low density simulations, photoevaporation is still very effective at destroying discs (Fig.~\ref{sim_4_2D_disc_10percent_5pc}), likely due to the still very high FUV radiation field (500\,$G_0$ in this particular simulation). From an initial population of 1465 discs, after 0.5\,Myr 1198 discs remain, reducing to 526 discs after 4\,Myr. As in the higher density simulations, we see very few instances of a positive correlation between disc mass and distance from the ionising stars in these simulations.

\begin{figure*}
  \begin{center}
\setlength{\subfigcapskip}{10pt}
\hspace*{-1.0cm}\subfigure[0\,Myr]{\label{sim_4_2D_disc_10percent_5pc-a}\rotatebox{270}{\includegraphics[scale=0.2]{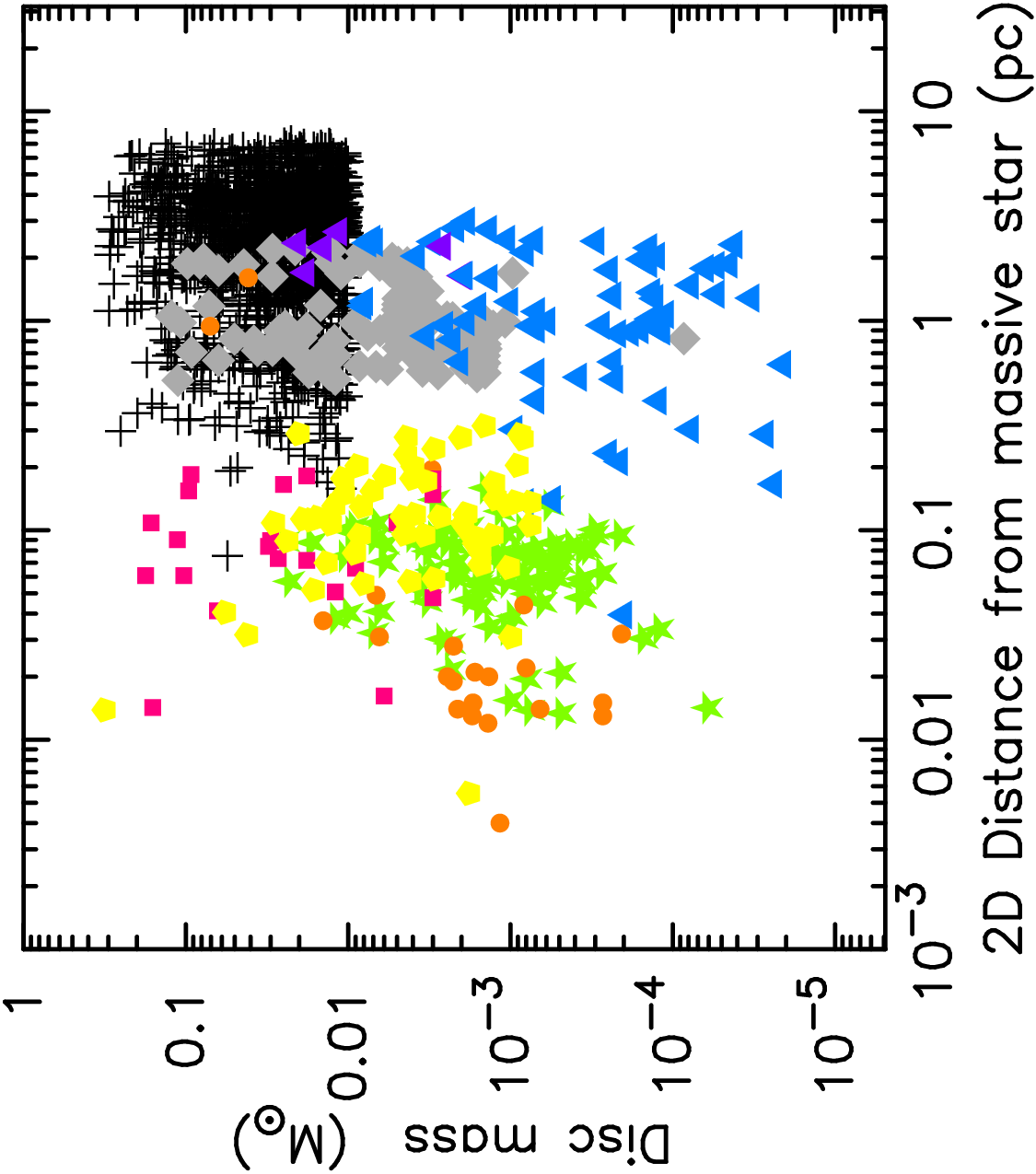}}}
\hspace*{0.3cm}
\subfigure[0.5\,Myr]{\label{sim_4_2D_disc_10percent_5pc-b}\rotatebox{270}{\includegraphics[scale=0.2]{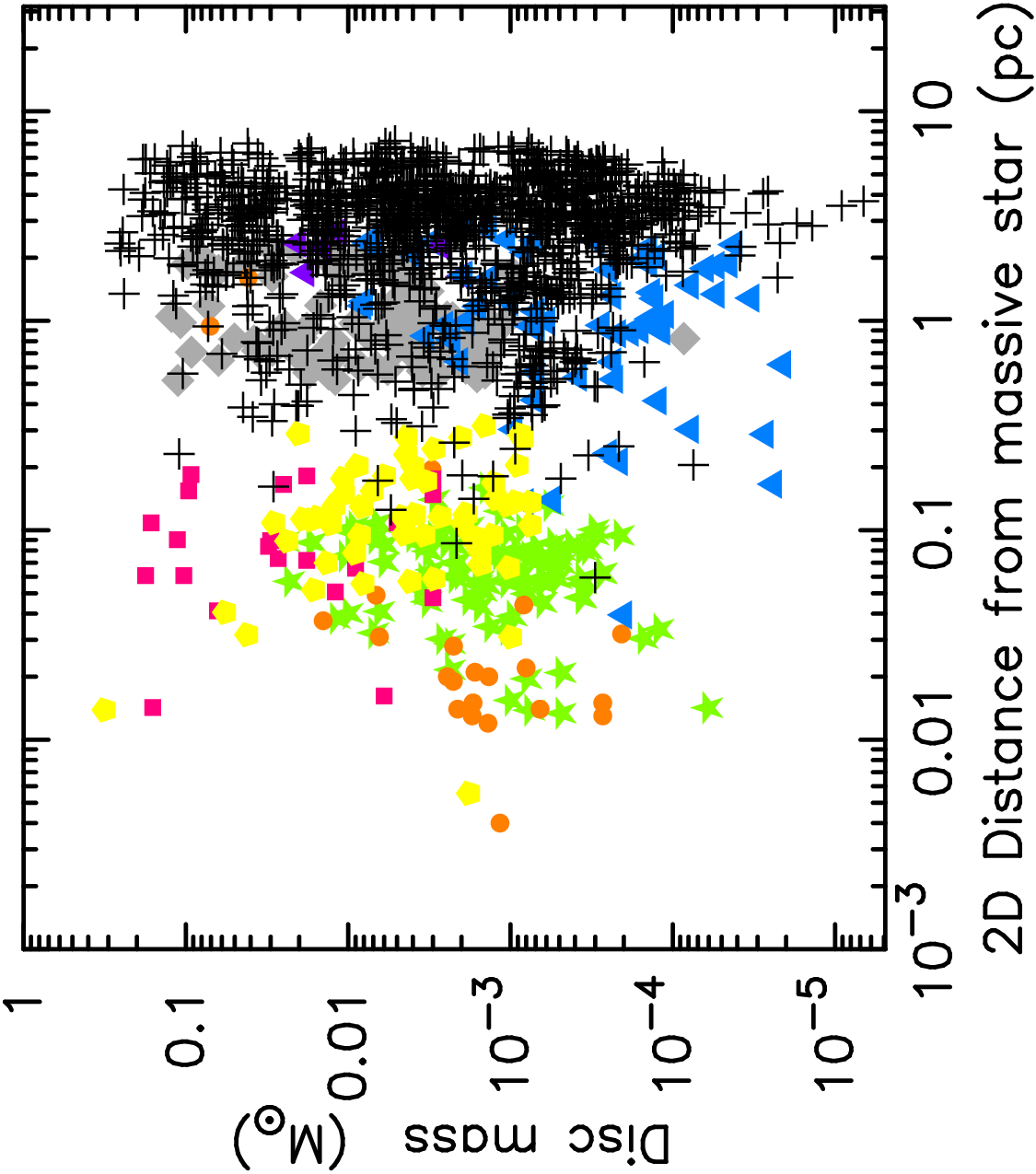}}}
\hspace*{0.3cm}\subfigure[1\,Myr]{\label{sim_4_2D_disc_10percent_5pc-c}\rotatebox{270}{\includegraphics[scale=0.2]{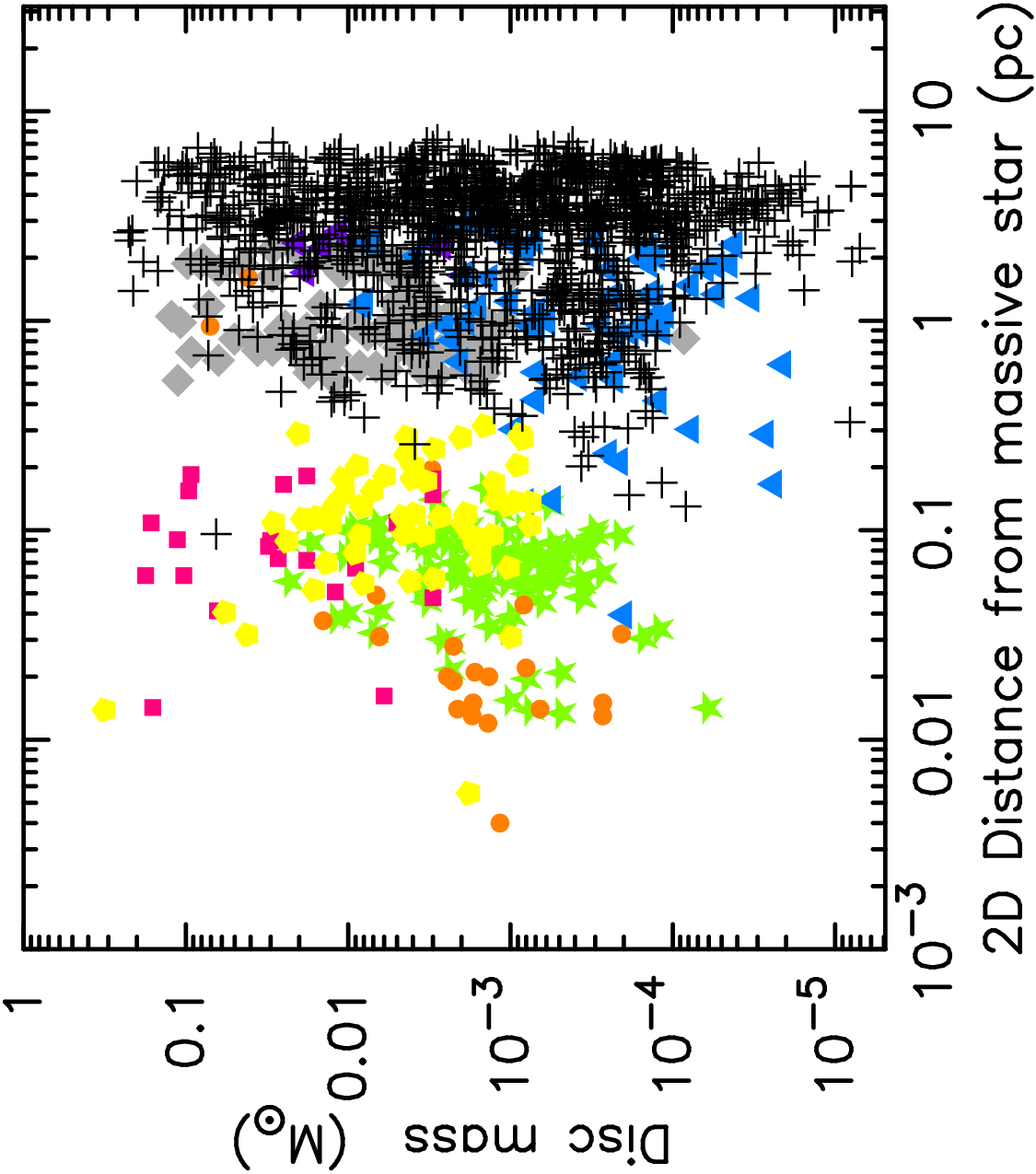}}}
\hspace*{0.3cm}\subfigure[4\,Myr]{\label{sim_4_2D_disc_10percent_5pc-d}\rotatebox{270}{\includegraphics[scale=0.2]{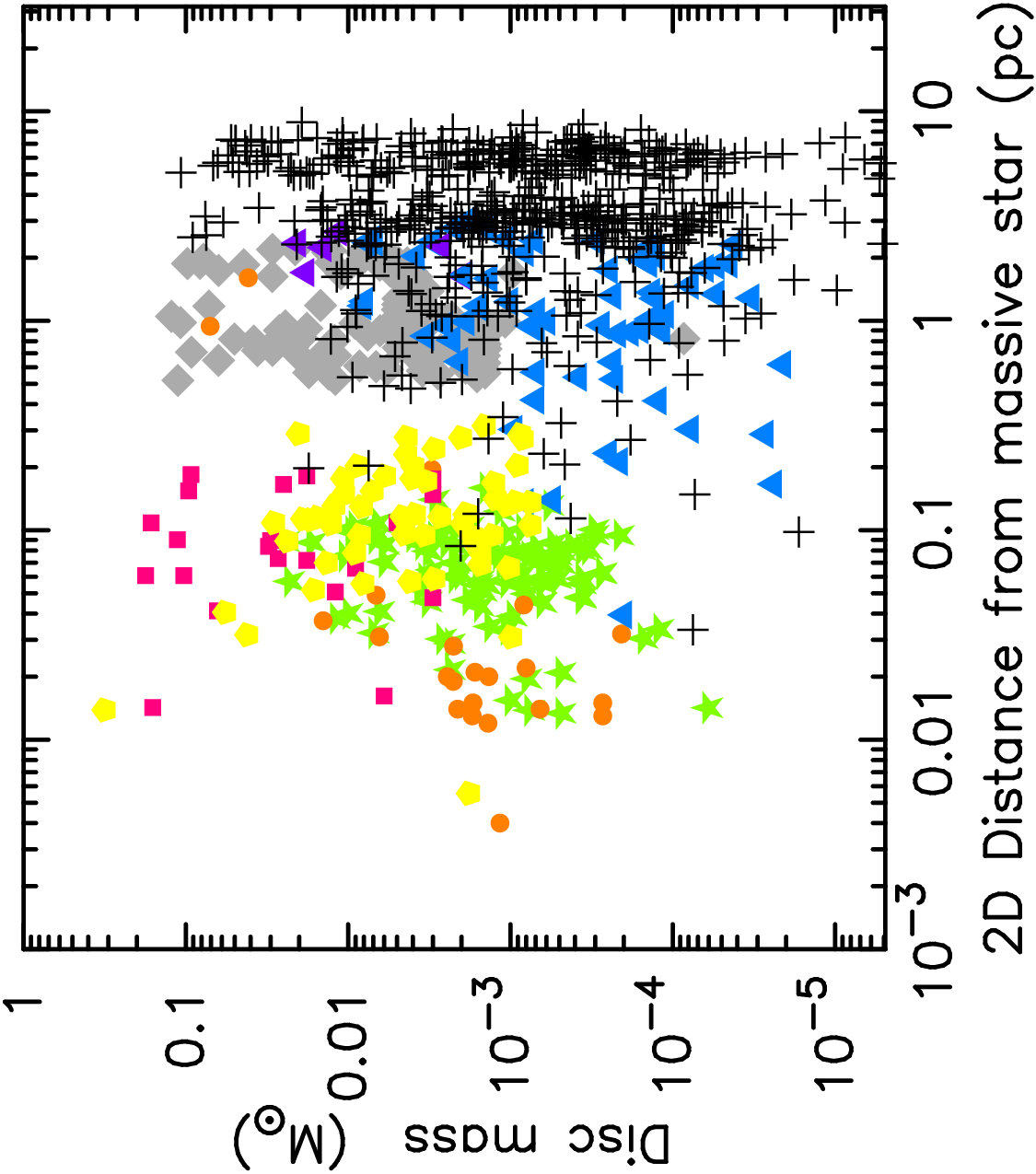}}}  
\hspace*{-1.0cm}\subfigure[0\,Myr]{\label{sim_4_2D_disc_10percent_5pc-e}\rotatebox{270}{\includegraphics[scale=0.2]{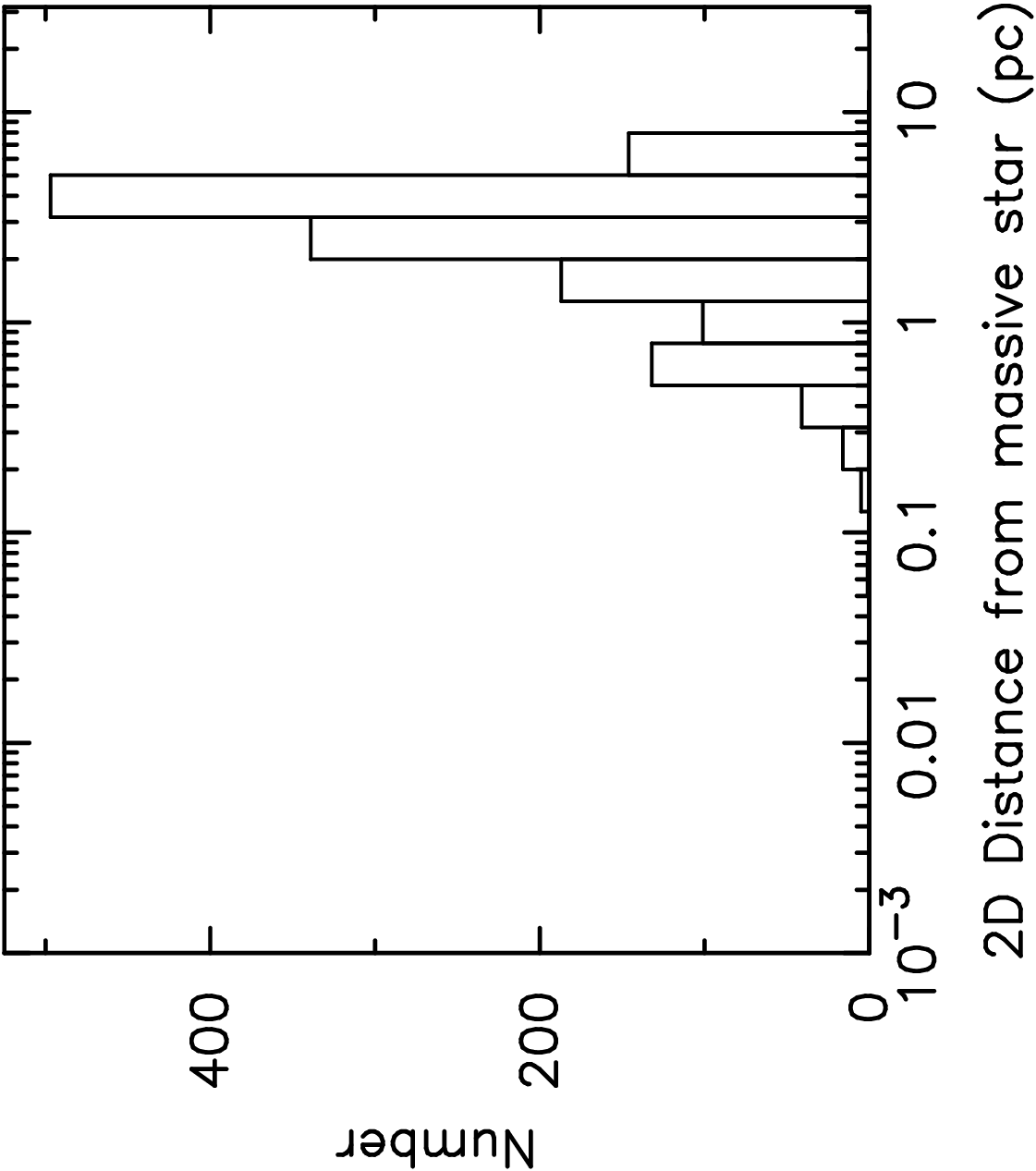}}}
\hspace*{0.3cm}
\subfigure[0.5\,Myr]{\label{sim_4_2D_disc_10percent_5pc-e}\rotatebox{270}{\includegraphics[scale=0.2]{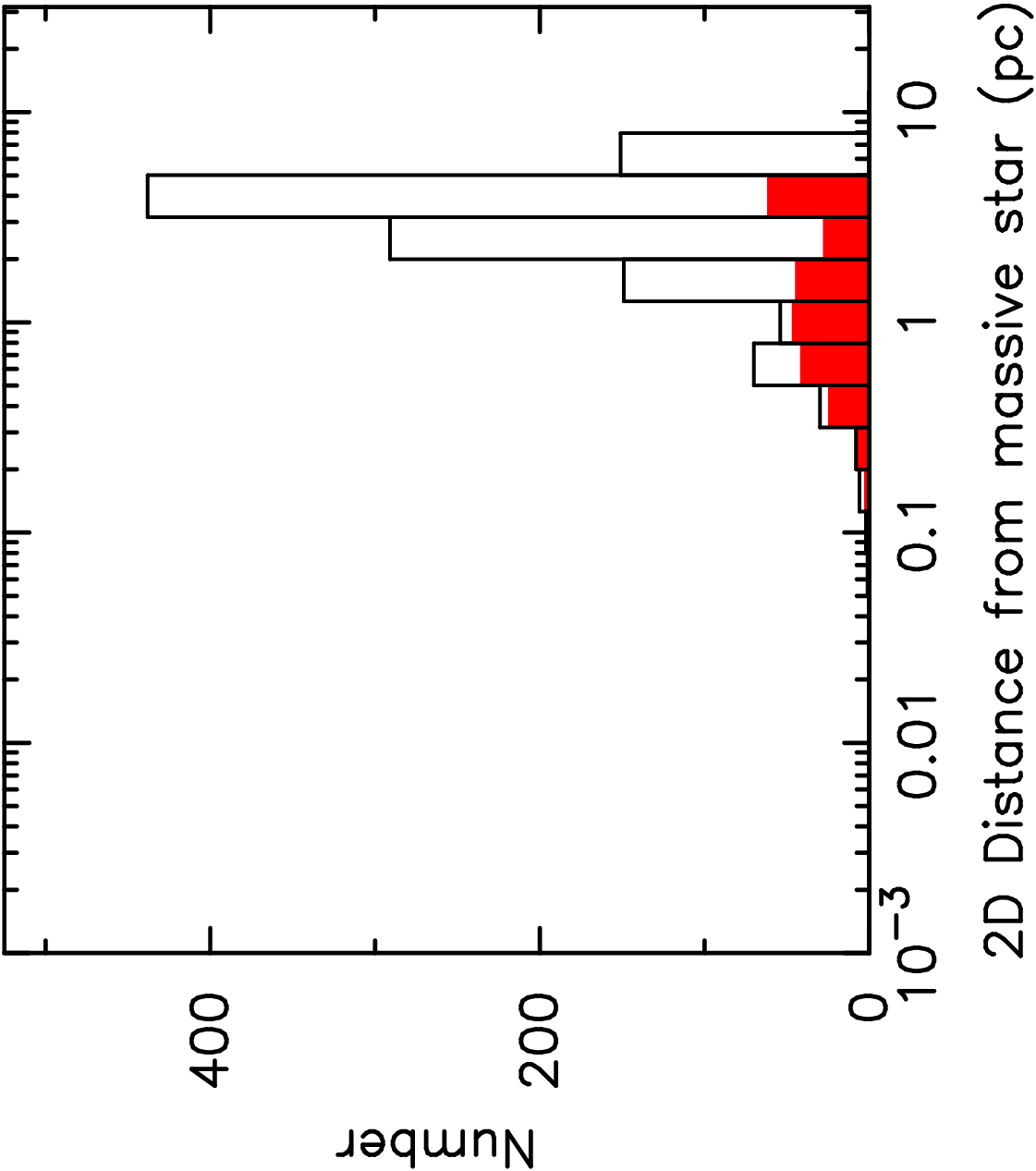}}}
\hspace*{0.3cm}\subfigure[1\,Myr]{\label{sim_4_2D_disc_10percent_5pc-g}\rotatebox{270}{\includegraphics[scale=0.2]{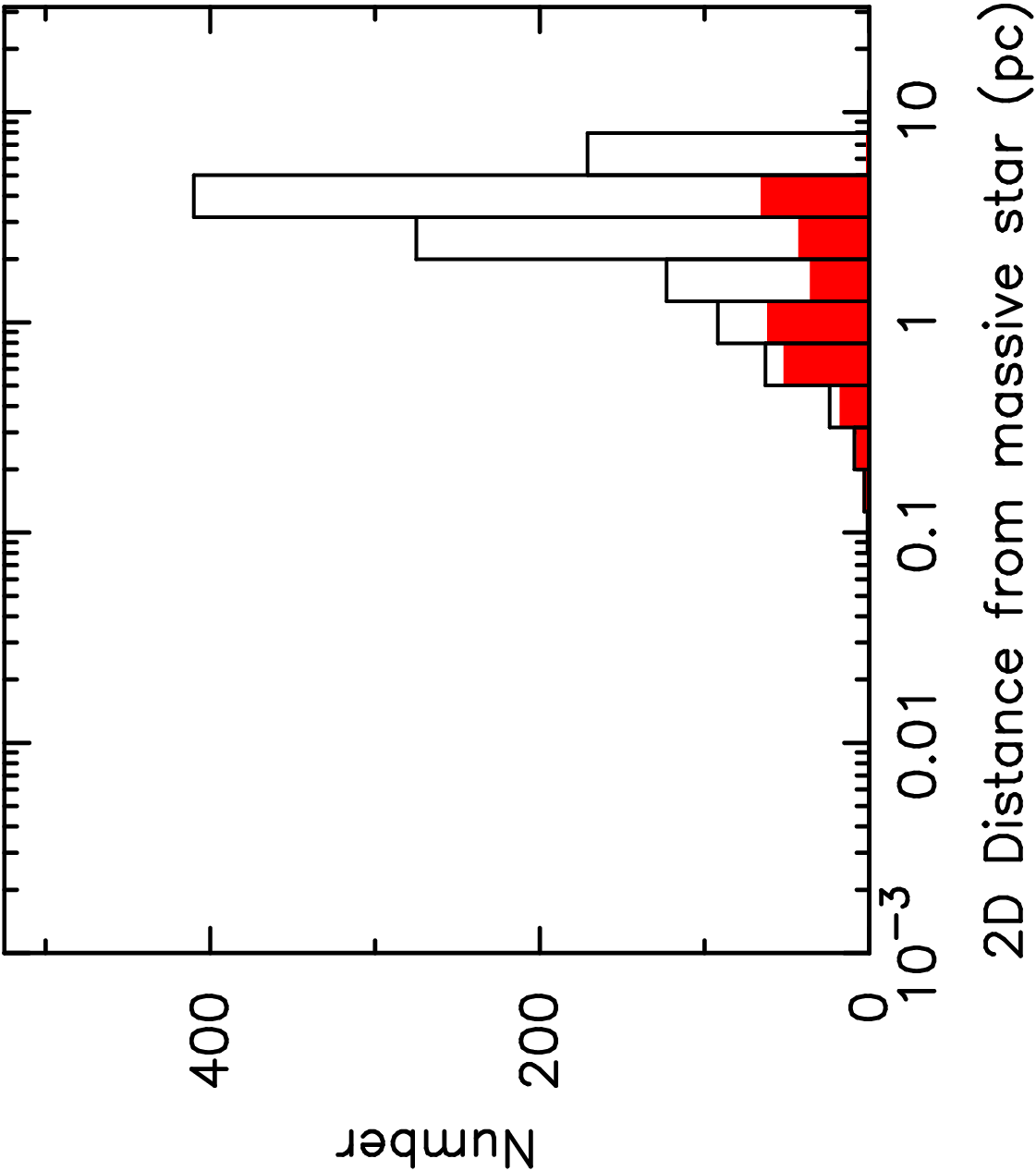}}}
\hspace*{0.3cm}\subfigure[4\,Myr]{\label{sim_4_2D_disc_10percent_5pc-h}\rotatebox{270}{\includegraphics[scale=0.2]{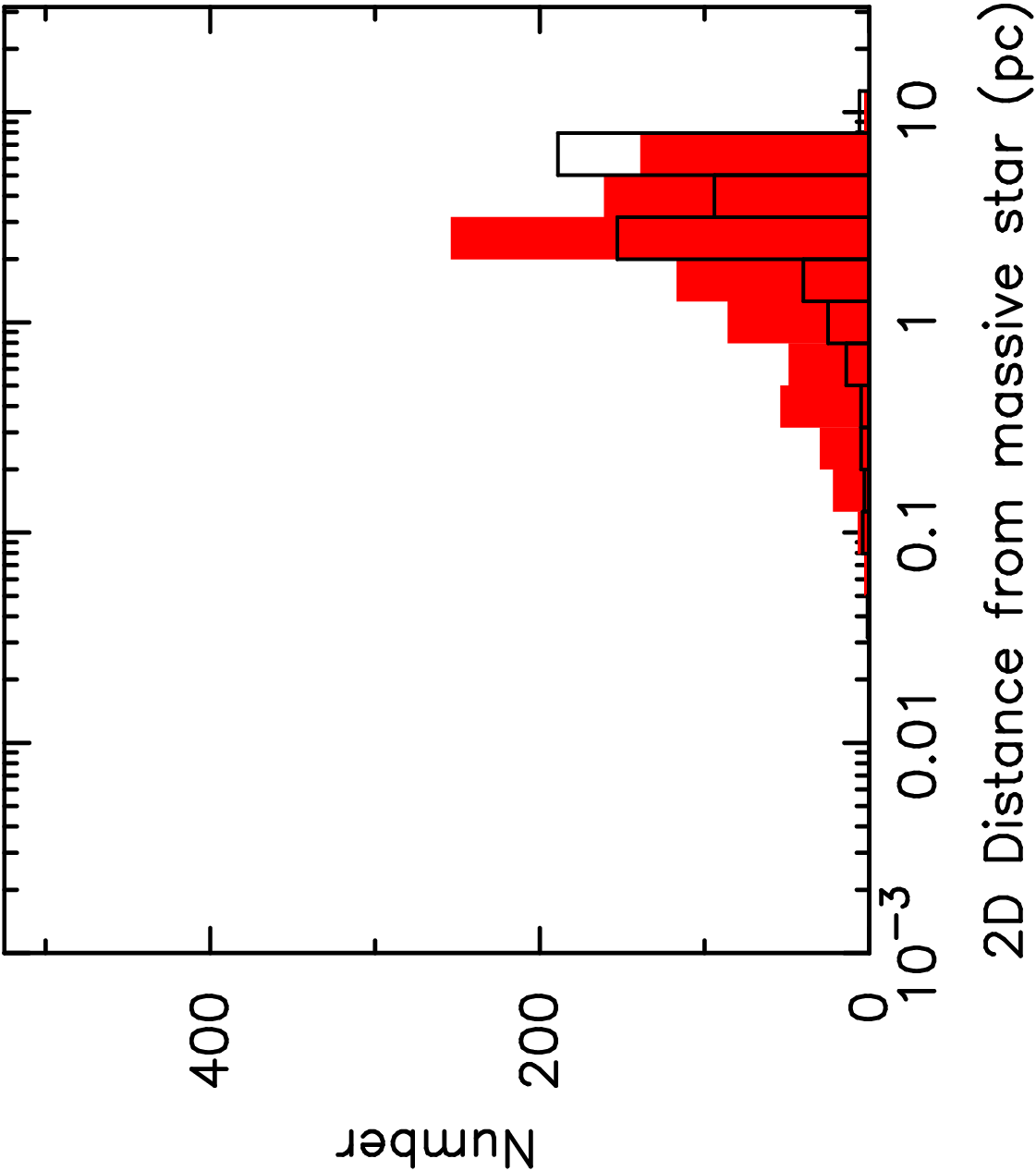}}}  
\caption{Photoevaporation of protoplanetary discs in low density star-forming regions ($10^2$\,stars\,pc$^{-3}$)  where the initial disc  mass is 10\,per cent of the host star's mass. The radius of the star-forming region is 5\,pc. {\it Top row:}  The plus symbols show the disc mass plotted against the distance to the massive ionising star for each low-mass star with a disc, using only two dimensions to calculate the projected distance. Stars whose discs have been completely evaporated are removed from the plot (hence the decreasing population of black points with increasing age). The masses (inferred from the dust content) and the respective distances from $\theta^1$~Ori~C of observed protoplanetary discs in the ONC \citep[][orange circles]{Mann14}, \citep[][green stars]{Eisner18}; and the Orion Molecular Cloud \citep[][the grey diamonds]{vanTerwisga19} are shown. We also show the masses and projected distances from IRS~2b of discs in NGC\,2024 by \citet[][the raspberry squares]{Mann15} and \citet[][the yellow pentagons]{vanTerwisga20}.  Finally, we show the masses and projected distances from the ionising star $\sigma$~Ori in this eponymous star-forming region \citep{Ansdell17} by the blue triangles.  Protoplanetary discs in  $\sigma$~Ori that have CO detections (i.e.\,\,gas content) are shown by the purple triangles. {\it Bottom row:} The number of remaining discs (open histogram) and the number of stars whose discs have been destroyed by photoevaporation (red histogram) as a function of the two dimensional projected distance from the ionising star.}
%Simulation 2, 10\% disc mass, 2D, distance to most massive star, SFR radius 5\,pc. 
\label{sim_4_2D_disc_10percent_5pc}
  \end{center}
\end{figure*}

In Fig.~\ref{sim_4_2D_disc_10percent_5pc} we also show the histograms of the distances to the most massive stars at 0, 0.5, 1 and 4\,Myr. These low-density regions take longer to dynamically relax than the higher density regions, and as such the histograms of remaining (open) and destroyed (red) discs differ to those in the high density regions until and age of $\sim$4\,Myr, when the region has dynamically mixed. In these regions the  destroyed discs tend to follow the same spatial distribution as the surviving discs until the region has collapsed to form a centrally concentrated cluster. At this point (after $\sim$4\,Myr), the stars with destroyed discs tend to be closer to the ionising stars than the stars with surviving discs.

We plot the average disc mass as a function of distance from the ionising star in Fig.~\ref{2D_disc_10percent_averages_LD}. As in Fig.~\ref{2D_disc_10percent_averages}, we show the rolling average of disc mass as a function of distance from the most massive stars for each simulation and again,  we only show the first ten realisations of the twenty simulations in plots for the sake of clarity. Because these simulations are lower density, the corresponding radiation fields will be smaller (by a similar factor to the decrease in density, i.e.\,\,a factor of $\sim$100 \citep{Parker21a}). This means that more discs survive for longer, and so we increase the size of our incremental bins to contain fifty stars. As before, where the Spearman rank test reports a significant ($p$-value $<0.1$) positive correlation, we show the rolling average by a coloured line. Where there is a significant negative correlation ($p$-value $<0.1$), we show the rolling average by a black line. Where there is no significant correlation, we show the rolling average by a light grey line. 

Our simulations are set-up such that we can keep the mass distributions of stars constant, but vary the initial radii of the star-forming regions after the masses have been drawn from the IMF. For this reason, we see the same positive correlation at 0\,Myr (before any photoevaporation has taken place) as in the high density simulations (the magenta line). Interestingly, this positive correlation remains throughout the simulation (see Figs.~\ref{2D_disc_10percent_av_LD-b}--\ref{2D_disc_10percent_av_LD-d}); however, some correlations that develop later in the simulation that could be attributed to distance from the ionising star are short-lived (e.g.\,\,the orange line present at 0.5 and 1\,Myr), again suggesting that this behaviour is rather stochastic. 

In the lower density simulations, at 0.5\,Myr, 4/20 simulations display a positive correlation (3/20 display a negative correlation), at 1\,Myr 5/20 display a positive correlation and 2/20 display a negative correlation. Finally, at 4\,Myr 7/20 show a positive correlation and none display a negative correlation. If we discount the simulation with the birth correlation (i.e.\,\,not attributable to photoevaporation) then again more than 70\,per cent of the simulations do not display a correlation of increasing disc mass with increasing distance from the ionising star.

\begin{figure*}
  \begin{center}
\setlength{\subfigcapskip}{10pt}

\hspace*{-1.0cm}\subfigure[0\,Myr]{\label{2D_disc_10percent_av_LD-a}\rotatebox{270}{\includegraphics[scale=0.2]{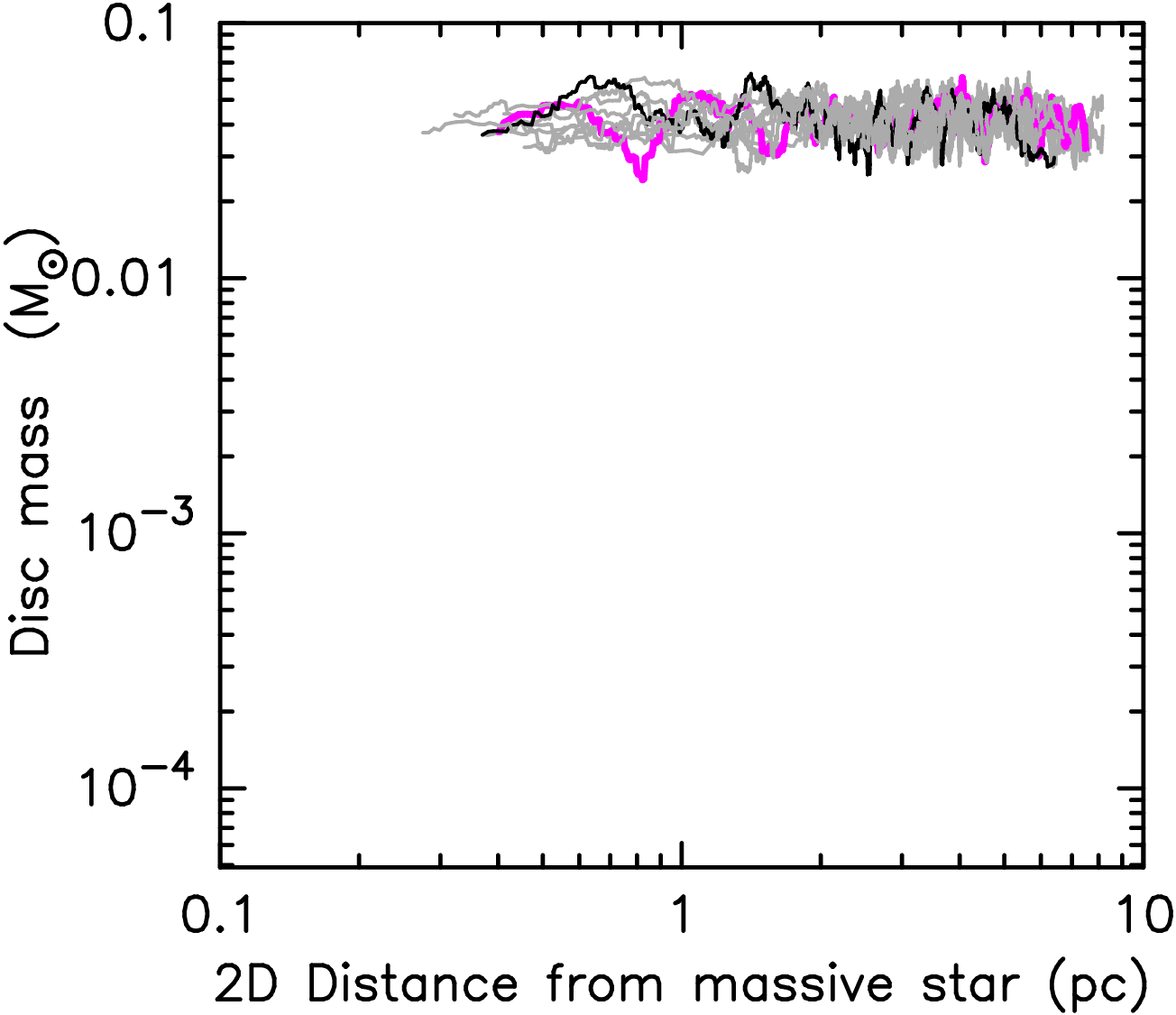}}}
\hspace*{0.1cm}
\subfigure[0.5\,Myr]{\label{2D_disc_10percent_av_LD-b}\rotatebox{270}{\includegraphics[scale=0.2]{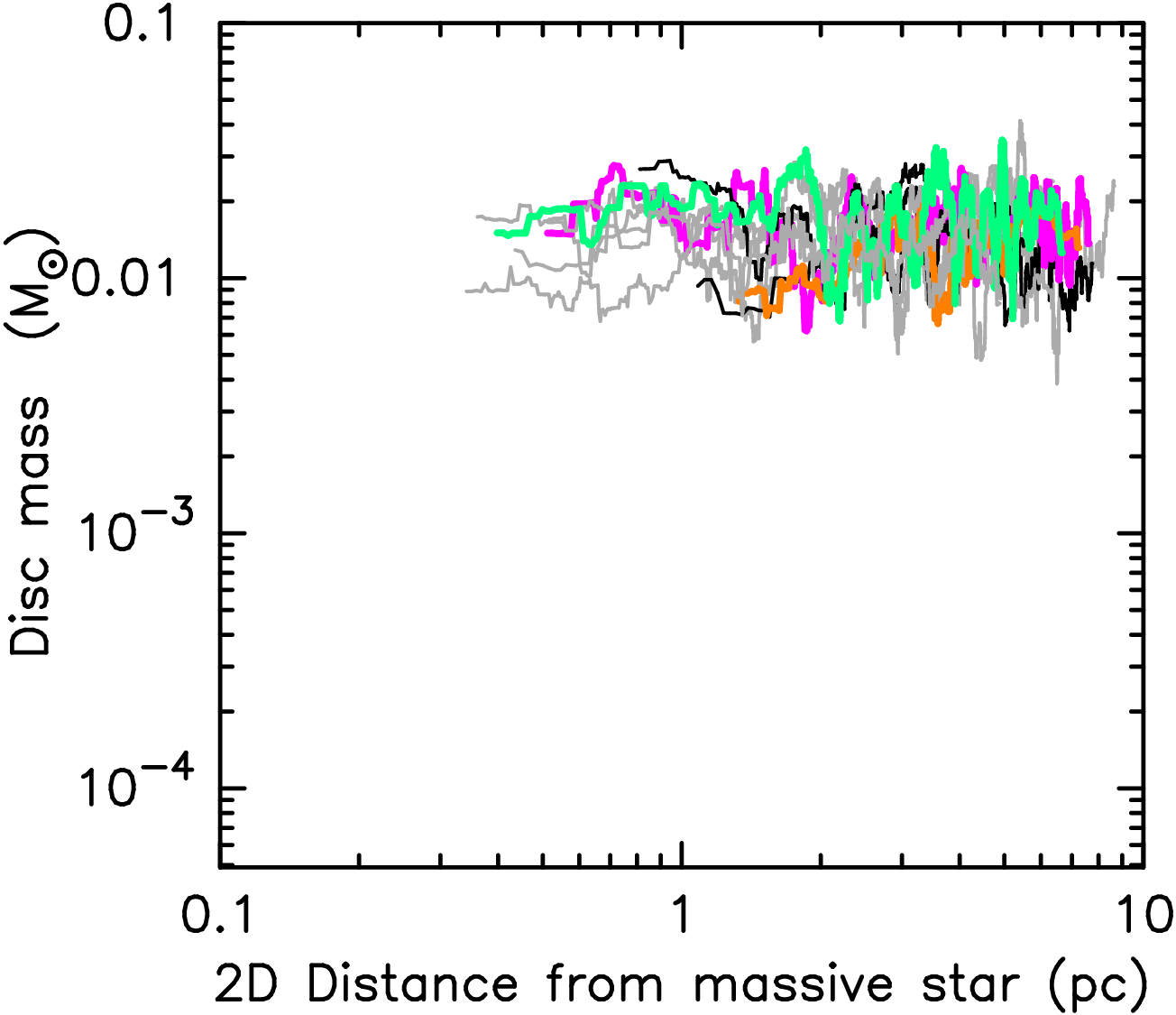}}}
\hspace*{0.1cm}\subfigure[1\,Myr]{\label{2D_disc_10percent_av_LD-c}\rotatebox{270}{\includegraphics[scale=0.2]{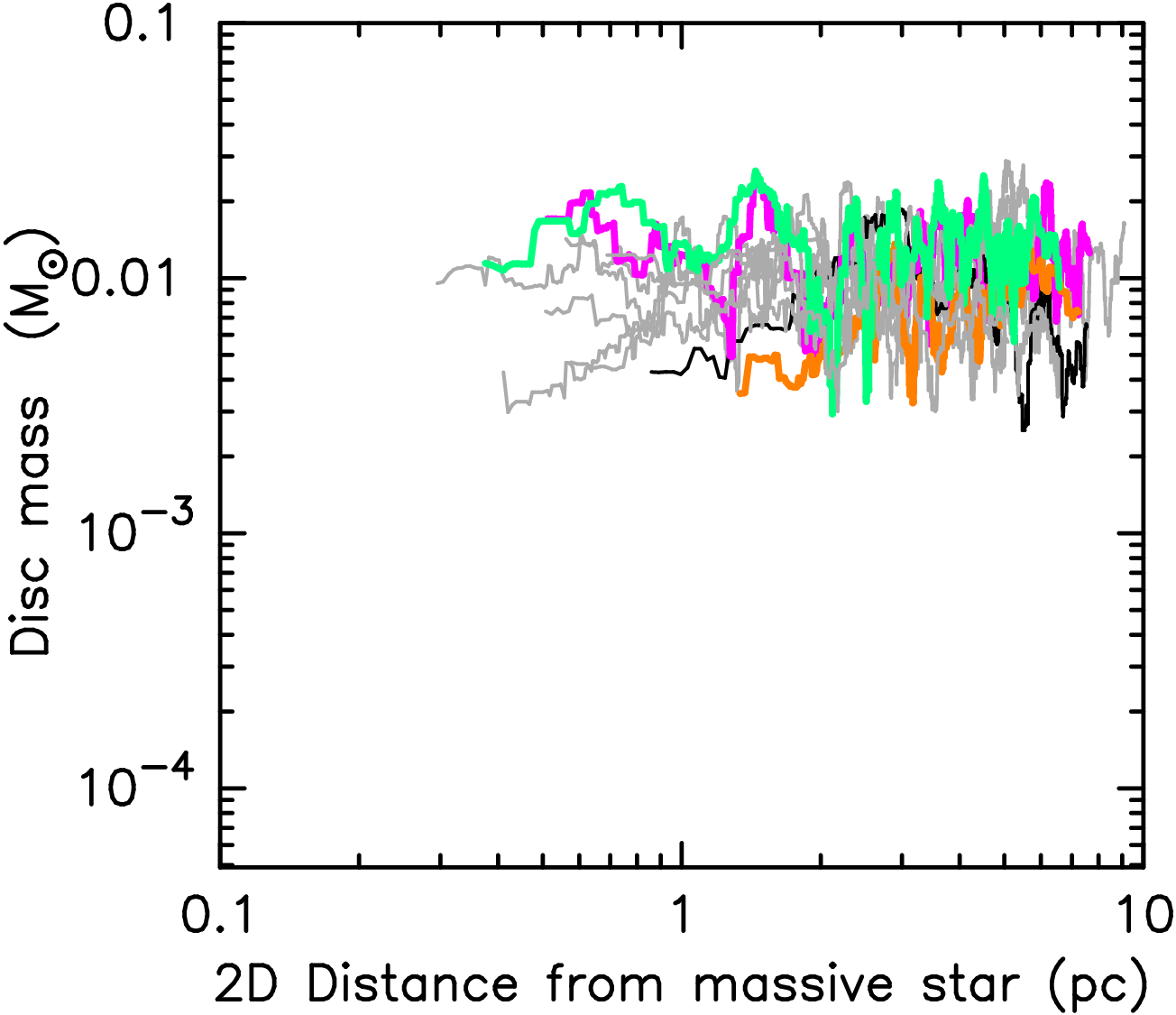}}}
\hspace*{0.1cm}\subfigure[4\,Myr]{\label{2D_disc_10percent_av_LD-d}\rotatebox{270}{\includegraphics[scale=0.2]{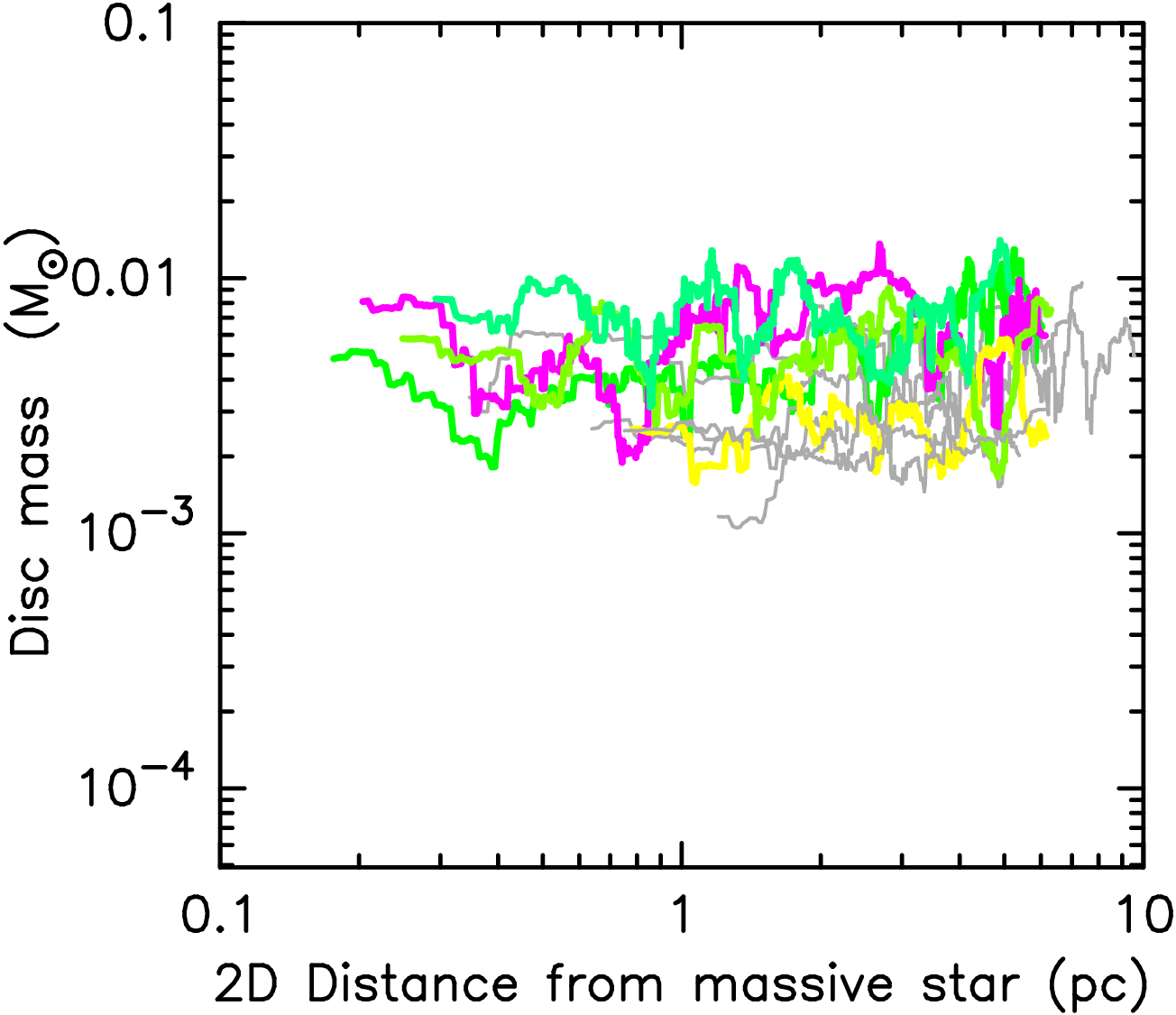}}}
\caption{Average disc mass as a function of distance from the most massive ionising star in the low-density simulations. Each line shows a rolling average of the disc masses of bins of fifty stars within the ordered list of distances to the ionising star. For clarity, we only show the first ten simulations (not all twenty) but we report on the statistics for all twenty in the text. Thin grey lines indicate simulations where there is no correlation according to the Spearman rank test. The thick coloured lines indicate a positive correlation according to the Spearman test (with a $p$-value $<$0.1), i.e.\,\,the disc masses increase with increasing distance from the ionising star. The thin black lines indicate a negative correlation according to the Spearman test (with a $p$-value $<$0.1), i.e.\,\,the disc masses decrease with increasing distance from the ionising star. We show snapshots at 0, 0.5, 1 and 4\,Myr, and colours are the same at different snapshots, i.e.\,\,any long-term correlation is indicated by similarly coloured lines.}
%Simulation 2, 10\% disc mass, 2D, distance to most massive star. 
\label{2D_disc_10percent_averages_LD}
  \end{center}
\end{figure*}

\subsubsection{1\,per cent disc masses, 2D projection}

\begin{figure*}
  \begin{center}
\setlength{\subfigcapskip}{10pt}
%\hspace*{-1.0cm}\subfigure[0\,Myr]{\label{sim_4_2D_disc_1percent_5pc-a}\rotatebox{270}{\includegraphics[scale=0.2]{disc_mass_OrBC0p3F2p05pSmFS10_01_Fe_fuv+euv_1percent_0Myr.ps}}}
%\hspace*{0.3cm}
%\subfigure[0.5\,Myr]{\label{sim_4_2D_disc_1percent_5pc-b}\rotatebox{270}{\includegraphics[scale=0.2]{disc_mass_OrBC0p3F2p05pSmFS10_01_Fe_fuv+euv_1percent_0p5Myr.ps}}}
%\hspace*{0.3cm}\subfigure[1\,Myr]{\label{sim_4_2D_disc_1percent_5pc-c}\rotatebox{270}{\includegraphics[scale=0.2]{disc_mass_OrBC0p3F2p05pSmFS10_01_Fe_fuv+euv_1percent_1Myr.ps}}}
%\hspace*{0.3cm}\subfigure[4\,Myr]{\label{sim_4_2D_disc_1percent_5pc-d}\rotatebox{270}{\includegraphics[scale=0.2]{disc_mass_OrBC0p3F2p05pSmFS10_01_Fe_fuv+euv_1percent_4Myr.ps}}}  

\hspace*{-1.0cm}\subfigure[0\,Myr]{\label{sim_4_2D_disc_1percent_5pc-a}\rotatebox{270}{\includegraphics[scale=0.2]{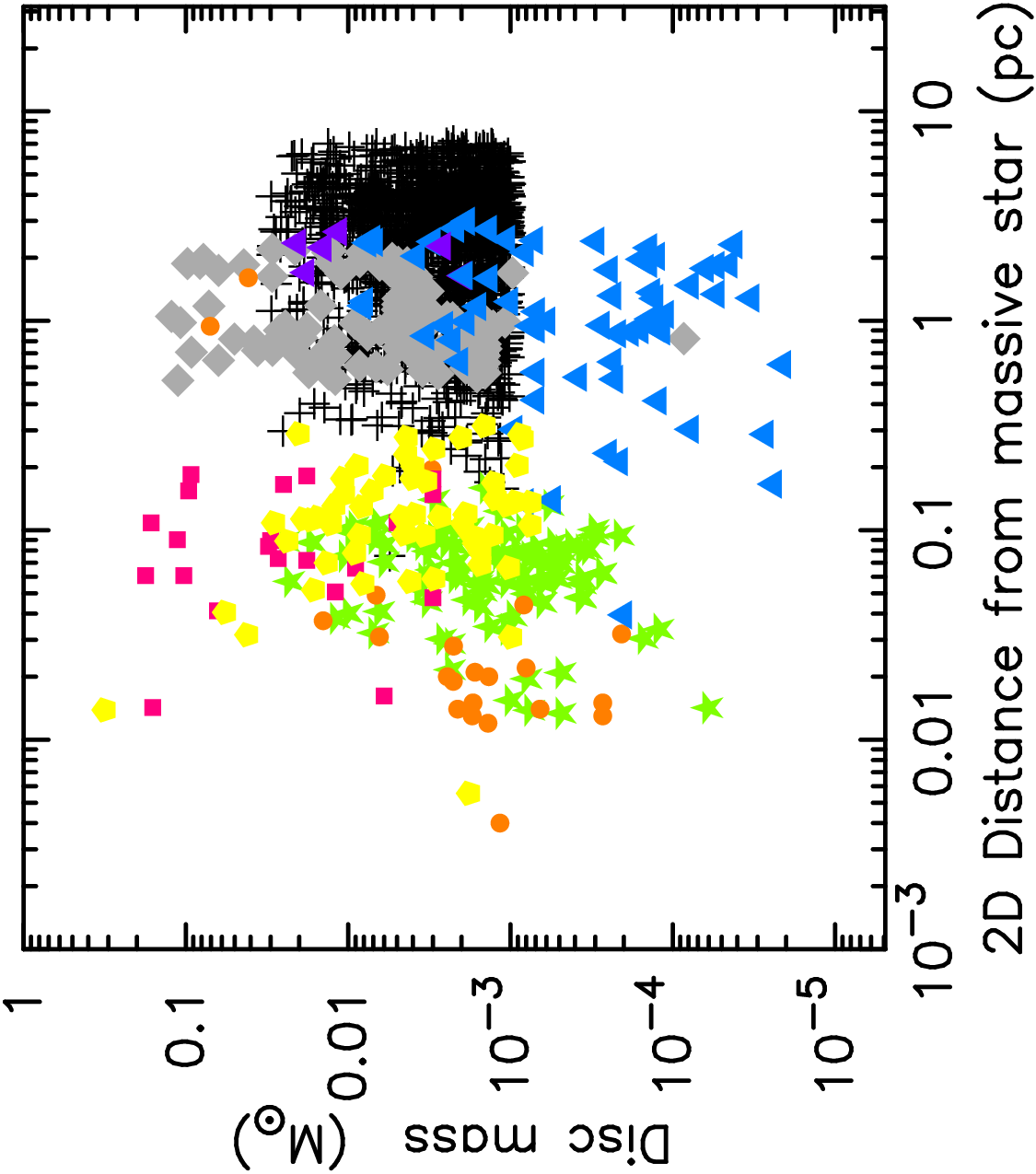}}}
\hspace*{0.3cm}
\subfigure[0.5\,Myr]{\label{sim_4_2D_disc_1percent_5pc-b}\rotatebox{270}{\includegraphics[scale=0.2]{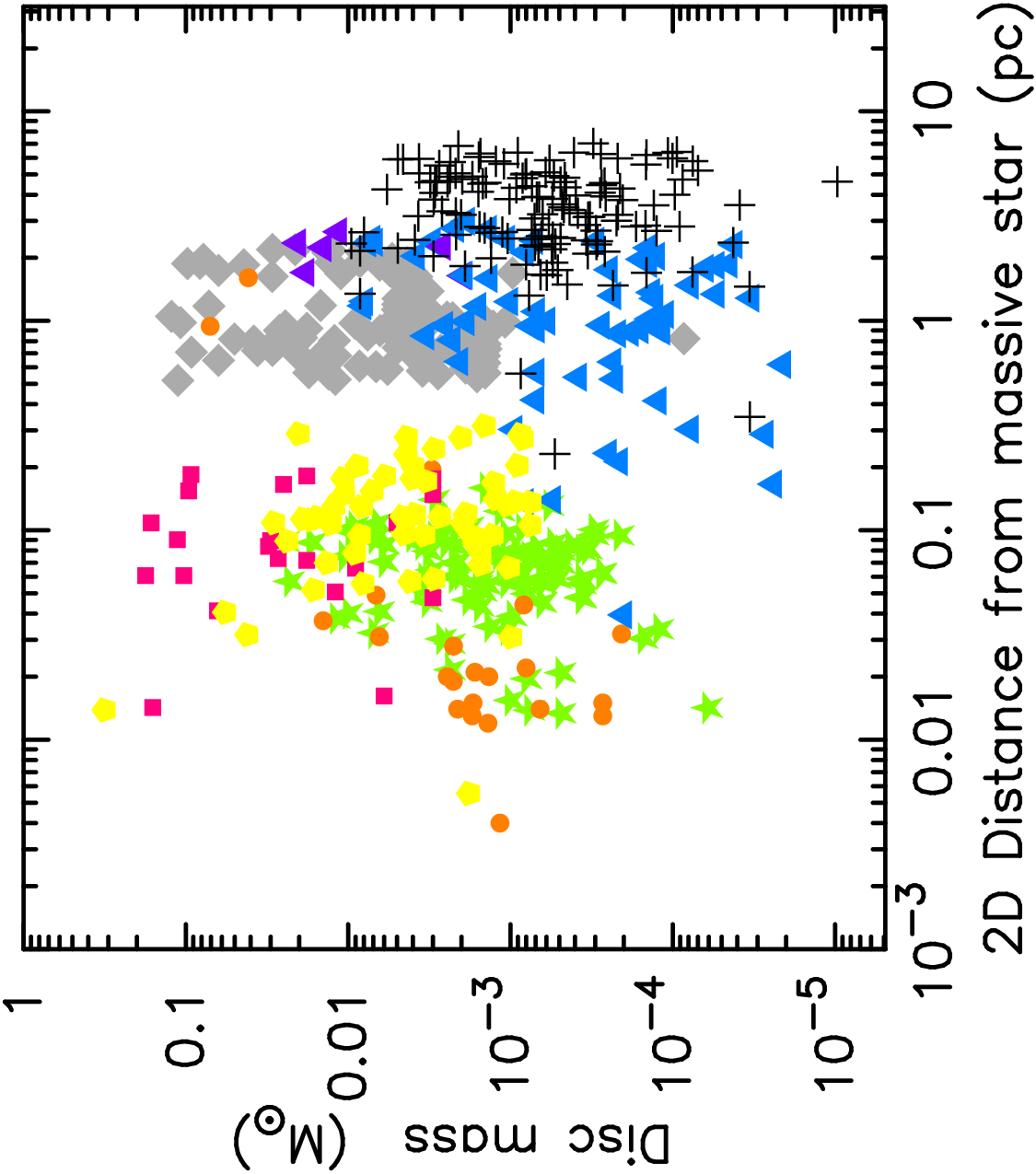}}}
\hspace*{0.3cm}\subfigure[1\,Myr]{\label{sim_4_2D_disc_1percent_5pc-c}\rotatebox{270}{\includegraphics[scale=0.2]{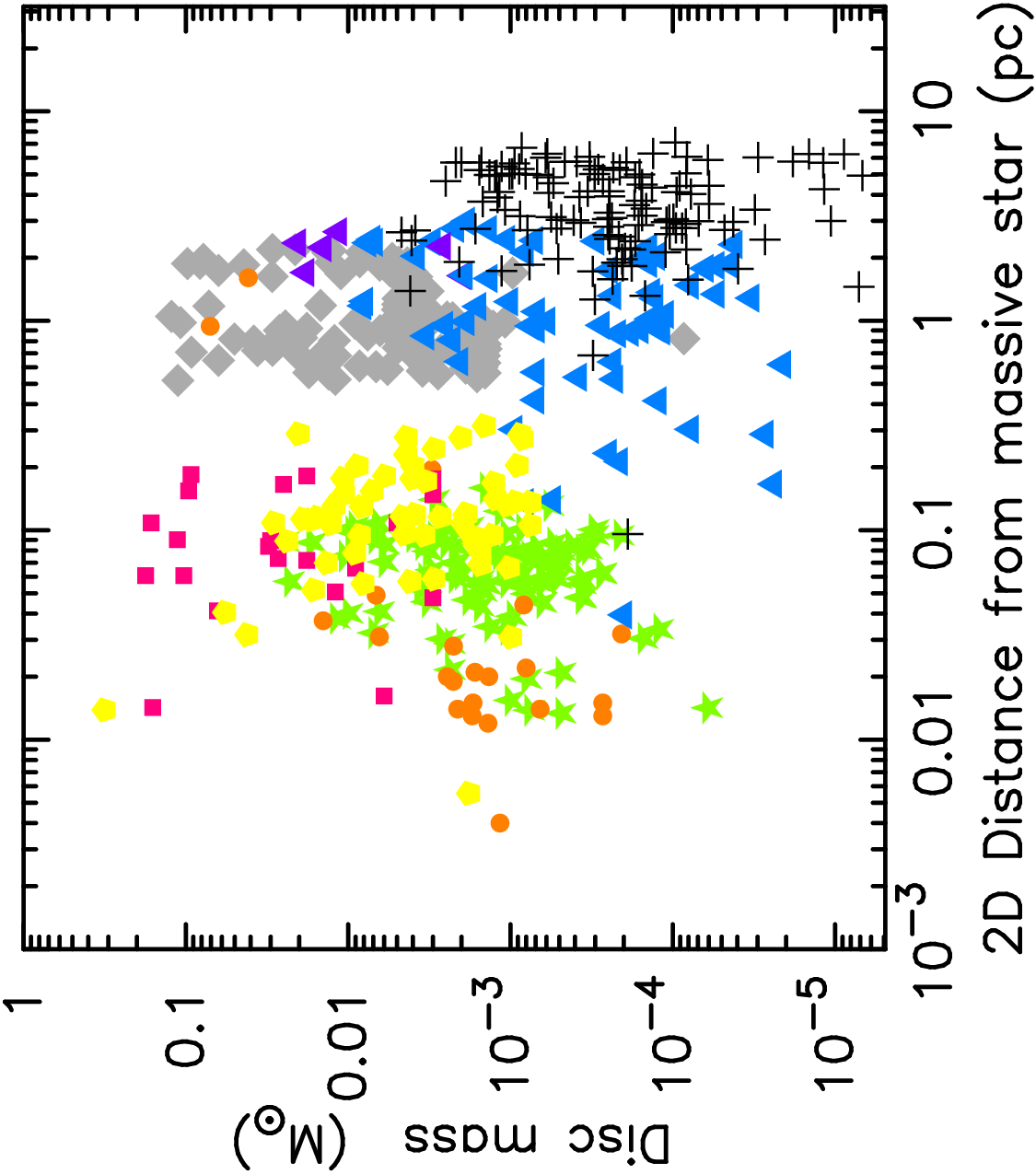}}}
\hspace*{0.3cm}\subfigure[4\,Myr]{\label{sim_4_2D_disc_1percent_5pc-d}\rotatebox{270}{\includegraphics[scale=0.2]{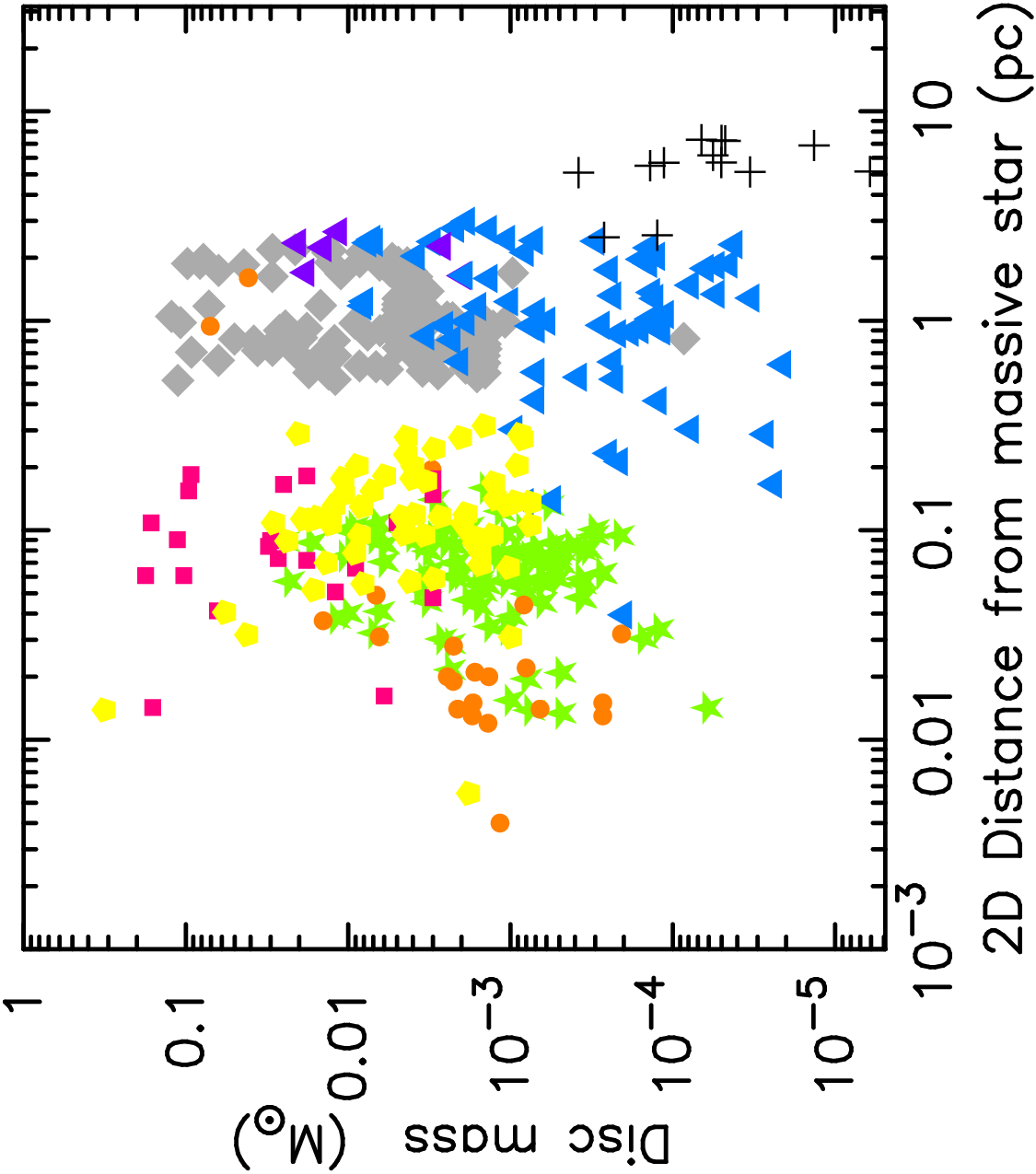}}}  
\caption{Photoevaporation of protoplanetary discs in low density star-forming regions ($10^2$\,stars\,pc$^{-3}$) where the initial disc mass is 1\,per cent of the host star's mass. The radius of the star-forming region is 5\,pc. The plus symbols show the disc mass plotted against the distance to the massive ionising star for each low-mass star with a disc, using only two dimensions to calculate the projected distance. Stars whose discs have been completely evaporated are removed from the plot (hence the decreasing population of black points with increasing age). The masses (inferred from the dust content) and the respective distances from $\theta^1$~Ori~C of observed protoplanetary discs in the ONC by \citet[][orange circles]{Mann14}, and \citet[][green stars]{Eisner18}; and the Orion Molecular Cloud \citep[][the grey diamonds]{vanTerwisga19} are shown. We also show the masses and projected distances from IRS~2b of discs in NGC\,2024  by \citet[][the raspberry squares]{Mann15}  and \citet[][the yellow pentagons]{vanTerwisga20}.  Finally, we show the masses and projected distances from the ionising star $\sigma$~Ori in this eponymous star-forming region \citep{Ansdell17} by the blue triangles.  Protoplanetary discs in  $\sigma$~Ori that have CO detections (i.e.\,\,gas content) are shown by the purple triangles.}
%Simulation 2, 1\% disc mass, 2D, distance to most massive star, SFR radius 5\,pc. 
\label{sim_4_2D_disc_1percent_5pc}
  \end{center}
\end{figure*}

If we reduce the initial disc masses to be only 1\,per cent of the host star (Fig.~\ref{sim_4_2D_disc_1percent_5pc}), then after 4\,Myr a higher fraction of the discs have lost mass or been destroyed due to photoevaporation, with only 11 discs from the initial population surviving to 4\,Myr. %However, a large number still survive, with masses from  $10^{-5}$M$_\odot$ and upwards.
In our simulations, the remaining disc distributions are inconsistent with the masses of discs, and the positions of their host stars with respect to the ionising sources, measured in \citet{Mann14} for the ONC, and \citet{Mann15} for NGC2024, but are consistent with the masses and positions in $\sigma$~Orionis \citep{Ansdell17}. \\

Interestingly, in both our low-density simulations the disc-hosting stars are never as close to the ionising stars as closest observed examples in both the ONC and NGC\,2024 at the start of the simulation (compare the black points in Figs.~\ref{sim_4_2D_disc_10percent_5pc-a}~and~\ref{sim_4_2D_disc_1percent_5pc-a} with the leftmost coloured points). This suggests that both of these star-forming regions must have been more dense in the past, and opens the promising avenue of the properties of disc-hosting stars being used as a diagnostic to predict the initial conditions of their host star-forming regions.

\subsection{Comparison with discs in star-forming regions}

Many authors have reported a correlation between the disc mass and the projected position of the host star in relation to ionising massive star(s) in young star-forming regions \citep{Guarcello07,Guarcello10,Fang12,Guarcello14,Mann14,Ansdell17,Eisner18,vanTerwisga19}, whilst some authors have found no correlation \citep{Mann15}. Our simulations span three orders of magnitude in initial stellar densities, and include discs that are both 10\,per cent, and 1\,per cent of the host star's mass.

In no simulations do we reproduce the strength of the observed trend of disc mass and distance to the ionising star in the \citet{Mann14} data for the ONC. In some of our simulations there is a correlation between the disc mass and the distance to the most massive star, but this appears to be coincidental and not dependant on the dynamical evolution of the star-forming region, or the strength of the FUV radiation field. A striking result that we will follow up in a future paper is that once the FUV field exceeds $\sim$100\,$G_0$, then the numbers of ionising stars present in the simulation does not strongly correlate with the number of discs that are destroyed. Therefore, the biggest factor in how many discs are destroyed is the initial stellar density of the star-forming region. Other changes to the initial conditions (different amounts of spatial and kinematic substructure, global virial ratio) have only a minimal effect on disc destruction rates \citep[see also][]{Nicholson19a}.

One feature of our simulations is that in dense star-forming regions like the ONC, we would expect photoevaporation to be extremely efficient at depleting the gas content of protoplanetary discs. Most observational studies that measure gas and dust masses in discs select targets based on known infrared excesses (i.e.\,\,the presence of a disc). However, in our simulations we quickly create a population of disc-less stars, and furthermore, the discless stars move due to dynamical evolution and so are not always the closest objects to the ionising sources.

Recently, \citet{Yao18} conducted a study of young stars in nearby regions and determined whether individual objects were disc-less, or disc bearing. We show histograms of their data for Orion in Fig.~\ref{Orion_discs}, where stars are binned depending on their projected distance from $\theta^1$~Ori~C. As in our simulated data, stars that still host discs are shown by the black (open) histogram, and stars without discs are included in the red (solid) histogram. The data in \citet{Yao18} are incomplete for the innermost regions of the ONC, but if anything there appear to be more disc-less stars further from the ionising stars than stars with discs. This is inconsistent with the results of our simulations, where -- on average -- stars that retain their discs are at similar distances from the ionising stars than those that lose their discs (compare the open and filled histograms, respectively, in panels (e)--(h) of Figs.~\ref{sim_2_2D_disc_10percent}~and~\ref{sim_2_3D_disc_10percent}).

\begin{figure}
\begin{center}
\rotatebox{270}{\includegraphics[scale=0.4]{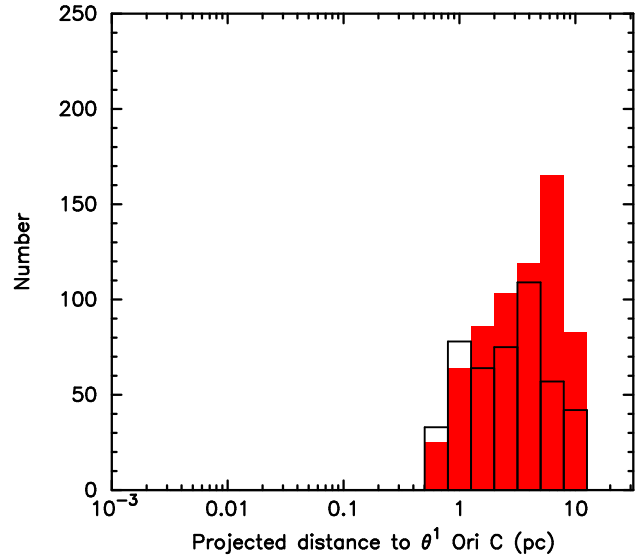}}
\end{center}
\caption{Disc demographics in Orion. A histogram of stars without a protoplanetary disc as a function of distance from $\theta^1$~Ori~C is shown by the filled red histogram, and a histogram of stars with a protoplanetary disc (i.e.\,\,an infrared excess) is shown by the open black histogram. Data taken from \citet{Yao18}. This can be directly compared with our simulation results in panels (e)--(h) in Figs.~\ref{sim_2_2D_disc_10percent}~and~\ref{sim_2_3D_disc_10percent}.}
\label{Orion_discs}
\end{figure}

Intriguingly, however, the observations of \citet{Yao18} are  consistent with our dynamical picture in which stars can lose their discs due to external photoevaporation, and then move significant distances within the star-forming region. The big caveat here is that our simulations do not include other disc depletion mechanisms (internal photoevaporation, truncation through encounters, or rapid planet formation) that may have occured within the observed discs. Whilst internal photoevaporation would be unlikely to provide a correlation between disc mass and distance to ionising stars, truncation of protoplanetary discs due to stellar encounters is efficient in very dense ($>10^4$\,stars\,pc$^{-3}$) regions \citep{Vincke15,Zwart16,Winter18b}. This could explain the low disc masses in the central regions of the ONC (although we reiterate that stars would move in and out of the central regions). On the other hand, the density of $\sigma$~Ori is probably too low for disc truncation to explain the correlation in that region. 

Finally, we note that our simulations assume only one epoch of star formation in a region, whereas recent work by \citet{Winter19b} assumes  multiple star formation episodes, which -- if prevalent in nearby star-forming regions -- means that any observed discs could have been subject to differing amounts of photoevaporation. 

Alternatively, if star formation is gradual then discs could have different ages and be in different stages of evolution \citet{Longmore14} propose a `conveyor belt' scenario for star formation, where clusters like the ONC could be gradually fed low-mass stars via filaments; this would be observed as both an age spread in the whole region with the younger stars being more central, and potentially with higher numbers of protoplanetary discs in the central regions \citep[see also][]{Hillenbrand97,Beccari17,Winter19b}.

\section{Discussion}
\label{discuss}

Several authors have found a trend that the masses of protoplanetary discs in star-forming regions increase the further the disc is (projected in 2D) from the massive star \citep{Guarcello07,Fang12,Guarcello14,Mann14,Ansdell17,Eisner18,vanTerwisga19}, which is explained as disc mass-loss due to photoevaporation from massive stars being more efficient the closer the the discs are to these ionising sources. Furthermore, this dependence on distance appears to be extremely strong. For example, in the ONC \citet{Mann14} report that disc masses increase from $10^{-4}$\,M$_\odot$ in close proximity ($<0.1$\,pc) to $\theta^1$~Ori~C, up to values around 0.1\,M$_\odot$ at distances of up to 1\,pc. Similar trends are reported in $\sigma$~Ori by \citet{Ansdell17}.

Recently, the advent of ALMA has enabled both the dust and gas components of protoplanetary discs to be probed, and the respective masses of the gas and dust content to be estimated. \citet{Mann14} do not detect any CO in any of the discs in their sample, suggesting that the mass of the gas is very low, or non-existent. \citet{Ansdell17} find that only six discs in $\sigma$~Ori contain significant amounts of gas, and all of them are more than 1\,pc away from the most massive star in projection (these discs are shown by the purple triangles in our plots in Figs.~\ref{sim_2_2D_disc_10percent},~\ref{sim_2_3D_disc_10percent},~\ref{sim_2_2D_disc_1percent},~\ref{sim_4_2D_disc_10percent_5pc}~and~\ref{sim_4_2D_disc_1percent_5pc}).  

If the trend in increasing disc mass with distance from the massive star is not explained by photoevaporation of the gas content, can dust be photoevaporated in significant amounts? In the FRIED grid models we use from \citet{Haworth18a}, the fraction of the dust mass in a disc that is photoevaporated is low. Even in an extremely strong ($10^3 - 10^4G_0$) FUV field, the dust mass only decreases by a factor of a few within 10\,Myr, assuming that the amount of mass in small dust grains, which could be entrained in the evaporative wind, is low. (In contrast, the gas component of the disc is reduced by several orders of magnitude, and is usually completely evaporated in such strong FUV fields.) The observed result that disc masses increase with increasing distance from the massive stars in several star-forming regions \citep{Guarcello07,Mann14,Ansdell17} would therefore only be consistent with external photoevaporation if these discs had a significant gas component by mass.

If photoevaporation did significantly reduce the dust content of protoplanetary discs, it is not clear that we would expect a trend in disc mass increasing the further the disc is from an ionising massive star. Dynamical evolution of star-forming regions depends on the initial stellar density of a star-forming region \citep{Parker14b,Parker14e}, but even low-density ($\geq$10\,M$_\odot$\,pc$^{-3}$) star-forming regions evolve on timescales similar to the photoevaporation timescales in the models of \citet{Storzer99}, \citet{Facchini16} and \citet{Haworth18a,Haworth18b} -- see \citet{Nicholson19a}.

Therefore, we would not expect a low-mass disc-hosting star to remain close to the ionising massive star for a sustained length of time, and the disc-hosting star would spend a considerable amount of time in different locations. This is evidenced in our simulations, where the histogram of stars with discs, compared to those whose discs have been destroyed, appears to be random, centred on the average interstellar distance between stars. In other words, the distributions are random because both sets of stars -- disc-hosting and disc-less -- have moved around in the star-forming region. The recent survey by \citet{Yao18} suggests there is very little difference in the distributions of disc-hosting and disc-less stars in Orion, although their data are incomplete in the immediate vicinity of $\theta^1$~Ori~C.

It is therefore unclear what is producing the observed trend of increasing disc mass with increasing distance from the massive stars in some star-forming regions. Mass segregation, where the most massive stars are preferentially found at the centre, is observed in many star-forming regions, including the ONC. This could result in a radially dependent spatial distribution of stellar masses, which could in turn result in a radial dependence of the disc masses. However, this would predict the opposite effect to what is observed, namely that the more massive stars (and hence discs) would be found closer to the centre of the star-forming region.

Finally, we reiterate that the timescales for photoevaporating the gas content of discs are extremely short \citep[see also][]{Nicholson19a,ConchaRamirez19}, depleting or destroying altogether the material from which gas giant planets form. This argues for either extremely rapid ($<<$1\,Myr) planet formation, or that giant planets exclusively form in star-forming regions like Taurus \citep[e.g.][]{Guedel07,Luhman09}, which contain no massive (O- and B-type) stars (unlike the ONC). The latter scenario is in tension with mounting evidence that the Sun experienced enrichment from short-lived radioisotopes in its natal star-forming region \cite[e.g.][]{Gounelle12,Boss17,Parker14a,Young16,Lichtenberg16b,Zwart19,Lichtenberg19}, which requires the presence of one or more massive stars \citep{Adams10,Lugaro18}. Alternatively, giant planet formation may need to exclusively occur within several au of the host star as \citet{Nicholson19a} demonstrate that discs with small radii ($\leq10$\,au) can survive in very strong radiation fields for several Myr.

%Our main result is that we cannot reproduce the observed trend that the mass of a %protoplanetary disc increases the further its host star is from the ionising source. If %we take the ONC as an example, \citet{Mann14} report disc masses ranging between $2% %\times 10^{-4}$M$_\odot$ close to the ionising star up to $\sim$0.5\,M$_\odot$. The %photoevaporation precription we adopt tends to 

\section{Conclusions}
\label{conclude}

We present $N$-body simulations of the dynamical evolution of star-forming regions in which protoplanetary discs lose mass due to external FUV and EUV radiation from the most massive stars. We have compared the positions of disc hosting stars and the disc masses to recent observations of discs in nearby star-forming regions. In particular, we have calculated the disc mass-loss due to photoevaporation as a function of projected distance from the ionising massive star(s). Our conclusions are the following.\\

(i) In agreement with other authors \citep[e.g.][]{Scally01,Adams04,Winter18b,ConchaRamirez19,Nicholson19a}, we find that external FUV and EUV radiation has a highly detrimental effect on protoplanetary discs; in many simulations very few discs with initial radii of 100\,au remain after only a few Myr.

(ii) In the observational data, the discs in the ONC sample published by \citet{Mann14}, the discs in the more extended Orion Molecular Cloud published by \citet{vanTerwisga20} and the discs in $\sigma$~Orionis published by \citet{Ansdell17} display a significant correlation of increasing disc mass with increasing distance from the ionising massive star. We note that if the initial disc dust masses are proportional to the host stars' masses, then such a trend may simply be due to the imprint on the disc masses of randomly sampling the stellar IMF.

(iii) Projection effects add significant noise to the spatial distribution of disc hosting stars. Stars with discs that appear to be close to the ionising sources may be fore- or background members of the star-forming region.

(iv) After accounting for projection effects, we find no evidence that the observed trend of increasing disc mass with increasing distance from ionising stars is due to external photoevaporation, unless the adopted photoevaporation models are very wrong. The reason for this is twofold. First, the gas content is uniformly destroyed on rapid timescales in the simulations. Secondly, the observed discs contain little or no gas, whereas even a strong FUV field will not result in significant mass-loss from the dust component of the disc.  

%(iii) Following on from (ii), the observed positions of disc hosting stars in the ONC are in such close proximity to the most massive star, $\theta$~Ori~C, that the photoevaporation models adopted in the literature are likely over-estimating the amount of mass-loss occuring from the discs, or these discs contain little or no gas.

(v) Even if the photoevaporation models are reasonably accurate, in a dense star-forming region there is no reason to expect a dependence of the disc properties on the distance to the ionising stars. Our simulations suggest that stars lose their discs, and then move around the star-forming region so that they may reside in far flung locations from the ionising stars.\\

A first step to resolving some of the issues we have identified would be to improve on existing observations of discs in star-forming regions, and to obtain observations of disc populations in other star-forming region to improve the statistical sample and any comparisons between regions.%Taken together, our conclusions suggest that the photoevaporation models for disc mass-loss in star-forming regions need to be revised, or discs predominately lose mass due to other mechanisms, such as internal accretion or truncations by stellar fly-bys.

\section*{Acknowledgements}

RJP and OP acknowledge support from the Royal Society in the form of Dorothy Hodgkin Fellowships. HLA was supported by the 2018 Sheffield university Undergraduate Research Experience (SURE) scheme. We thank Yuhan Yao for providing their observational data ahead of final publication. 

%\section*{Data availability statement}

%The data used to produce the plots in this paper will be shared on reasonable request to the corresponding author.

%\newpage
\appendix

\section{Different realisations of the initial mass function}
\label{appendix:mass-function}

In this section of the Appendix we show the non-dependence of stochastic sampling on the initial mass function (IMF) on our results.  In each realisation of a star-forming region, we draw masses randomly from the IMF in the mass range 0.1 -- 50\,M$_\odot$. Due to the modest numbers of stars in our simulations ($N = 1500$), this means that the high-mass end of the IMF is not fully sampled, which in turn results in different numbers of massive stars ($>17.5$\,M$_\odot$) in each simulation.

The different numbers of massive stars result in different strength radiation fields, and as we also change the initial positions and velocities of stars in the different realisations of the simulations, the effect of changing the radiation field is difficult to predict analytically. For this reason, we present the results of four additional simulations of star-forming regions with high initial stellar densities ($10^4$\,stars\,pc$^{-3}$)  and these are shown in Fig.~\ref{mass_function_high_density}. For brevity, we only show the results at 1\,Myr, and we indicate the masses of the five most massive stars in the panel captions. These plots should be directly compared with Figs.~\ref{sim_2_2D_disc_10percent-c}~and~\ref{sim_2_2D_disc_10percent-g}, which show the results for a simulation where the most massive stars are 23, 18, 17, 15 and 15\,M$_\odot$.

The different numbers (and masses) of these ionising stars does change the number of surviving discs after 1\,Myr, but our conclusions are unchanged. There is still no correlation between the mass of the disc and the distance to the most massive ionising star, and the process of dynamical mass segregation means that all of the ionising stars are in central locations. Furthermore, the stars that lose their discs move around the star-forming region following disc destruction, and so the locations of stars with and without discs present a wide distribution in distance from the most massive star.

In Fig.~\ref{mass_function_low_density} we present the results of four additional simulations for our low-density simulations. Here, we compare different simulations after 4\,Myr of dynamical evolution and disc photoevaporation, and so these plots should be directly compared with  Figs.~\ref{sim_4_2D_disc_10percent_5pc-d}~and~\ref{sim_4_2D_disc_10percent_5pc-h}, whose five most massive stars were 19, 19, 19, 12 and 11\,M$_\odot$. As with the higher density simulations, the numbers of surviving discs differs depending on the mass function (and stochastic differences in the evolution of star-forming regions), but there is no dependence of the mass of a disc on the distance of its host star to the most massive star(s).

\begin{figure*}
  \begin{center}
\setlength{\subfigcapskip}{10pt}

\hspace*{-1.0cm}\subfigure[34, 18, 11, 8 \& 7\,M$_\odot$]{\label{mass_function_high_density-a}\rotatebox{270}{\includegraphics[scale=0.2]{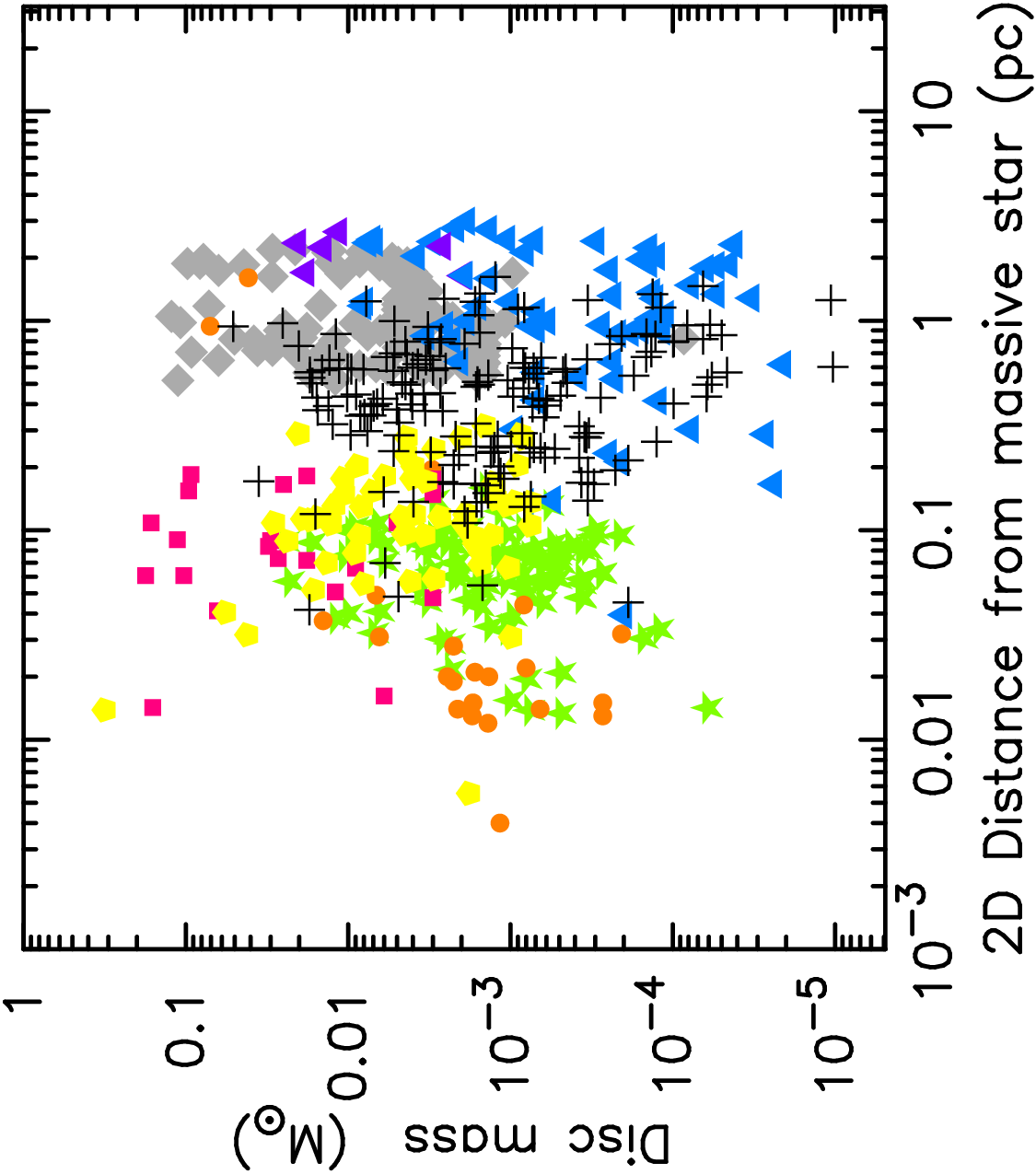}}}
\hspace*{0.3cm}
\subfigure[19, 11, 11, 11 \& 10\,M$_\odot$]{\label{mass_function_high_density-b}\rotatebox{270}{\includegraphics[scale=0.2]{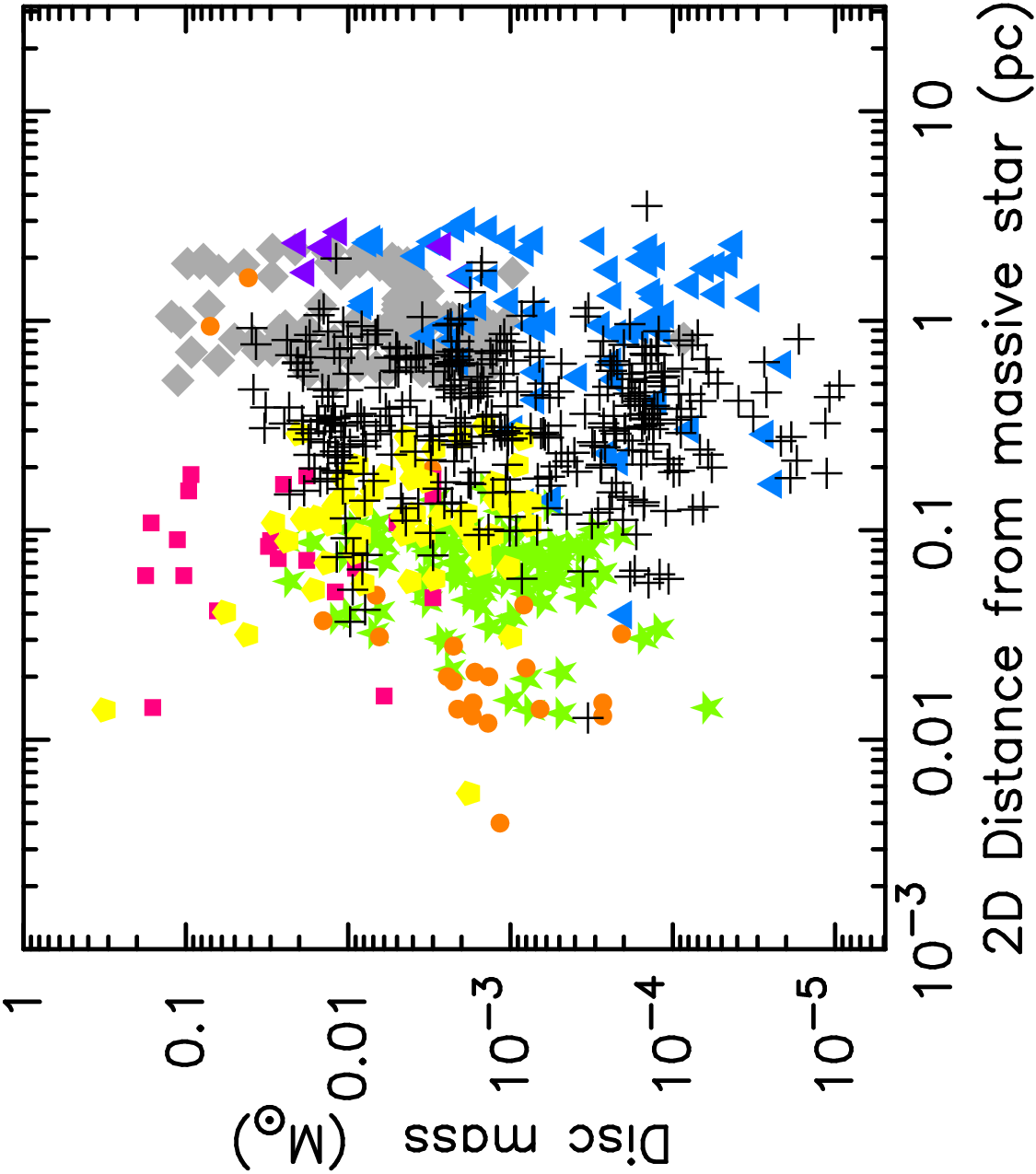}}}
\hspace*{0.3cm}\subfigure[49, 39, 36, 28 \& 27\,M$_\odot$]{\label{mass_function_high_density-c}\rotatebox{270}{\includegraphics[scale=0.2]{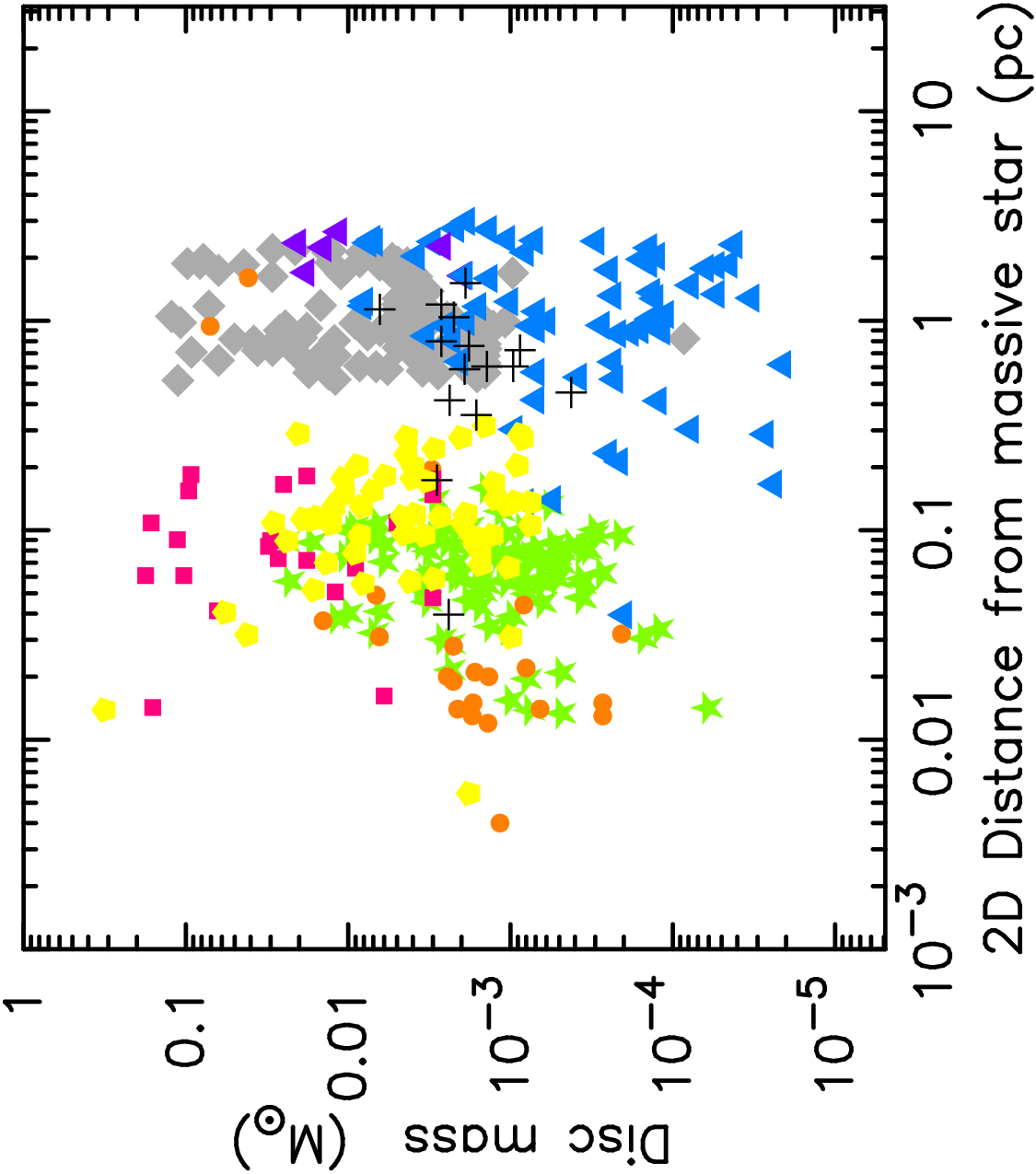}}}
\hspace*{0.3cm}\subfigure[42, 27, 16, 14 \& 13\,M$_\odot$]{\label{mass_function_high_density-d}\rotatebox{270}{\includegraphics[scale=0.2]{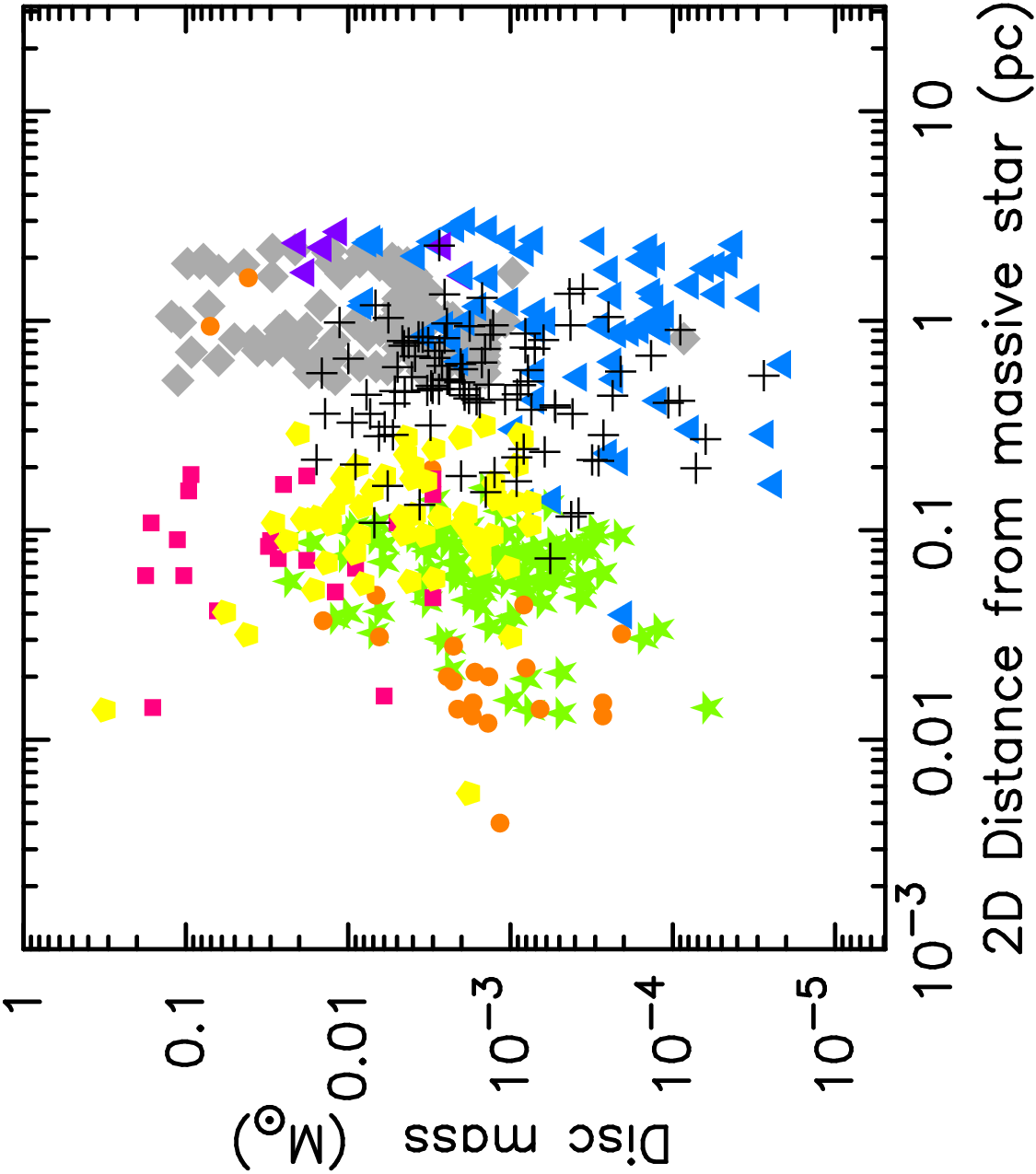}}}
\hspace*{-1.0cm}\subfigure[34, 18, 11, 8 \& 7\,M$_\odot$]{\label{mass_function_high_density-e}\rotatebox{270}{\includegraphics[scale=0.2]{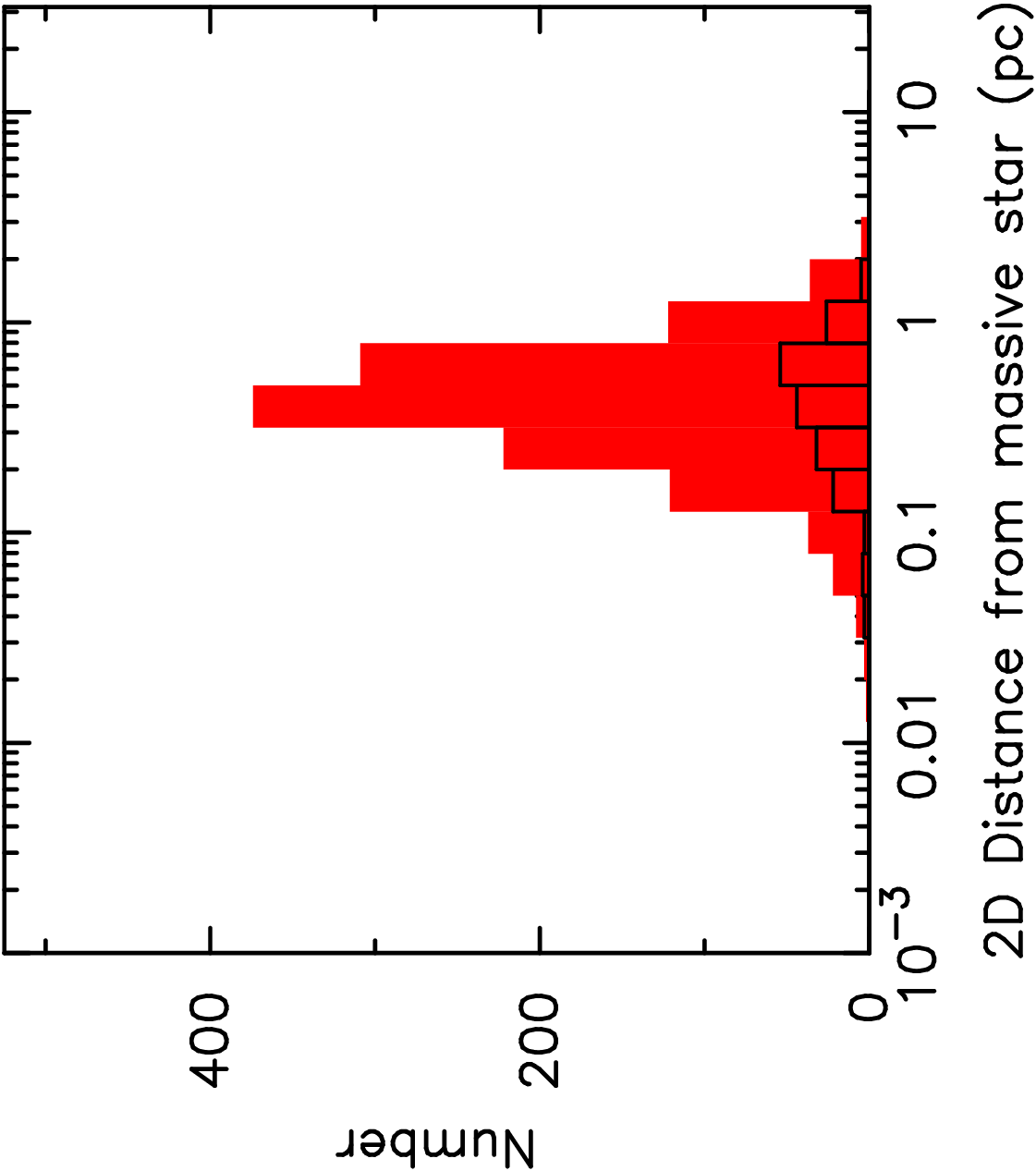}}}
\hspace*{0.3cm}
\subfigure[19, 11, 11, 11 \& 10\,M$_\odot$]{\label{mass_function_high_density-f}\rotatebox{270}{\includegraphics[scale=0.2]{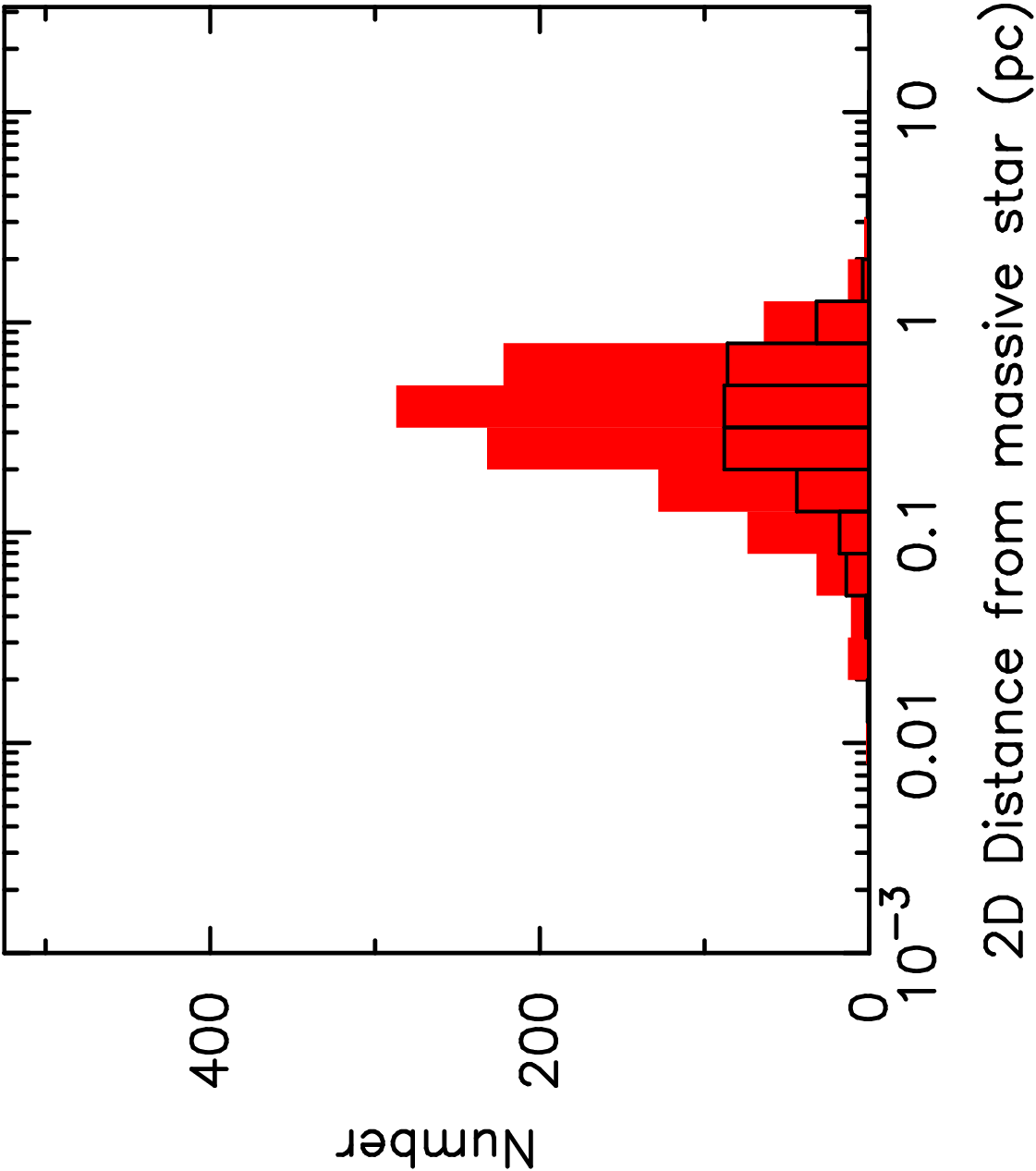}}}
\hspace*{0.3cm}\subfigure[49, 39, 36, 28 \& 27\,M$_\odot$]{\label{mass_function_high_density-g}\rotatebox{270}{\includegraphics[scale=0.2]{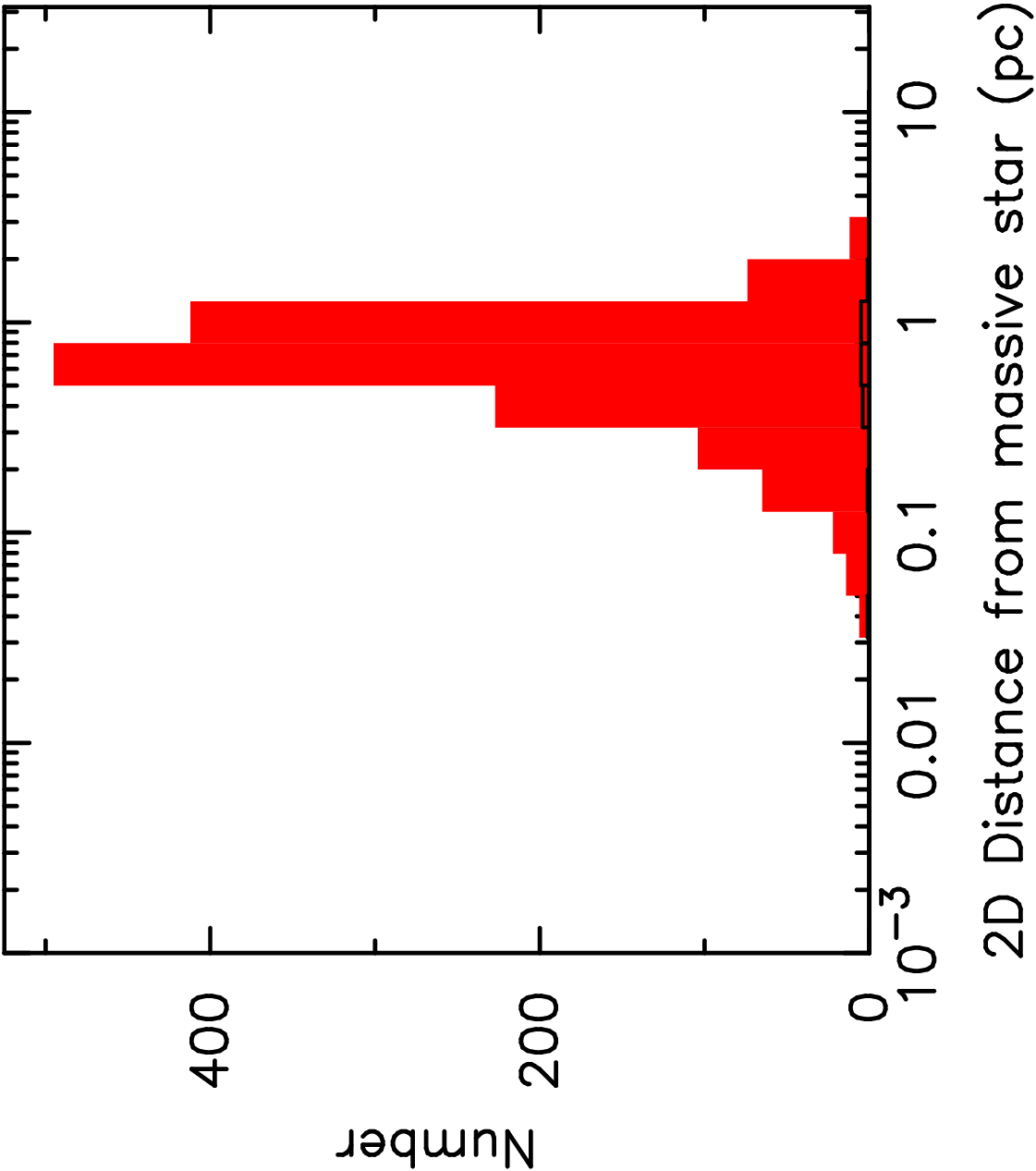}}}
\hspace*{0.3cm}\subfigure[42, 27, 16, 14 \& 13\,M$_\odot$]{\label{mass_function_high_density-h}\rotatebox{270}{\includegraphics[scale=0.2]{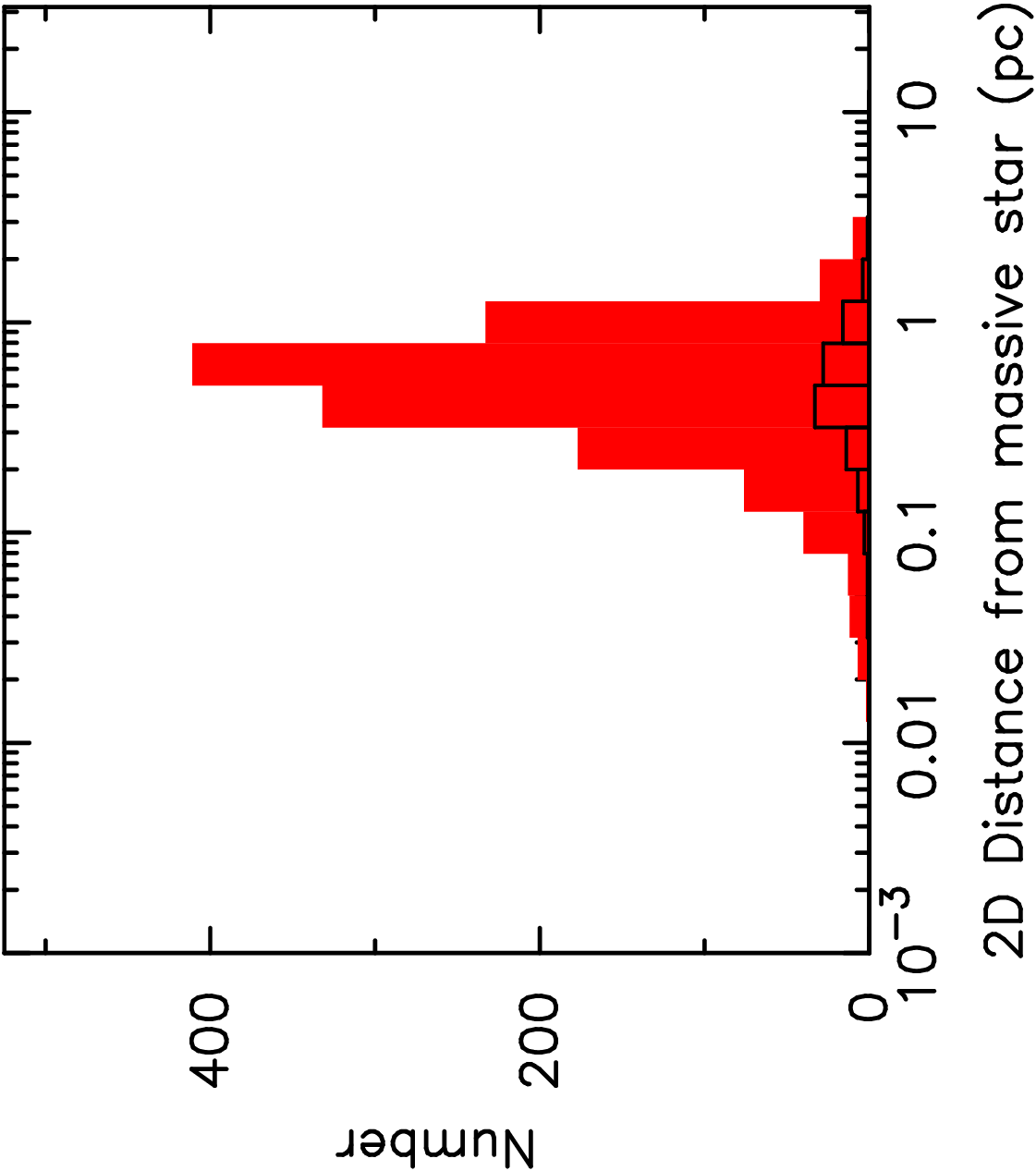}}}  
\caption{The effect of stochastic sampling of the initial mass function on the results for high density star-forming regions ($10^4$\,stars\,pc$^{-3}$) where the initial disc mass is 10\,per cent of the host star's mass, and the distance to the ionising star is calculated in two dimensions. The radius of the star-forming region is 1\,pc, and tha age of thew star-forming region in each panel is 1\,Myr. {\it Top row:} The plus symbols show the disc mass plotted against the distance to the massive ionising star for each low-mass star with a disc, using only two dimensions to calculate the projected distance. The masses and the respective distances from $\theta^1$~Ori~C of observed protoplanetary discs in the ONC by \citet[][orange circles]{Mann14}, and \citet[][green stars]{Eisner18}; and the Orion Molecular Cloud \citep[][the grey diamonds]{vanTerwisga19} are shown. We also show the masses and projected distances from IRS~2b of discs in NGC\,2024 by \citet[][the raspberry squares]{Mann15} and \citet[][the yellow pentagons]{vanTerwisga20}.  Finally, we show the masses and projected distances from the ionising star $\sigma$~Ori in this eponymous star-forming region \citep{Ansdell17} by the blue triangles. Protoplanetary discs in  $\sigma$~Ori that have CO detections (i.e.\,\,gas content) are shown by the purple triangles. {\it Bottom row:} The number of remaining discs (open histogram) and the number of stars whose discs have been destroyed by photoevaporation (red histogram) as a function of the two dimensional projected distance from the ionising star. }
%Simulation 2, 10\% disc mass, 2D, distance to most massive star. 
\label{mass_function_high_density}
  \end{center}
\end{figure*}

\begin{figure*}
  \begin{center}
\setlength{\subfigcapskip}{10pt}

\hspace*{-1.0cm}\subfigure[34, 18, 11, 8 \& 7\,M$_\odot$]{\label{mass_function_low_density-a}\rotatebox{270}{\includegraphics[scale=0.2]{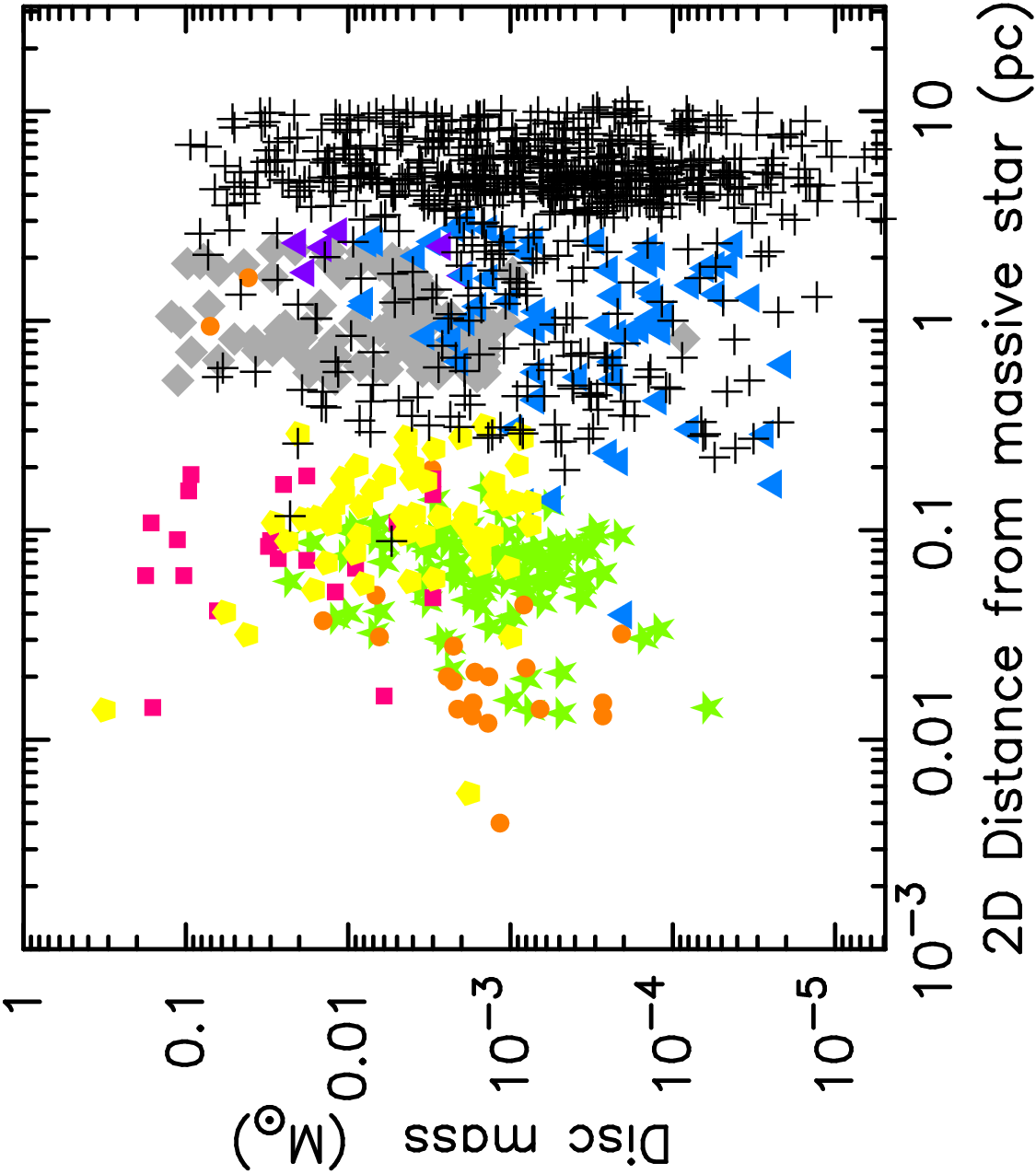}}}
\hspace*{0.3cm}
\subfigure[19, 11, 11, 11 \& 10\,M$_\odot$]{\label{mass_function_low_density-b}\rotatebox{270}{\includegraphics[scale=0.2]{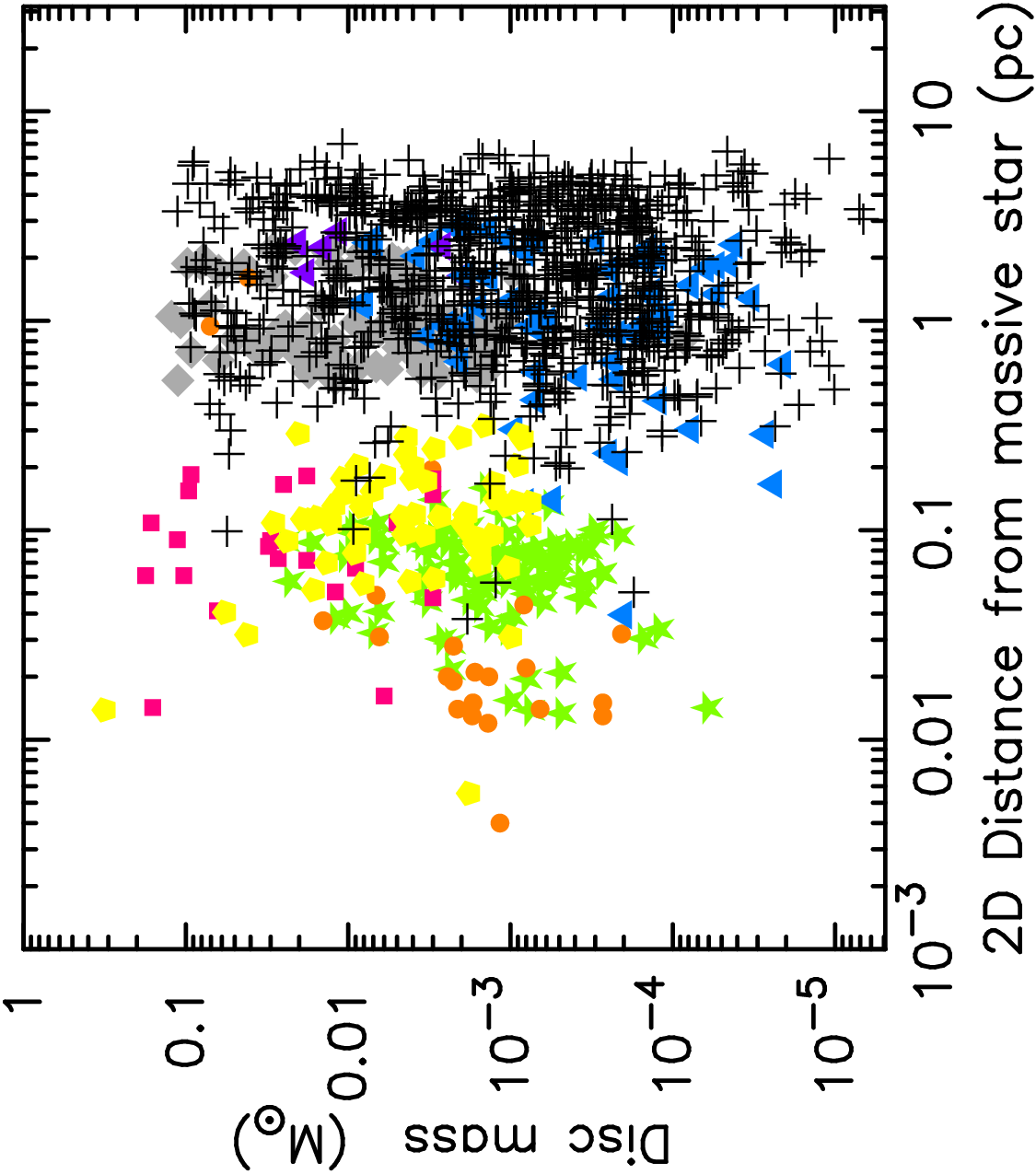}}}
\hspace*{0.3cm}\subfigure[49, 39, 36, 28 \& 27\,M$_\odot$]{\label{mass_function_low_density-c}\rotatebox{270}{\includegraphics[scale=0.2]{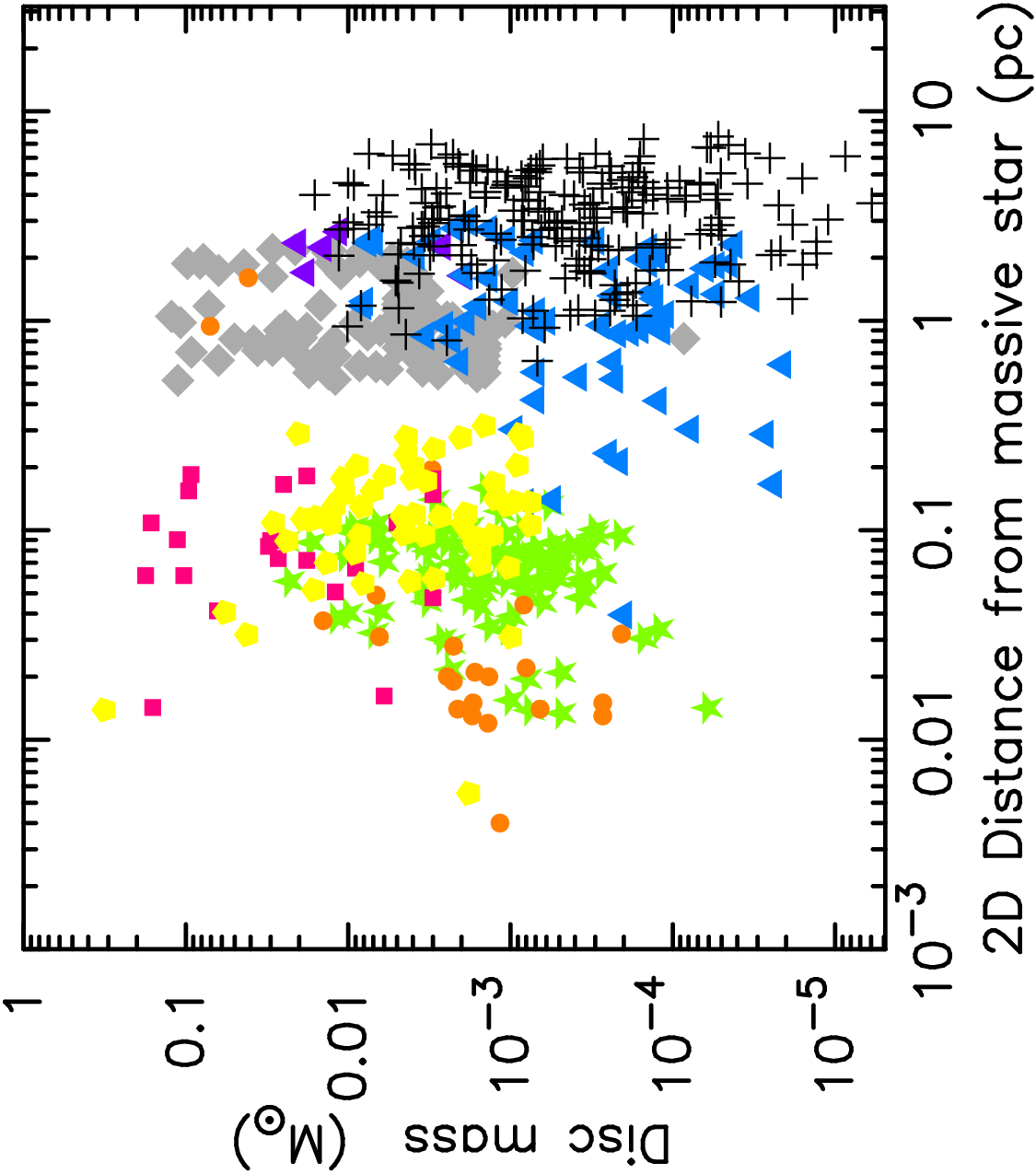}}}
\hspace*{0.3cm}\subfigure[42, 27, 16, 14 \& 13\,M$_\odot$]{\label{mass_function_low_density-d}\rotatebox{270}{\includegraphics[scale=0.2]{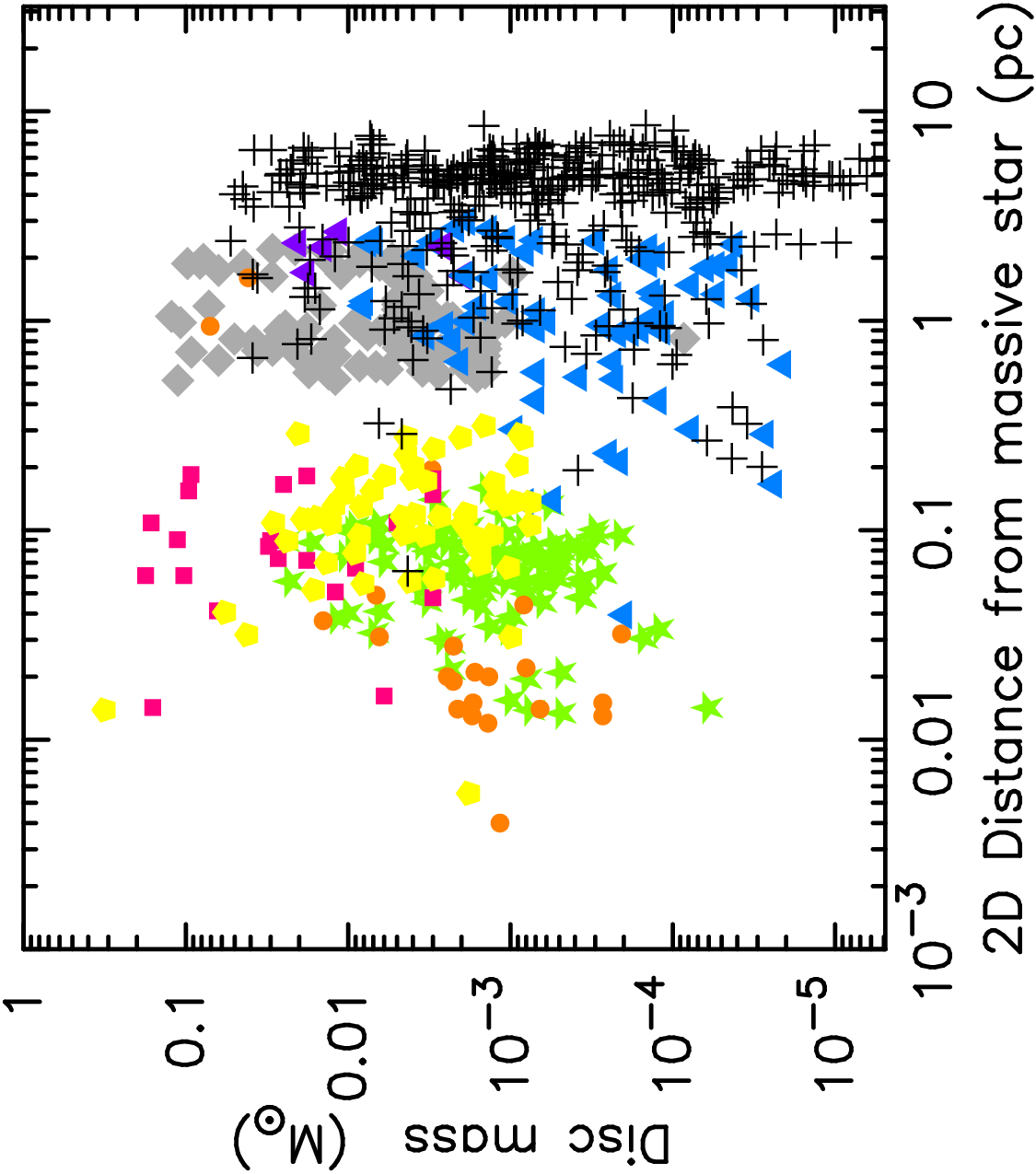}}}
\hspace*{-1.0cm}\subfigure[34, 18, 11, 8 \& 7\,M$_\odot$]{\label{mass_function_low_density-e}\rotatebox{270}{\includegraphics[scale=0.2]{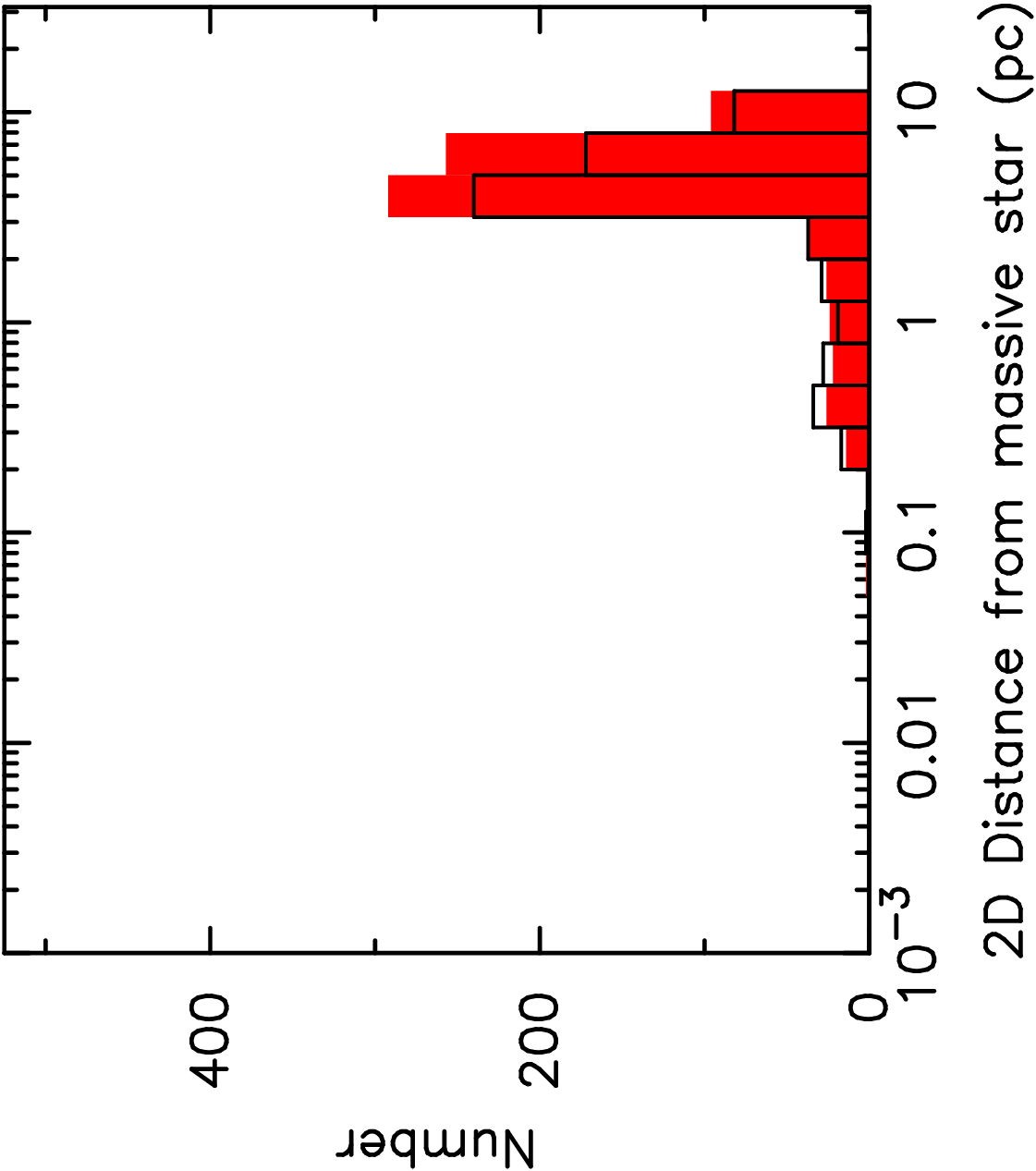}}}
\hspace*{0.3cm}
\subfigure[19, 11, 11, 11 \& 10\,M$_\odot$]{\label{mass_function_low_density-f}\rotatebox{270}{\includegraphics[scale=0.2]{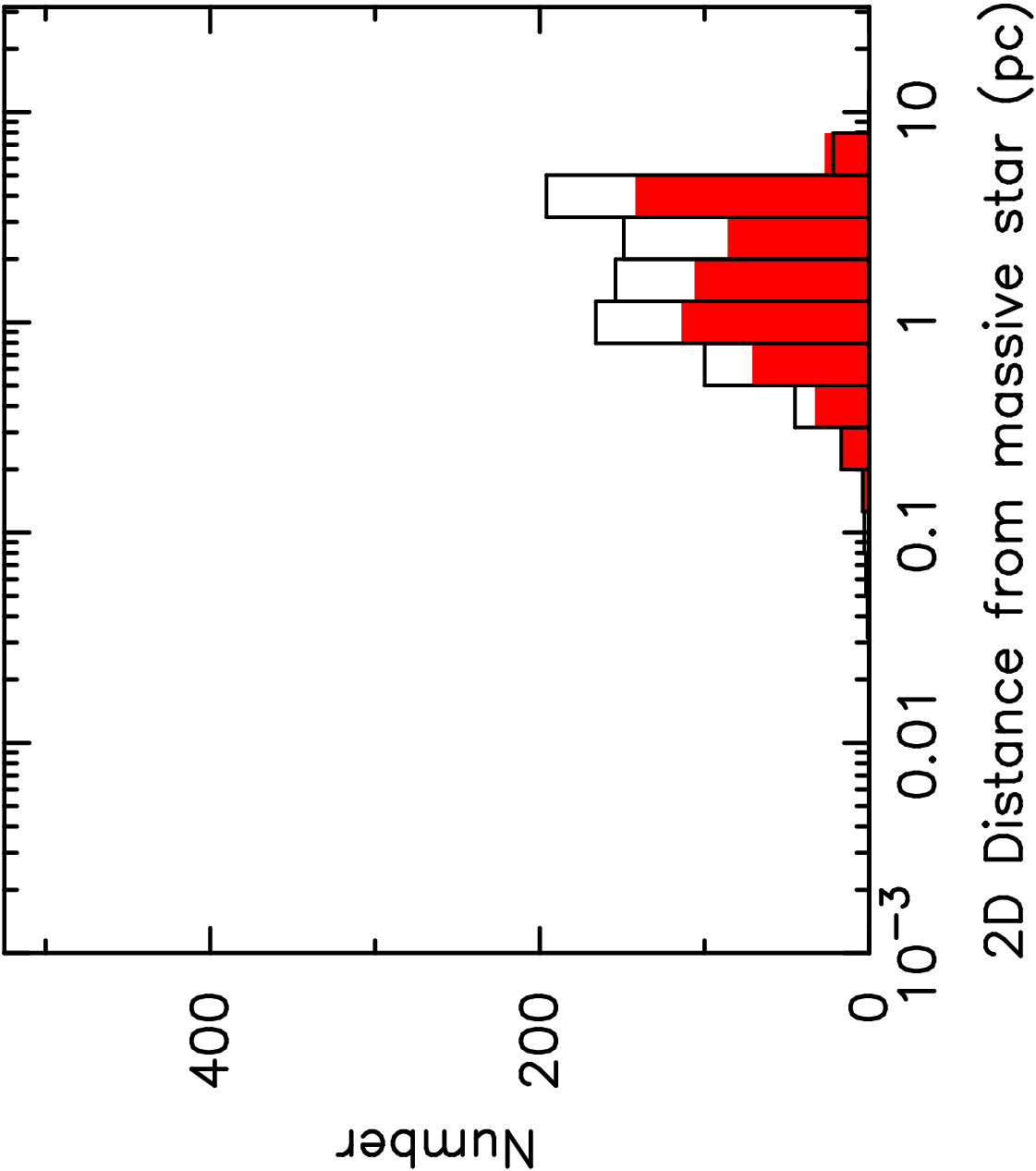}}}
\hspace*{0.3cm}\subfigure[49, 39, 36, 28 \& 27\,M$_\odot$]{\label{mass_function_low_density-g}\rotatebox{270}{\includegraphics[scale=0.2]{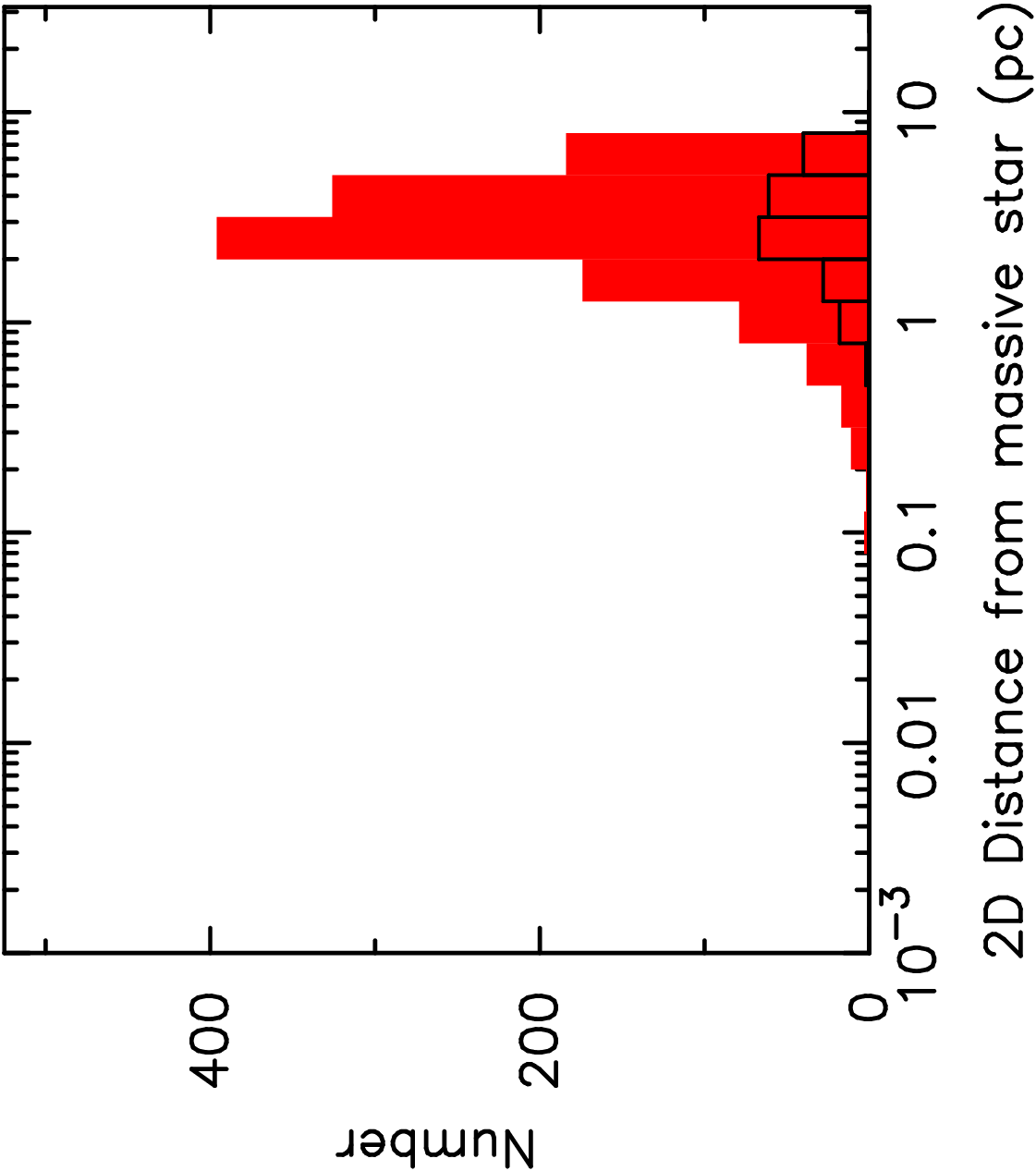}}}
\hspace*{0.3cm}\subfigure[42, 27, 16, 14 \& 13\,M$_\odot$]{\label{mass_function_low_density-h}\rotatebox{270}{\includegraphics[scale=0.2]{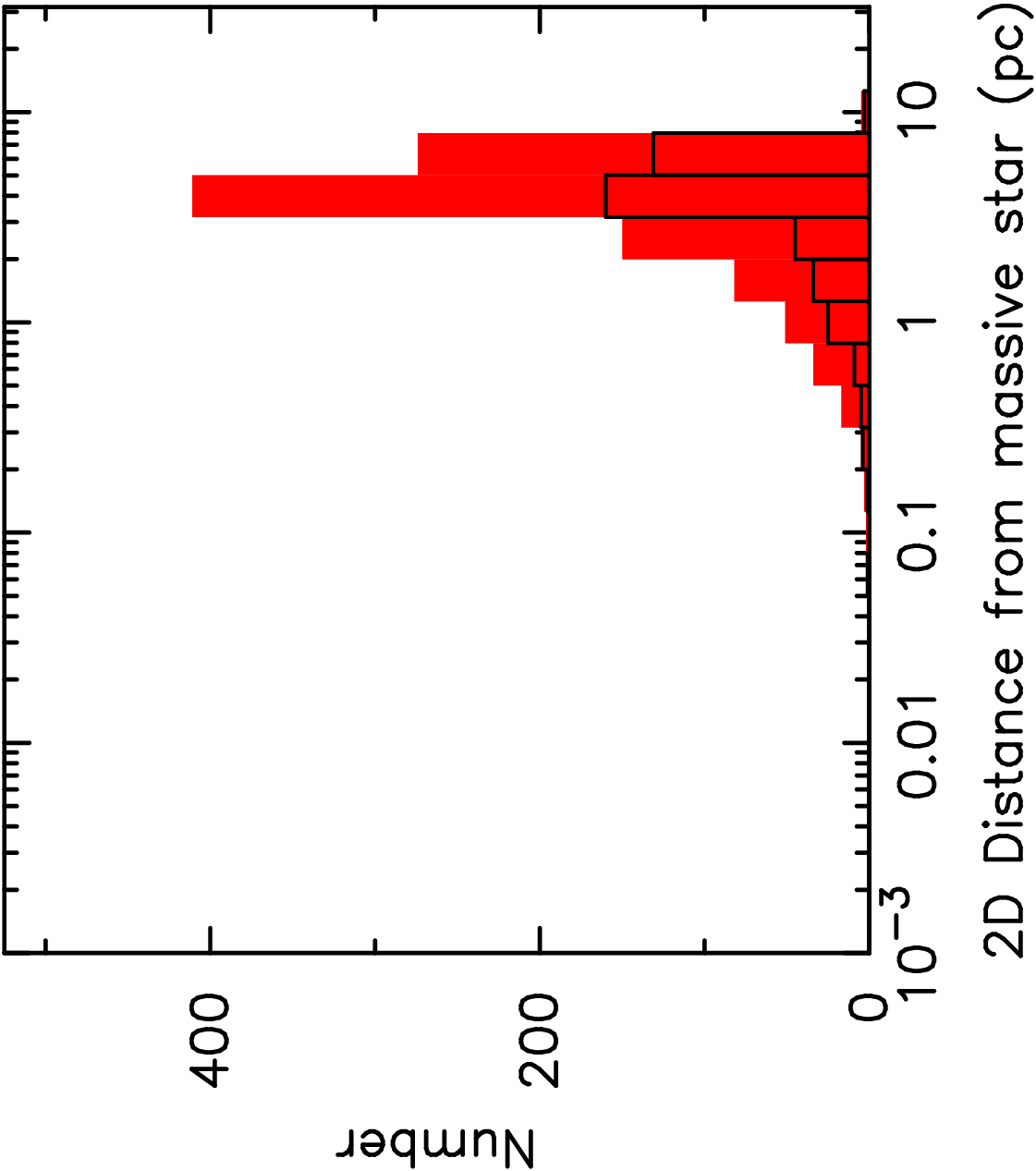}}}  
\caption{The effect of stochastic sampling of the initial mass function on the results for low density star-forming regions ($10^2$\,stars\,pc$^{-3}$) where the initial disc mass is 10\,per cent of the host star's mass, and the distance to the ionising star is calculated in two dimensions. The radius of the star-forming region is 5\,pc, and the age of the star-forming region in each panel is 4\,Myr. {\it Top row:} The plus symbols show the disc mass plotted against the distance to the massive ionising star for each low-mass star with a disc, using only two dimensions to calculate the projected distance. The masses and the respective distances from $\theta^1$~Ori~C of observed protoplanetary discs in the ONC by \citet[][orange circles]{Mann14}, and \citet[][green stars]{Eisner18}; and the Orion Molecular Cloud \citep[][the grey diamonds]{vanTerwisga19} are shown. We also show the masses and projected distances from IRS~2b of discs in NGC\,2024  by \citet[][the raspberry squares]{Mann15} and \citet[][the yellow pentagons]{vanTerwisga20}.  Finally, we show the masses and projected distances from the ionising star $\sigma$~Ori in this eponymous star-forming region \citep{Ansdell17} by the blue triangles. Protoplanetary discs in  $\sigma$~Ori that have CO detections (i.e.\,\,gas content) are shown by the purple triangles. {\it Bottom row:} The number of remaining discs (open histogram) and the number of stars whose discs have been destroyed by photoevaporation (red histogram) as a function of the two dimensional projected distance from the ionising star. }
%Simulation 2, 10\% disc mass, 2D, distance to most massive star. 
\label{mass_function_low_density}
  \end{center}
\end{figure*}

\bibliography{general_ref}
\bibliographystyle{aasjournal}

\end{document}